%% file: main.tex
\documentclass[a4paper,11pt]{article}
\pdfoutput=1
\usepackage{jheppub}

\usepackage[T1]{fontenc}
\usepackage{slashed}
\usepackage{bbm}
\usepackage{bm}
\usepackage{comment}
\usepackage{graphicx}
\usepackage{subcaption}
\usepackage{physics}

\usepackage{feynmp-auto}
\usepackage{cancel}
\usepackage{mathtools}
\usepackage{soul}
\usepackage{todonotes}
\usepackage{empheq}
\usepackage[normalem]{ulem}
\usepackage[dvipsnames]{xcolor}
\allowdisplaybreaks
\usepackage{hyperref}
\hypersetup{
    colorlinks=true,
    linkcolor=blue,
    filecolor=magenta,
    urlcolor=blue,
    citecolor=blue,
}
\newcommand{\cu}[1]{\mathcal{#1}}

\preprint{\vbox{
\hbox{MIT-CTP/6080}
\hbox{DESY-26-101}
}
}

\title{EFT Approaches to Sommerfeld Enhancement and Bound States in Singular Potentials}

\author[a, b]{Arindam Bhattacharya}
\author[b]{, Rashmish K.~Mishra}
\author[c,d]{, and Tracy R.~Slatyer}

\affiliation[a]{Deutsches Elektronen-Synchrotron DESY, Notkestra\ss e 85, 22607 Hamburg, Germany}
\affiliation[b]{Department of Physics, Harvard University, Cambridge, MA 02138, USA}
\affiliation[c]{Center for Theoretical Physics -- a Leinweber Institute, Massachusetts Institute of Technology, Cambridge, Massachusetts 02139, USA}
\affiliation[d]{Radcliffe Institute for Advanced Study at Harvard University, Cambridge, Massachusetts 02138, USA}

\emailAdd{arindam.bhattacharya@desy.de}
\emailAdd{rashmishmishra@fas.harvard.edu}
\emailAdd{tslatyer@mit.edu}

\abstract{
The Sommerfeld enhancement (SE) from long-range interactions, and the related bound state formation rate, can be important non-perturbative inputs to dark matter (DM) annihilation signals. In the presence of singular interaction potentials, the conventional boundary conditions of the Schr\"odinger equation fail, obfuscating the computation of SE, its physical origin, and its relation to bound states in such potentials. In this work, we clarify the origin of SE in singular potentials in a two-fold manner: using the framework of velocity power counting in non-relativistic effective field theory (NREFT), and via position-space regularization of singular potentials at short distances to compute SE and bound states. We illustrate our findings through the case of pseudoscalar mediators interacting with massive Dirac DM. We find that when such a system arises from a UV-complete theory at weak coupling, no SE is generated. However, in the case of a derivatively coupled pseudoscalar, where the interaction is described by a higher-dimension operator in an effective theory, SE (and bound states) can occur for weak couplings if there is a large hierarchy between the cutoff scale of the effective theory and the dark matter mass. This SE is sensitive to the choice of the UV completion of the potential at short distances, but a non-negligible SE can persist even when the UV physics alone would not generate any SE; we elucidate the interplay of UV and IR physics in this case.
}

\begin{document}
\maketitle
\flushbottom
\begin{fmffile}{feyngraph}
\unitlength = 1mm

\section{Introduction}\label{sec:intro}

Long-range interactions between particles can lead to large (non-perturbative) deformations of their wavefunctions at low velocities, even when the underlying interaction is quite weak. In the context of particle-antiparticle annihilation, the enhancement to the annihilation cross-section from an attractive long-range interaction is often termed {\it Sommerfeld enhancement} (SE) \cite{Sommerfeld:1931qaf}. The same term has been used to characterize enhancements to the elastic scattering cross section relative to the first term in the perturbative expansion (obtained from the Born approximation or a tree-level computation in quantum field theory) (e.g.~\cite{Kahlhoefer:2017umn,Agrawal:2020lea}), and both phenomena arise from the non-perturbative deformation of the wavefunction. For the Coulomb-like and Yukawa-like potentials commonly considered in the literature, the SE is also closely related to the presence of bound states (see e.g.~Ref.~\cite{Kamada:2023iol} for an in-depth discussion).

While SE already occurs in Standard Model processes, the possibility that dark matter (DM) might interact either through Standard Model force carriers or new mediators offers a richer range of phenomenology involving SE. Heavy DM particles charged under the $SU(2)_L$ gauge symmetry of the Standard Model can experience SE mediated by the electroweak gauge bosons \cite{Hisano:2003ec, Hisano:2004ds};  more generally, if DM is embedded in a ``dark sector'' of new particles, one or more of these particles may mediate SE and bound-state formation.

In this DM-related context, a number of papers in the literature have explored the possible quantum numbers of these mediators and the operators through which they could couple to the DM. The situation for scalar and vector mediators, which generate manifestly non-singular tree-level Yukawa potentials, is relatively well-understood; however, pseudoscalar and pseudovector mediators generally lead to spin-dependent interactions with apparently singular non-relativistic (NR) potentials (with the magnitude of the potential growing faster than $1/r^2$ at small $r$), and how to model SE in these cases has been more debated (see e.g.~\cite{Bedaque:2009ri,Liu:2013vha,Bellazzini:2013foa,Kahlhoefer:2017umn,Agrawal:2020lea}), as the NR quantum mechanics (QM) calculation typically requires some regulation of the UV/short-distance behavior.

However, Ref.~\cite{Agrawal:2020lea} argued that in the case of elastic scattering in a weakly-coupled theory, for all the cases with singular potentials, applying the first Born approximation in the QM approach gives a finite amplitude that is consistent with the tree-level scattering amplitude computed in the full theory, and in particular does not require any regularization of the singular potential (at this order). Furthermore, both these amplitudes are perturbatively small, suggesting that resummation of higher-order terms associated with the singular potentials is not required.

In this work we seek to clarify the degree to which pseudoscalar- or axial-vector-mediated interactions can contribute to elastic scattering, Sommerfeld-enhanced annihilation, and bound state formation, and to understand the relationship between perturbative calculations in the full QFT and the QM treatment. We will consider two kinds of coupling for the pseudoscalar mediator with the DM fermions: $i)$ a renormalizable non-derivative coupling, and $ii)$ a dimension-5 derivative coupling like the pion. We will refer to the second case as the pseudo-Goldstone case, when the nature of the interaction is relevant. The plan of the rest of the paper is as follows. In Sec.~\ref{sec:litreview} we review relevant literature on the SE and bound-state formation, singular potentials (in particular those generated by pseudoscalar/axial vector exchange), and NR effective field theory (NREFT) methods for tracking velocity scaling. We also provide dimensional analysis arguments for why the SE and bound states are expected to be UV-sensitive for singular NR potentials. We then present three complementary analyses:
\begin{enumerate}
\item Sec.~\ref{sec:eft_pc_se}: From the QFT perspective, we develop a systematic framework, using NREFT, to identify the scaling of the effective interaction terms with coupling and velocity, in the case where the mass of the force carrier can be ignored. This allows us to specifically identify classes of diagrams which are enhanced at low velocity and may require resummation. This approach allows us to identify the diagrams that can source non-perturbative modifications to elastic scattering, large Sommerfeld enhancement to annihilation, and bound-state formation, in the case of scalar/vector exchange, and to show that there are no such comparable contributions from pseudoscalar exchange. This diagrammatic approach circumvents discussing the singular nature of the QM potential, and allows us to understand the presence/absence of SE directly. In the process, one can also identify the requirements on the parameters to give a SE (within the regime of validity of the specific NREFT).
\item Sec.~\ref{sec:bound_sq_well}: In the QM approach, we develop and discuss the generalization of the treatment of elastic scattering given in Ref.~\cite{Agrawal:2020lea}, for the cases of Sommerfeld-enhanced annihilation and bound-state formation. We show that the tree-level matching performed in Ref.~\cite{Agrawal:2020lea} is not generally sufficient to remove dependence on the UV regularization in these cases. However, we confirm that this approach also implies there is no substantial enhancement to annihilation and no bound-state formation from exchange of a pseudoscalar coupled via a weak renormalizable interaction, including in the regime where the NREFT from our first approach becomes invalid.
\item Sec.~\ref{sec:numerics}: In the case where the interaction term is  non-renormalizable and arises from an effective theory valid up to a scale $\Lambda$, the effective IR interaction strength may be enhanced by a large hierarchy between $\Lambda$ and the DM mass. The UV physics above the scale $\Lambda$ may also itself support bound states, Sommerfeld enhancement, etc. We use our QM framework to examine numerically and with analytical approximations the interplay between the long-range potential induced by pseudo-Goldstone exchange and a parameterized short-distance interaction, similarly to the studies in Refs.~\cite{Bedaque:2009ri,Bellazzini:2013foa}, but now separating the UV and IR physics.
\end{enumerate}

We present our conclusions in Sec.~\ref{sec:con}. The appendices describe possible UV completions of the pseudo-Goldstone case, details of the analytic approximation we use to study the Sommerfeld enhancement, some alternative rescalings of the coordinates to demonstrate various aspects of the problem, Sommerfeld results for the $\ell=1$ sector in the renormalizable pseudoscalar case, a discussion of the loop-level potentials, and an explicit example to demonstrate the failure of tree-level matching for Sommerfeld enhancement.

\section{General principles and literature review}
\label{sec:litreview}

\subsection{The Sommerfeld enhancement}

From a diagrammatic perspective, the large SE observed for scalar and vector exchange arises dominantly from a set of Feynman diagrams called \emph{ladder diagrams} where the mediator is exchanged between the
interacting particles. We will expand on this point and describe the diagrams in more detail in Sec.~\ref{sec:eft_pc_se}. These ladder diagrams arise in both scattering and annihilation processes, and correspond to iterating the tree-level $t$-channel mediator exchange. At low relative velocities between the DM particles, the momentum transfer is spacelike and leads to an instantaneous interaction. In this limit, the ladder diagrams can be numerically enhanced (depending on the details of the interaction) and if so, one needs to sum them to all orders in perturbation theory, to obtain reliable predictions about the interaction rates of the DM particles.
More practically, it has been shown that the aforementioned resummation is equivalent to solving the Bethe-Salpeter equation~\cite{Salpeter:1951sz}, which reduces in the NR limit to the time-independent Schr\"odinger equation for
the static potential generated by the tree-level mediator exchange between the DM particles~(see for review~Ref.~\cite{Kaplan:1996xu}). Bound states appear as negative-energy solutions of the same equation.

In the case where the process experiencing SE (such as annihilation) can be approximated as a contact interaction, the SE can be quantified by the ratio $\text{lim}_{r\rightarrow 0}|u(r)/u_0(r)|^2$, where $u(r)$ is the reduced wavefunction in the presence of long-range interactions, $u_0(r)$ is the corresponding reduced wavefunction without interactions, and $r$ is the separation between the interacting particles. The wavefunction can be obtained by solving the appropriate Schr\"odinger equation; this procedure goes beyond the perturbative approach and can thus describe bound states and large Sommerfeld enhancements, albeit it neglects higher-order corrections to the potential (coming from relativistic effects and from  loops beyond the ladder diagrams). One can also perform a partial-wave decomposition and define a SE factor for each partial wave individually; for higher partial waves $\ell > 0$, the reduced wavefunction falls more rapidly to zero at the origin, and so the Sommerfeld enhancement probes higher derivatives of the wavefunction at the origin. In cases where multiple states are coupled by the potential, fully describing the SE requires more than one parameter (see e.g.~\cite{Hisano:2004ds}), but the essential idea that the SE encodes modifications to the wavefunction as $r\rightarrow 0$ (which in turn modify the amplitude for a contact interaction between the particles) still holds.

As the energy of a bound state approaches zero, it induces a resonant enhancement in the SE, leading to an even faster scaling of the SE at low velocity. For a Yukawa potential, for example, the non-resonant SE for annihilation is $\propto 1/v$ down to some saturation velocity, for the $s$-wave ($\ell=0$) partial wave, while on resonances the $s$-wave SE is instead $\propto 1/v^2$ \cite{Arkani-Hamed:2008hhe}. Both bound state formation and significant SE generally require a shared condition on the strength and range of the potential; if this condition is not met (the potential is too weak or too short-range), the system will not support either bound states or large SE, due to relatively small perturbations to the two-particle wavefunction. In this circumstance we also do not expect very large corrections to the elastic scattering cross section, relative to the leading-order perturbative result.

\subsection{Singular potentials}
\label{sec:dim_anal_se}

The Schr\"{o}dinger equation method appears to break down when the NR tree-level potential is {\it singular}: it grows faster than $r^{-2}$ (in magnitude) as $r\rightarrow 0$. In this case, the solutions to the Schr\"{o}dinger equation no longer take a simple universal form at the origin controlled by the centrifugal term,\footnote{For the free-particle solution decomposed into partial waves, and also for potentials with $1/r$ behavior at the origin, the regular solution scales as $r^{\ell+1}$ at small $r$. For $V(r) \sim r^{-(n+1)}, n > 1$, in the $r\to0$ limit, the kinetic term $p^2/2m_\chi  \sim r^{-2}$ and the centrifugal term $\ell(\ell+1) r^{-2}$ become subleading, and the reduced wavefunction $u_{\ell}$ for the $\ell^\text{th}$ angular momentum no longer has the typical $r^{\ell+1}$ scaling.} and there may not be any regular solution; furthermore, attempting to solve for bound states generally leads to the result that the energy of the system is unbounded below.
The mathematical aspects of singular potentials have been explored in great detail very early on, see Ref.~\cite{RevModPhys.43.36} for a review.

The presence of singular potentials and their irregular short-distance behavior might na\"ively be somewhat surprising, especially in cases when the underlying UV physics is described by a perturbative renormalizable QFT whose short-distance behavior is under good control. The root issue is that the construction of an effective low-energy QM description from an underlying QFT requires an upper limit on the momentum transfer: the QM description requires going to the NR limit and for an NREFT, the magnitude of the momentum transfer must be bounded above by the rest mass of the DM particles. When Fourier transformed to position space, this limits the validity of the resulting potential at distances smaller than the inverse rest mass~\cite{Lepage:1997cs} (since smaller distances correspond to larger momenta, which are not in the NR limit). This introduces a cutoff to the potential, of the order of inverse rest-mass. In principle this is an issue regardless of whether the potential is singular or not. However, one can show by dimensional analysis that for non-singular potentials, the characteristic scale of the wavefunction deformation due to the potential (corresponding to the length scale associated with the bound states) is large compared to the cutoff scale, justifying neglect of the short-range physics.

For this discussion, it suffices to take the force mediator between the DM particles to be massless.\footnote{For a non-zero mediator mass, the potential has a Yukawa suppression, which makes the size of the bound state smaller. There can also be additional prefactors that depend on the ratio of the mediator mass to the DM mass.} The only relevant scale is then the mass $m_\chi$ of DM. Restricting to purely radial potentials $V(r)$, the QM Hamiltonian is
\begin{align}
    H = -\frac{\nabla^2}{m_{\chi}} + V(r)\:.
\end{align}
We parametrize the potential as
\begin{align}
    V(r) = -\frac{\alpha}{r(m_{\chi}r)^n}\:,
\end{align}
where $\alpha$ is a positive dimensionless coupling and we have taken the potential to be attractive (negative) so it can support a bound state. The quantity $n$ parametrizes the singular nature of the potential: for $n > 1$, the potential is singular, whereas for $n=0$ it is regular (of Coulomb type). $n=1$ corresponds to the marginal case where the potential scales in the same way as the centrifugal term.

If this potential supports bound states, one can estimate the size $r_b$ of this bound state by balancing the contributions from the kinetic and the potential terms, and using that the typical momentum is of order $1/r_b$, by the uncertainty principle. We get
\begin{align}
    &\frac{\nabla^2}{m_{\chi}}
    \sim
    \frac{1}{m_{\chi}r_{b}^2} \sim |V(r_{b})| = \frac{\alpha}{r_b (m_{\chi}r_b )^n}\:,
\end{align}
which gives
\begin{align}
    r_{b}
    \sim
    (1/m_\chi)\,
    \alpha^{\frac{1}{n-1}}\: .
\end{align}
Note that $m_{\chi}\sim r_{c}^{-1}$ is the inverse Compton length of the DM particle. To describe the dynamics at length scales shorter than the Compton length, a QM treatment breaks down, since one can now probe energies of the order of the rest mass. Thus, to have a bound state that is fully described by QM, one needs $r_b/r_c \gg 1$ which translates to the condition
\begin{align}
    \alpha^{\frac{1}{n-1}} \gg 1\:.
\end{align}
For perturbative couplings $\alpha\ll1$, this condition can only be satisfied if $n<1$. In particular, for the Coulomb/Yukawa case ($n=0$), there is a $1/\alpha$ enhancement to the ratio $r_b/r_c$, which keeps the size of the bound state larger than the Compton length and justifies the QM treatment.

For $n > 1$, we see that we generally expect bound states and large potential effects to be confined to the relativistic region (where the QM treatment is invalid) unless $\alpha \gtrsim 1$. If $\alpha$ corresponds to an effective coupling in the QFT used to derive the potential, so the large$-\alpha$ regime corresponds to the QFT being strongly coupled, then one has to be cautious in matching the amplitude from the QFT to the NREFT, since one cannot  just use the tree-level perturbative QFT amplitude. In other words, in general we expect the QM calculation of effects from a tree-level singular potential to be internally inconsistent, signaling sensitivity to UV physics that is not captured by the QM approach. Of course, this conclusion can be avoided if the potential is not actually singular in the regime of interest. This could occur due to relativistic corrections \cite{Bastai:1963,Mangin-Brinet:2003vmv,Dorkin:2007wb,Hall:2014dua}, or if there is another intermediate scale -- such as the mass of a light mediator with $m_\phi \ll m_\chi$ (e.g.~Refs.~\cite{Xu:2021daf,Coy:2022cpt}) -- which renders the potential non-singular.

The question of how to deal with singular potentials consistently in a QM framework has been studied in the nuclear theory literature (e.g.~\cite{Beane:2000wh,Long:2007vp,Barford:2002je}), and a common approach is to replace the singular potential within some short-distance region ($r < a$) with an appropriate QM potential which is {\it not} singular, or equivalently with boundary conditions at an appropriate matching radius. In principle, the observable consequences of this potential may either be computed from first principles (given a UV theory) or treated as a phenomenological input (or set of phenomenological inputs). For scattering problems, the low-energy scattering phase sourced by the UV potential (in each partial wave) is typically treated as a phenomenological input. The overall scattering phase shift is independent of the details of the UV potential so long as the scattering energy is sufficiently low and so it does not probe the UV region.

\subsection{Tree-level potentials from pseudoscalar and axial vector interactions}
\label{sec:potentials}

We will be particularly interested in the singular potentials arising from pseudoscalar and axial vector couplings. We review the key aspects of these potentials here.

In general, we can write the leading-order NR potential in the form (following \cite{Bedaque:2009ri}):
\begin{align} V(r) = V_0(r) + V_C(r) \vec{S}_1 \cdot \vec{S}_2 + V_T(r) (3 (\hat{r} \cdot \vec{S}_1) (\hat{r} \cdot \vec{S}_2) - \vec{S}_1 \cdot \vec{S}_2),\end{align}
where $\vec{S}_1$ and $\vec{S}_2$ denote the spins of the two interacting fermions. We will focus on the cases of renormalizable pseudoscalar and non-renormalizable pseudo-Goldstone couplings, and briefly discuss the axial vector coupling.

As discussed in Refs.~\cite{Agrawal:2020lea,Parikh:2020ggm}, the primary difference between the potentials for $\chi \chi$ and $\bar{\chi}\chi$ scattering is the presence of a $s$-channel contact interaction in the latter case, which contributes delta-function terms to $V_0(r)$ and $V_C(r)$.

For $\chi\chi$ scattering, the potentials for all these cases are summarized in Ref.~\cite{Parikh:2020ggm} and can be decomposed solely in terms of $V_C(r)$ and $V_T(r)$ terms (i.e. for these cases $V_0(r) = 0$). The coefficients taken from that paper are given by:

\begin{enumerate}
\item Pseudoscalar ($\mathcal{L}  \supset i g  \phi \bar{\chi} \gamma^5 \chi$):
\begin{align}
    V_C(r) & = \alpha \frac{e^{-m_\phi r}}{3 m_\chi^2} \left[ -4\pi \delta^{(3)}(\vec{r})   +  \frac{m_\phi^2}{r} \right], \quad V_T(r)  = \alpha \frac{e^{-m_\phi r}}{m_\chi^2 r^3} \left(1 + m_\phi r + \frac{m_\phi^2 r^2}{3}\right).
    \label{eq:potential-pseudoscalar}
\end{align}
\item Pseudo-Goldstone ($\mathcal{L} \supset \frac{g}{\Lambda}  \bar{\chi}\gamma^\mu \gamma^5 \chi \partial_\mu \phi$):
\begin{align} V_C(r) &  = \frac{G e^{-m_\phi r}}{m_\chi^2} \left[ -4\pi \delta^{(3)}(\vec{r})   +  \frac{m_\phi^2}{r} \right], \quad V_T(r)  = \frac{3 G e^{-m_\phi r}}{m_\chi^2 r^3}  \left(1 + m_\phi r + \frac{m_\phi^2 r^2}{3}\right)\ . \label{eq:potential-goldstone} \end{align}
\item Axial vector ($\mathcal{L} \supset g  A_\mu \bar{\chi}\gamma^\mu \gamma^5 \chi$):
\begin{align}
    V_C(r) & = 4\alpha e^{-m_A r} \left[-\frac{1}{r} +  \frac{1}{3 m_A^2} \left( -4\pi \delta^{(3)}(\vec{r})   +  \frac{m_A^2}{r} \right)\right] =- 4\alpha e^{-m_A r} \left[\frac{(2/3)}{r}  + \frac{4\pi}{3 m_A^2}  \delta^{(3)}(\vec{r})\right] , \nonumber \\
      V_T(r)  & = 4 \alpha \frac{e^{-m_A r}}{m_A^2 r^3} \left(1 + m_A r + \frac{m_A^2 r^2}{3}\right).
    \label{eq:potential-pseudovector}
\end{align}
\end{enumerate}
Here we have defined $\alpha \equiv \frac{g^2}{4 \pi}$, and for the pseudo-Goldstone case, $G\equiv \frac{g^2 m_\chi^2}{3 \pi \Lambda^2}$.

Note that under this sign convention, a positive potential corresponds to a repulsive interaction and vice versa for a negative potential. Note also that in the axial vector case, the UV completion of Ref.~\cite{Agrawal:2020lea} implies that $g \propto (m_A/m_\chi)$, so the axial vector is intrinsically weakly coupled and the potential does not diverge as $m_A \rightarrow 0$. We expect this to be a generic feature of viable UV completions, as from the EFT perspective, taking $m_A\rightarrow 0$ while keeping $g$ finite would imply a massless vector coupled to an anomalous current, i.e.~we would have gauged an anomalous symmetry.

For the $\chi\bar{\chi}$ case, the additional $s$-channel contact-interaction contributions are given as

\begin{enumerate}
    \item Pseudoscalar
    \begin{align}
        \Delta V(r) = \delta^3(\vec{r}) \frac{g^2}{4m_\chi^2 -m_\phi^2} \left(-2\vec{S}_{1} \cdot \vec{S}_{2} + \frac{1}{2}\right)\label{eq:chibarchi_pseudoscalar}
    \end{align}
    \item Pseudo-Goldstone
    \begin{align}
        \Delta V(r) = \delta^3(\vec{r}) \frac{4m_\chi^2}{\Lambda^2} \frac{g^2}{4m_\chi^2 -m_\phi^2} \left(-2\vec{S}_{1} \cdot \vec{S}_{2} + \frac{1}{2}\right) \label{eq:chibarchi_goldstone}
    \end{align}
    \item Axial Vector
    \begin{align}
        \Delta V(r) = \delta^3(\vec{r}) \left(\frac{4m_\chi^2}{m_A^2}-1\right) \frac{g^2}{4m_\chi^2 -m_A^2} \left(-2\vec{S}_{1} \cdot \vec{S}_{2} + \frac{1}{2}\right)\:. \label{eq:chibarchi_axial}
    \end{align}
\end{enumerate}
The results for the pseudoscalar and axial vector cases were derived in~\cite{Agrawal:2020lea}.

These potentials can be projected onto the space of states of fixed total angular momentum but varying spin and orbital angular momentum. Here we follow the discussion in Ref.~\cite{Agrawal:2020lea}. The operator $3(\hat{r} \cdot \vec{S}_1) (\hat{r} \cdot \vec{S}_2) - \vec{S}_1 \cdot \vec{S}_2$ couples states with different orbital angular momentum, but the total angular momentum (characterized by quantum numbers $J^2$ and $J_z$) remains conserved. We can work in the basis of states characterized by $(j,m_j, \ell, s)$ and defined by:
\begin{itemize}
    \item Spin-singlet: $|j,\sigma,j,0\rangle_{j,m_j,\ell,s}$
    \item Spin-triplet: $|j,\sigma,j+1,1\rangle_{j,m_j,\ell,s}$, $|j,\sigma,j,1\rangle_{j,m_j,\ell,s}$, $|j,\sigma,j-1,1\rangle_{j,m_j,\ell,s}$
\end{itemize}
We note that the spin-singlet state is an eigenvector of $\vec{S}_1\cdot \vec{S}_2 = (1/2)\left[ (\vec{S}_1 + \vec{S}_2)^2 - \vec{S}_1^2 - \vec{S}_2^2\right]$ with eigenvalue $-(1/2)(3/2) = -3/4$, and the spin-triplet state is an eigenvector with eigenvalue $(1/2) (2 - 2(1/2)(3/2)) = 1/4$. Note that this implies that the additional $s$-channel contact-interaction contributions given in Eqs.~\eqref{eq:chibarchi_pseudoscalar}-\eqref{eq:chibarchi_axial} vanish for the spin-triplet state (this can be explained by conservation of total angular momentum, as the contact interaction selects the $L=0$ state and the $s$-channel mediator in this case is spin-zero).

For the more complicated operator $\mathcal{O}_T \equiv 3(\hat{r} \cdot \vec{S}_1) (\hat{r} \cdot \vec{S}_2) - \vec{S}_1 \cdot \vec{S}_2$, Ref.~\cite{Agrawal:2020lea} shows that:
\begin{align} & \mathcal{O}_T |j,\sigma,j,0\rangle_{j,m_j,\ell,s} = 0, \nonumber \\
& \mathcal{O}_T |j,\sigma,j,1\rangle_{j,m_j,\ell,s} = (1/2) |j,\sigma,j,1\rangle_{j,m_j,\ell,s}, \nonumber \\
&  \mathcal{O}_T \begin{pmatrix}  |j,\sigma,j-1,1\rangle_{j,m_j,\ell,s} \\ |j,\sigma,j+1,1\rangle_{j,m_j,\ell,s} \end{pmatrix} = \begin{pmatrix} - \frac{(j-1)}{2(2j+1)}  & \frac{3 \sqrt{j(j+1)} }{2(2j+1)} \\ \frac{3 \sqrt{j(j+1)} }{2(2j+1)} & -\frac{(j+2)}{2 (2j+1)} \end{pmatrix} \begin{pmatrix}  |j,\sigma,j-1,1\rangle_{j,m_j,\ell,s} \\
|j,\sigma,j+1,1\rangle_{j,m_j,\ell,s}\end{pmatrix}.  \end{align}

We will focus in this work on the lowest-$\ell$ states. For $\ell=0$, we see that the $s=0$ state experiences a scalar potential given by $V(r) = -(3/4) V_C(r)$. The $1/r$ part of this potential is attractive for the pseudoscalar cases and repulsive for the axial vector case (note this reflects a difference from Ref.~\cite{Bedaque:2009ri}, which says the potential should be repulsive in the pseudo-Goldstone case).  However, we see the effective coupling in these cases (for the $1/r$ potential) is controlled by $\alpha (m_\phi/m_\chi)^2$, $G(m_\phi/m_\chi)^2$ and $\alpha$ for the pseudoscalar, pseudo-Goldstone and axial vector cases respectively, and for a non-singular Yukawa potential with range $1/m_\phi$, a necessary condition for bound states and/or a large Sommerfeld enhancement is that the coupling should be $\gtrsim m_\phi/m_\chi$ (e.g.~\cite{Arkani-Hamed:2008hhe}). We see that this would require $\alpha, G \gtrsim m_\chi/m_\phi \gtrsim 1$ in the pseudoscalar/pseudo-Goldstone cases, calling the perturbative expansion into doubt.\footnote{In Sec.~\ref{sec:numerics} we will discuss the regime where $g$ is small but $m_\chi/\Lambda \gg 1$, allowing for $G \gtrsim 1$. See also appendix~\ref{app:UV-completion-Goldstone-interaction} for a discussion of UV completions that can allow this hierarchy.} For the axial vector case, in the setup of Ref.~\cite{Agrawal:2020lea} which we have employed, $\alpha$ is determined by a Yukawa coupling multiplied by $m_A/m_\chi$, so the condition for substantial Sommerfeld enhancement or bound-state formation again requires a  (Yukawa) coupling that is not perturbatively small. Thus even if the interaction is attractive, we do not expect formation of bound states or appreciable Sommerfeld enhancement for weakly-coupled theories of this type.

The $j=1,\ell=1,s=1$ state also experiences a scalar potential, given by $V(r) = (1/4) V_C(r) + (1/2) V_T(r)$. However, in all the cases given above, the $1/r^3$ piece of the potential is purely repulsive, and as discussed above we do not expect a large enhancement or bound-state formation from the $1/r$ contributions. There is also the $j=0, \ell=1, s=1$ channel, which does not have any $\ell$ mixing and experiences a potential $V(r)=(1/4) V_C(r) - V_T(r)$. This is an attractive singular potential, and the simplest case for which we observe an attractive $1/r^3$ potential. In this work, our emphasis is on the behavior of SE in the coupled-channel case that contains the $\ell=0$ mode; however, our arguments as to the absence of SE in the weakly coupled regime also  extend to this attractive single channel, as we show explicitly in appendix~\ref{app:p_wave_sing_chan}.

The $j=1,s=1$ state does have $\ell=0$ and $\ell=2$ components which mix under the action of the $\mathcal{O}_T$ operator, giving rise to a matrix potential coupling these states, of the form:
\begin{align}
V(r) & = (1/4) V_C(r)
+
    \begin{pmatrix}
        0 & \frac{ \sqrt{2} }{2} \\ \frac{ \sqrt{2} }{2}  & -\frac{1}{2}
    \end{pmatrix}
    V_T(r) = \frac{1}{4}
    \begin{pmatrix}
        V_C(r) & \sqrt{8} V_T(r) \\ \sqrt{8} V_T(r) & V_C(r) - 2 V_T(r)
    \end{pmatrix}
    \label{eq:potential-longRange}
\end{align}
where the first row/column corresponds to the $\ell=0$ state and the second to the $\ell=2$ state.
(Any $V_0(r)$ term, as is relevant in the case of $\bar{\chi}\chi$ scattering, can be multiplied by the identity matrix and added directly to $V(r)$.) This is the same form as given in Ref.~\cite{Bedaque:2009ri}, up to an overall factor of $1/4$. This factor of $1/4$ simply reflects a notation difference in the definition of $V_T(r)$ and $V_C(r)$; we have defined them as coefficients of operators written in terms of $\vec{S}_1=(1/2)\vec{\sigma}_1$, $\vec{S}_2=(1/2)\vec{\sigma}_2$, whereas Ref.~\cite{Bedaque:2009ri} defines operators in terms of $\vec{\sigma}_1$ and $\vec{\sigma}_2$, such that the operators are a factor of 4 larger and their coefficient functions are a factor of 4 smaller.

We see that among these examples, the only situation where a bound state might be induced by the $1/r^3$ part of the potential is in the mixed $\ell=0$ and $\ell=2$ case (at least for $\alpha, G \lesssim 1$), so we will focus on this example in this work.

\subsection{Modeling dark-sector Sommerfeld enhancement from singular potentials}

Several approaches to tackle singular potentials, in the specific case of dark sector interactions mediated by pseudoscalars/axial vectors, have already been proposed~\cite{Bedaque:2009ri,Bellazzini:2013foa,Agrawal:2020lea,Parikh:2020ggm,Bottaro:2023wjv}. The possibility of large Sommerfeld enhancement for DM annihilation due to a singular potential (and specifically for the case of a Goldstone  mediator) was studied early on in Ref.~\cite{Bedaque:2009ri}. That work first studied a simple $1/r^3$ attractive potential with normalization controlled by a dimensionless coupling $G$. To regularize the short-distance singularity, that study replaced the singular potential with a finite spherical well within some radius $R$, and related the depth $V_0$ and range $R$ of the short-range potential by the requirement that the low-energy scattering amplitude should be invariant under changes in $R$ (accompanied by appropriate changes in $V_0$). The parameter $V_0$, controlling the short-range physics, was then treated as an adjustable parameter governing the UV behavior of the theory. This prescription led to a large Sommerfeld enhancement with a characteristic resonance pattern as a function of $G$, indicating the presence of bound states; this resonance pattern varied as a function of the short-distance coefficient $V_0$. That analysis focused primarily on large values of $G \gtrsim 1$; $V_0$ was also of the order at which bound states would be expected purely from the short-range spherical well. The authors of that work also calculated the potential coupling different partial waves for a pseudo-Goldstone mediator, regulated at short distances with a spherical well purely in the $s$-wave channel. In this case, they found large non-perturbative corrections (both enhancing and suppressing the signal, at different parameter points), but again focused primarily on the case with a dimensionless coupling $\ge 1$, and with the short-distance spherical well being close to the threshold to support bound states on its own.

Ref.~\cite{Bellazzini:2013foa} explored a broader range of NR potentials, including terms matching those generated by pseudoscalar or pseudovector exchanges. The authors discussed the approach of zeroing out the NR potential for $r < a$, and augmenting it with a series of contact interaction potentials (built from delta functions and derivatives of delta functions) to capture the unknown UV physics. However, in practice, that work also regulated the potential by replacing it with a spherical well at short distances. They also identified the need for a ``wavefunction renormalization'' step, which consisted of rescaling the Sommerfeld enhancement to ensure that it approaches 1 at relativistic speeds, once the short-distance spherical well was fixed. Again, this work found that large enhancements could be produced, but for its numerical examples focused on an attractive $1/r^3$ potential without the spin-dependent structure present in the pseudoscalar case, and did not attempt to separate the contributions to the enhancement from UV vs IR physics.

Ref.~\cite{Agrawal:2020lea} focused on the case of non-perturbative enhancements to elastic scattering, as relevant to DM self-interactions, and argued that pseudoscalar- and pseudovector-mediated interactions are well described by perturbation theory (i.e.~we expect no large Sommerfeld enhancement). The distinction with earlier work was that rather than treat the normalization of the short-distance regulator as a free parameter, the authors of Ref.~\cite{Agrawal:2020lea} required that the short-distance scattering amplitude match that predicted from perturbation theory. They also considered only weak dimensionless couplings for the IR physics, and found that in these circumstances, the long-range potential does not lead to a significant deformation of the wavefunction. While this work did not focus on annihilation or bound-state formation, one would expect that if the wavefunction is well-described by small perturbations to a plane wave, neither significantly enhanced annihilation nor bound-state formation would be present. A key aspect of this argument, extended to other singular potentials in Ref.~\cite{Parikh:2020ggm}, is that the first Born approximation to the scattering amplitude in a singular potential is generally finite for potentials derived from QFT, and so to compute the first-order scattering amplitude it is not necessary to regulate the short-distance physics or worry about dependence on this regulator. Provided there is no large enhancement that appears for the first time at higher order, the QM calculation becomes equivalent to simply calculating the tree-level scattering amplitude in the low-energy limit of the UV theory.

Recent work focusing on two-fermion exchange has provided another avenue where apparently-singular potentials appear \cite{Xu:2021daf,Coy:2022cpt,Ferrante:2025lbs}. In this case, the potential can take the form $V(r) \propto -1/(\Lambda^2 r^3)$ or $V(r) \propto -1/(\Lambda^4 r^5)$ for some substantial range in $r$. Ref.~\cite{Ferrante:2025lbs} finds that even in the absence of a short-range spherical well or similar regulator, there can be a large Sommerfeld enhancement in these models, but only in the case where there is a hierarchy $\Lambda \ll m_\chi$ (the further hierarchy $\Lambda \gtrsim m_\chi v_\text{rel}$ is required for the EFT to be valid). This hierarchy may imply further Sommerfeld enhancement (and potentially bound state formation) due to physics at the scale $\Lambda$ that has been integrated out, e.g.~due to on-shell exchange of mediators with mass $\sim \Lambda \ll m_\chi$.

As well as these works focusing specifically on DM scenarios with novel forces, there have been many studies on interactions mediated by {\it composite} pseudoscalars in the context of the Standard Model, where pions can play the role of the pseudoscalar mediator. In this case there is a natural UV cutoff scale associated with the QCD scale, and so we do expect new physics (beyond the pseudoscalar effective theory) at sufficiently small radial separations. Even in the absence of such new physics, relativistic corrections are expected to soften the singular behavior of the potential at short distances \cite{Bastai:1963}. The possible presence of relativistic bound states, supported by pure pseudoscalar exchange, has been explored in the context of the Bethe-Salpeter equation \cite{Dorkin:2007wb} and Light-front Dynamics \cite{Mangin-Brinet:2003vmv}, with the conclusion that ``there is no relativistic
bound state in a deuteron-like system with pure pseudoscalar exchange'', although the numerical calculation becomes unstable for sufficiently high coupling.

\section{Enhancement of scattering with NREFT and power counting}\label{sec:eft_pc_se}
In this section, we illustrate how the NR limit of DM interactions can be codified into an EFT treatment, enabling one to understand the presence or absence of Sommerfeld enhancement for low energy DM scattering. We borrow heavily from the existing NREFT treatment for QCD bound states in NRQCD~\cite{Caswell:1985ui,Bodwin:1994jh,Labelle:1996en,Grinstein:1997gv,Luke:1997ys,Luke:1996hj,Griesshammer:1997wz,Brambilla:1999xf}, which has also been applied to DM physics~\cite{Beneke:2012tg,Hellmann:2013jxa,Biondini:2018ovz,Biondini:2021ccr,Biondini:2021ycj,Baumgart:2017nsr,Bauer:2014ula,Baumgart:2015bpa,Baumgart:2014saa,Ovanesyan:2014fwa,Baumgart:2014vma,Biondini:2023ksj}. These applications of NREFT to DM phenomenology have often focused on providing resummation-improved estimates for annihilation and bound state formation, with the long-range interaction potential being primarily Coulombic (or Yukawa-like) in nature, mediated by vector or scalar exchange. These resummations captured the enhancements coming from large Sudakov logs (large logs of ratio of mediator to DM mass) or small DM velocity, and the focus was primarily on Coulomb/Yukawa-like interactions. In this work, we emphasize the role of velocity power counting in a general DM NREFT, in order to explain the presence (absence) of the Sommerfeld enhancement. To do so, we shall be completely general about the nature of the DM-mediator interaction in the UV, to generalize beyond the well studied case of Coulomb/Yukawa-like interactions. While a derivation of Sommerfeld enhancement in the Coulomb potential using NREFT Green's functions has been done earlier~\cite{Beneke:2013jia,Binder:2026fwe}, to our knowledge, such a general analysis has not been done in the literature thus far.

In what follows, we shall first review the setup of an NREFT to explain the DM scattering and thereafter demonstrate how velocity power counting explains the presence (absence) of Sommerfeld enhancement in pure scalar (pseudoscalar) mediator exchange. We emphasize that the construction of the NREFT is not new (see Refs~\cite{Biondini:2021ccr,Biondini:2023ksj} which had constructed the NREFT for scalar and pseudoscalar mediators in position space and applied it to indirect detection); we simply rederive the NREFT Lagrangian in label-space formalism to emphasize the role of velocity power counting, and hence explicate when and how SE arises in NREFTs.

NREFTs describe the low energy physics of DM particles where the dispersion relation $E = p^2/2m_\chi = \frac{1}{2}m_\chi v^2$ holds. Thus, the typical momentum of DM particles in this regime is $p = m_\chi v \ll m_\chi$, i.e. their speed is NR, and the energy scale hierarchy goes as
\begin{align}
    m_\chi \:\: \gg \:\:   m_\chi v = p \:\: \gg \:\:  m_\chi v^2 = p^2/m_\chi = 2E\ .
\end{align}
NREFTs are similar to heavy quark EFTs in the aspect that they have a $1/m_\chi$ expansion, but differ in that they additionally power count operators in the velocity parameter $v$~\cite{Manohar:1997qy}. We shall illustrate their use for DM scattering for scalar and pseudoscalar mediators, and review what is known for vector couplings and other mediators. Technical details about the matching of the underlying UV QFTs to the corresponding NREFTs can be reviewed in Refs.~\cite{Biondini:2021ccr,Luke:1996hj,Luke:1997ys}.

\subsection{Scalar mediator}
We first illustrate the use of NREFT velocity power counting for Sommerfeld enhancement for  fermionic DM interacting via a massless scalar mediator. Specifically, in the UV we have the following renormalizable QFT Lagrangian
\begin{align}\label{eq:l_uv}
    \mathcal{L}_{\rm scalar}^{\rm UV} = \overline{\chi}(i\slashed{\partial}-m_{\chi})\chi + g \overline{\chi}\chi \phi + \frac{1}{2}(\partial_\mu \phi)(\partial^{\mu} \phi)\ ,
\end{align}
where $\chi$ denotes a Dirac dark fermion and $\phi$ denotes the massless scalar mediator. Note that parity is a well-defined symmetry for $\cu{L}_{\rm scalar}^{\rm UV}$ and $\phi$ is a $P$ even field. Inclusion of a mediator mass $m_{\phi}$ introduces an additional scale into the relevant energy scales of the NREFT, and necessitates a comparison between two power counting parameters: velocity $v$ and $\epsilon_{\phi} \equiv m_{\phi}/m_{\chi}$. This complicates the matching of the NREFT~\cite{Biondini:2021ccr,Biondini:2023zcz}, and we leave a full exploration of the velocity power counting including a mediator mass to future work. In what follows, we shall treat only the massless mediator case\footnote{The massless mediator assumption can be rephrased as the statement that $m_{\phi}$ is the smallest energy scale in the problem. Thus, one can still include a small mediator mass in our analysis as long as it is much smaller than the typical kinetic energy scale. A one-loop analysis comparing the effect of mediator mass across scalar and pseudoscalar mediators has been done before~\cite{Agrawal:2020lea}, in the context of DM-DM elastic scattering.}.

Given the IR scales in the NR limit, one can investigate the possible energy and momentum scaling of the modes in the NREFT. The hierarchy between energy and momentum in the NR limit is typically $E\ll p$ (e.g. for the DM field, as discussed earlier), and $E\lesssim p$ for mediator field. Therefore one has to consider several modes or active degrees of freedom in the NREFT, with different hierarchy between $E, p$, as shown in Table~\ref{tab:modes}.

\begin{table}[!htb]
\centering
\begin{tabular}{|c|c|c|}
    \hline
     \textbf{Mode} & \textbf{Energy} & \textbf{Momentum}  \\
     \hline
     Soft & $m_{\chi} v$ & $m_{\chi} v$ \\
     \hline
     Potential & $m_{\chi} v^2$ & $m_{\chi} v$ \\
     \hline
     Ultrasoft & $m_{\chi} v^2$ & $m_{\chi} v^2$ \\
    \hline
\end{tabular}
\caption{Modes in the NREFT}
\label{tab:modes}
\end{table}

For on-shell massive excitations, the dispersion relation $E=p^2/2m_\chi$ necessitates that their momentum scaling be that of potential modes. The typical momentum of the DM fermion field falls into the aforesaid category. For on-shell massless modes such as that of the mediator, the dispersion relation $E=p$ holds, making the typical momentum scaling either soft or ultrasoft in nature. Off-shell modes of the mediator can have the potential scaling, and are eponymous because they mediate the instantaneous interaction between the DM particles.

There are several formulations of NREFT that have been used in the earlier literature. Some popular ones involve a two-stage matching, involving successive integrating out of the soft and potential modes as described in Refs.~\cite{Brambilla:1999xf,Pineda:1997bj,Pineda:1997ie,Pineda:1998kj,Biondini:2018ovz,Biondini:2021ycj,Biondini:2023zcz}, or a single-stage matching using the label-formalism split of the momentum into large components ($\sim m_\chi v$) and small components ($\sim m_\chi v^2$) ~\cite{Luke:1999kz,Hoang:2001rr,Fleming:2000ib,Manohar:2000cg}. We find the latter formulation useful for power counting purposes and shall demonstrate our arguments using it. In the label formalism, we formally split the total energy and  momenta $(E,\vec{P})$~\cite{Georgi:1990um,Luke:1999kz} into a large label component and a small residual component. This split for the DM field $\chi$ is
\begin{align}
    E &= \underbrace{k^0}_{\sim m_\chi v^{2}}\ , \qquad  \vec{P} = \underbrace{\vec{p}}_{\sim m_\chi v} + \underbrace{\vec{k}}_{\sim m_\chi v^{2}}\ ,
\end{align}
and for the mediator field $\phi$ is
\begin{align}
    E &= \underbrace{p^0}_{\sim m_\chi v} +  \underbrace{k^0}_{\sim m_\chi v^{2}}\ , \qquad \vec{P} = \underbrace{\vec{p}}_{\sim m_\chi v} + \underbrace{\vec{k}}_{\sim m_\chi v^{2}}\ \ ,
\end{align}
where the first (second) term in the RHS refers to the label (residual) component.
The label-formalism implements the multipole expansion of the DM-mediator interaction in the NREFT, and enables a homogeneous velocity power counting of each term in the Lagrangian.

We denote the DM field in the NREFT by $\chi_{\vec{p}}$, which has the dispersion relation of the potential mode, since it is a propagating degree of freedom. We denote the soft mode of the mediator by $\phi_{s,\vec{p}}$ where $\vec{p}$ denotes its label momentum, and the ultrasoft mode of the mediator by $\phi_{\rm us}$ (which has zero label momentum). It is useful to work with the fields Fourier transformed w.r.t. the residual momenta, namely
\begin{align}
    \chi_{\vec{p}}(\vec{x},t) &= \int \frac{d^3 \vec{k}}{(2\pi)^3} \frac{dk^{0}}{2\pi} e^{-i\vec{k}\cdot \vec{x}+i k^{0} t} \chi_{\vec{p}}(\vec{k}, k^{0})\ .
\end{align}
Thus, the full field can then be expressed as a sum of label momenta as
\begin{align}
    \chi(\vec{x},t) &= \sum_{\vec{p}\neq 0} \chi_{\vec{p}}(\vec{x},t)\ .
\end{align}

The label-space formalism makes a clean separation between the relevant energy scales of the interacting fields, and makes it easier to compute the momentum space Feynman rules in the NREFT. When computing Feynman diagrams in the NREFT, one ensures conservation of momentum for both label and residual momentum separately. Any loop integration over the overall momenta occurs via a sum over the label momenta $\vec{p}$, and an integration over the residual momenta $\vec{k}$. For more details regarding the label formalism we recommend the interested reader to Ref.~\cite{Luke:1999kz,Manohar:2006nz}.\footnote{Another approach to systematizing NREFTs is the method of regions expansion of Feynman diagrams ~\cite{Griesshammer:1997wz,Beneke:1997zp,Beneke:1999qg,Beneke:2019qaa,Beneke:2020vff}. It provides an equivalent alternative to diagram computations in the label formalism, with appropriate removal of possible double counting of regions (zero bins).}

By matching amplitudes from the UV Lagrangian in Eq.~\eqref{eq:l_uv} in the $v\ll 1$ limit, one can show that the Lagrangian for the NREFT with a scalar mediator in the label formalism takes the form:
\begin{align}
    \cu{L}_{\rm S}^{\rm NREFT} &= \sum_{\vec{p}}\chi_{\vec{p}}(x)^{\dagger}\left(i\partial_t - \frac{\vec{p}^{\ 2}}{2m_{\chi}}\right)\chi_{\vec{p}}(x) + \sum_{\vec{q}} \phi_{s,\vec{q}}(x) \left[q_{0}^2-\vec{q}^{\ 2}\right] \phi_{s,\vec{q}}(x) + \frac{1}{2}\partial_{\mu}\phi_{\rm us}(x)\partial^{\mu}\phi_{\rm us}(x) \nonumber \\
    & +\ \sum_{\vec{p}, \vec{q}}\frac{g^2}{(\vec{p}-\vec{q})^2} \chi_{\vec{q}}^{\dagger}~\chi_{\vec{p}}~\chi_{-\vec{q}}^{\dagger}~\chi_{-\vec{p}}\ +\ \frac{g^2}{2} \sum_{\vec{p}, \vec{q}, \vec{p}^{\prime}, \vec{q}^{\prime}} \chi_{\vec{p}}^{\dagger} \chi_{\vec{p}^{\ \prime}} \phi_{s,\vec{q}}~\phi_{s,\vec{q}^{\ \prime}}\left(\frac{1}{q^{0}}+\frac{1}{q^{\prime\ 0}}\right) \nonumber\\
    & + g\sum_{\vec{p}}\chi_{\vec{p}}^{\dagger}\chi_{\vec{p}} ~\phi_{\rm us}(0,t)+\cdots \label{eq:nreft_scalar}
\end{align}
where $\chi_{\vec{p}}$ and $\phi_{s, \vec{q}}$ denote the DM and soft mediator field with label momenta $\vec{p}, \vec{q}$ which are of order $m_\chi v$. Here the Lagrangian above is expressed using $x=(\vec{x}, t)$ as the conjugate coordinates for the residual momenta, and as a sum over terms specified by their label momenta\footnote{We omit writing the DM anti-DM potential $V_{\chi\overline{\chi}}$, which differs from the DM-DM potentials by contact $s$-channel terms which are generally $v$ suppressed. For the pseudoscalar, such a term is technically the same order as the $t$-channel scattering, but it does not alter the $v$ counting needed to explain the absence of SE.}.

We will now explain the origin of the terms in Eq.~\eqref{eq:nreft_scalar}. Consider first the kinetic terms in the first line of Eq.~\eqref{eq:nreft_scalar} in the NREFT Lagrangian. One can obtain them by expanding the full theory propagators in the $v\ll 1$ limit.  In doing so, one finds that the DM particle $\chi$ propagator takes the form
\begin{align}
    D_{\chi}(E,\vec{p})=\frac{i}{E-\frac{\vec{p}^2}{2m_{\chi}}}\sim v^{-2}\ , \label{eq:prop_dm}
\end{align}
which is the Green's function for the free Schr\"odinger's equation. Similarly, the propagator for the mediator modes is given by
\begin{align}
    D_{\phi}(E,\vec{p}) &= \frac{i}{E^2-\vec{p}^{\ 2}}. \label{eq:prop_med}
\end{align}
which scales as $v^{-2}$ for the soft modes and $v^{-4}$ for the ultrasoft modes.

The $v$ scaling of the propagators (Eqs~\eqref{eq:prop_dm},\eqref{eq:prop_med}) along with the requirement that all kinetic terms be leading power in $v$ suffices to determine the power counting of the fields themselves. Noting that the coordinate measure scales as
\begin{align}
    dt~d^{3}x \sim v^{-5}\ ,
\end{align}
for massive modes, and using the scaling of the propagator in Eq.~\eqref{eq:prop_dm}, we see that the field $\chi$ must scale as $v^{3/2}$ for its kinetic term to be leading power in $v$. Similarly for massless modes, the measure scales as
\begin{align}
    dt~d^{3}x \sim v^{-4a}\ ,
\end{align}
where $a=1,2$ respectively for soft and ultrasoft modes. Using the scaling of the propagator in Eq.~\eqref{eq:prop_med} thereby gives us the $v$ scaling of $\phi_{s}$ and $\phi_{\rm us}$ to be $v$ and $v^2$ respectively. The power counting of the modes is summarized in Table~\ref{tab:pc_nreft} .

\begin{table}[!h]
\centering
\begin{tabular}{|c|c|}
    \hline
     \textbf{Mode} & \textbf{Power Counting}\\
     \hline
     $\chi_{\vec{p}}$ & $v^{3/2}$ \\
     \hline
     $\phi_{s, \vec{q}}$ & $v$ \\
     \hline
     $\phi_{\rm us}$ & $v^{2}$ \\
    \hline
\end{tabular}
\caption{\small{Power counting of NREFT Modes}}
\label{tab:pc_nreft}
\end{table}

Having established the kinetic terms and their power counting, we now explicate the interaction terms in the second line of Eq.~\eqref{eq:nreft_scalar}. The potential modes have been encoded in the 4-pt fermion interaction term of the NREFT Lagrangian in Eq.~\eqref{eq:nreft_scalar}. The NR limit of the interaction scales as shown in Eq.~\eqref{eq:pot_diag},

\input{tikz_figs/pot_1}

The scaling of the tree-level potential interaction can be understood as follows. By virtue of the external fermion legs  being on-shell, and satisfying $E=p^2/(2m_{\chi})$, the exchanged mediator momentum $l^{\mu}$ scales as $(v^2, v)$, i.e. a potential mode. Each interaction vertex also brings in a factor of $g$, and the $v$ scaling of the mediator propagator being $1/(E^2-\vec{p}^{\ 2})\sim v^{-2}$, brings the interaction to $g^2/v^2$. The interaction mediated by potential mode exchanges (the first term in second line of Eq.~\eqref{eq:nreft_scalar}) is the most dominant interaction in $v$ scaling for scalar mediators.
\\ \\
Now consider the second term in the second line of Eq.~\eqref{eq:nreft_scalar}, which encodes the DM-soft mediator interactions. When a soft mediator interacts with an on-shell DM field, the resultant momentum of the DM scales as $(v,v)$, making it off-shell. The intermediate off-shell DM particle can then re-emit a soft mediator and return on-shell. Thus, the lowest order interaction preserving the label momenta of soft mediators and DM fields is via a four point interaction, which arises from the full theory diagrams as shown in Eq.~\eqref{eq:comp_scalar_nreft}.
\\ \\
\input{tikz_figs/soft_1}
Since the intermediate fermion propagator goes as $\slashed{\ell}/\ell^2 \sim v^{-1}$, and the interaction vertices bring a factor of $g^2$, the resulting 4 point interaction term of the soft mediator-DM fields scales as $g^2/v$. Thus, the resultant $v$ scaling makes it subdominant to DM-DM-potential interaction. The exact Wilson coefficient for the 4-pt soft mediator-DM term in Eq.~\eqref{eq:nreft_scalar} can be obtained by taking the $v\ll 1$ limit of the full theory amplitude and accounting for NR normalization of states in the NREFT. Thus for leading $v$ power considerations, soft mode interactions are irrelevant for a scalar mediator. However at loop level, soft mediator modes produce the leading quantum correction to the DM interaction potential, and are necessary for the consistency of the NREFT~\cite{Beneke:1997zp,Beneke:2019qaa,Beneke:2020vff,Luke:1999kz,Manohar:2000cg,Hoang:2001rr,Fleming:2000ib}.
\\ \\
The label formalism makes it clear that ultrasoft fields can only change the residual momenta $(\sim m_\chi v^2)$. Thus, it is possible for a DM particle to emit ultrasoft mediator modes and retain its $v$ power scaling. The effect of ultrasoft mediator emission in the NREFT is thus obtained by systematic multipole expansion of the interaction term as~\cite{Luke:1997ys}
\begin{align}
\overline{\chi}(\vec{x}, t)\chi(\vec{x}, t)\phi_{\rm us}(\vec{x}, t) = \overline{\chi}(\vec{x}, t)\chi(\vec{x}, t)\left[\phi_{\rm us}(0, t) + \vec{x} \cdot \vec{\nabla}\phi_{\rm us}(0, t) +\cdots \right].
\end{align}
When Fourier transformed into momentum-space one can verify that each term in the expansion above gives a homogeneous power counting in $v$. Only the leading power interaction is listed in line 3 of the NREFT Lagrangian in Eq.~\eqref{eq:nreft_scalar}, and the subleading terms are relegated to the dots. This completes our short explication of the NREFT scalar mediated Lagrangian in Eq.~\eqref{eq:nreft_scalar}. Using the NREFT modes, along with their $v$ counting, one can systematically match the NR limit of the amplitudes given by the UV Lagrangian in Eq.~\eqref{eq:l_uv} to operators in the NREFT. Combined with the power counting of the label and residual momenta, the NREFT allows one to velocity power count diagrams in the NR limit.

We now demonstrate how the $v$ power counting allows one to infer the existence of SE in the NREFT. Sommerfeld enhancement in the DM scattering rate can be understood as the low velocity $(v\ll 1)$ behaviour of the elastic scattering in the NREFT. Since we are interested in the most singular behaviour, we see that iteration of the tree-level potential in Eq.~\eqref{eq:nreft_scalar} scales the strongest ($\sim v^{-2}$) giving us the leading behaviour from

\input{tikz_figs/ladder_chain_snippet}
Here, the dashed {\color{OliveGreen}green} lines represent the full theory diagrams where the mediator momentum scales like a potential mode, which is reproduced in the NREFT using iteration of the four point fermion interaction (first term on second line of Eq.~\eqref{eq:nreft_scalar}). Consider the scattering process $\chi_{s_{1},~\vec{p}}~\chi_{s_{2},-\vec{p}}\rightarrow \chi_{s^{\prime}_{1},~\vec{q}}~\chi_{s^{\prime}_{2},-\vec{q}}$, where the first subscript denotes spin and the second denotes the label momenta. The tree-level amplitude $\cu{A}_{\rm tree}$ in the NREFT can thus be seen to be\footnote{We ensure that there is a NR normalization when matching from the full theory, i.e. the matching is done to $\lim_{v \rightarrow 0}(2m_\chi)^{-2}\cu{A}^{\rm UV}_{\rm tree}$ from the UV QFT, where the first prefactor is just accounting for NR normalization of states in the NREFT.}

\begin{align}
    i\cu{A}_{\rm tree} = \frac{ig^2}{(\vec{p}-\vec{q})^2} (\xi_{s_{1}^{\prime}}^{\dagger}\xi_{s_{1}})(\xi_{s^{\prime}_{2}}^{\dagger}\xi_{s_{2}}) \sim \frac{g^2}{v^2}
\end{align}
making it enhanced in the NR limit. Similarly, the one-loop amplitude $\cu{A}_{1-\rm loop}$, given by a twice iteration of the tree-level potential is given by
\begin{align}
    i\cu{A}_{1-\rm loop}
    &= \frac{1}{2} \int \underbrace{ \frac{d\tilde{E}}{2\pi}}_{v^{2}}~\underbrace{\frac{d^3\vec{l}}{(2\pi)^3}}_{v^{3}} \underbrace{ \frac{ig^2}{(\vec{p}-\vec{l})^{\ 2}} \frac{i}{E+\tilde{E}-\frac{\vec{l}^2}{2m_{\chi}}+i0^{+}}\frac{ig^2}{(\vec{q}-\vec{l})^{\ 2}} \frac{i}{E-\tilde{E}-\frac{\vec{l}^{\ 2}}{2m_{\chi}}+i0^{+}} }_{v^{-8}}~(\xi_{s_{1}^{\prime}}^{\dagger}\xi_{s_{1}})(\xi_{s^{\prime}_{2}}^{\dagger}\xi_{s_{2}})
\end{align}
Rather than compute the full integral\footnote{For brevity, we skip writing the full result for the integral. The interested reader is pointed to Ref.\cite{Beneke:1997zp}, Eq.~(31) for the full result. The integral is formally IR divergent with zero-mediator mass, a manifestation of the IR divergent Coulomb phase. The IR divergence can be neglected for the purposes of velocity power counting, which is what we are after here.}, the $v$ counting of the amplitudes shows that
\begin{align}
    \cu{A}_{1-\rm loop}&\sim \frac{g^4}{v^3}\: , \qquad \frac{\cu{A}_{1-\rm loop}}{\cu{A}_{\rm tree}} \sim \frac{g^2}{v}\ .
\end{align}
Thus, in the NR limit with weak coupling where $g^2\sim v \ll 1$, the one-loop diagram is formally the same power counting order as that of the tree-level exchange. In fact, at $L$ loops, power counting in $v$ shows that the iterated diagrams scale as $\cu{A}_{\rm tree}(g^2/v)^L$, making it necessary to resum them. Thus, the necessity of resumming these diagrams in the NREFT is the manifestation of the Sommerfeld enhancement of the cross section in the $v\ll 1$ limit. We also point out that it has been demonstrated that resumming these ladder diagrams in the NREFT is equivalent to solving the Schr\"odinger equation with the tree-level potential~\cite{Kaplan:1996xu,Luke:1996hj}, thus making contact with the QM method of computing the Sommerfeld enhancement for the interaction~\cite{Iengo:2009ni,
Slatyer:2009vg}.

Before explicating the NREFT for the other forms of mediators, we mention that the cross box diagram in the NREFT can be safely neglected for leading power considerations in $v$ counting. In fact,

\input{tikz_figs/crossed_exchange_snippet}

in the NREFT is given by
\begin{align}
    i\cu{A}_{\mathrm{cross\ box}}
    &= \int \frac{d\tilde{E}}{2\pi}~\frac{d^{3}\vec{l}}{(2\pi)^3}~\frac{ig^2}{\vec{l}^{\ 2}}\frac{ig^2}{(\vec{q}-\vec{p}+\vec{l})^{2}}
    \frac{i}{E-\tilde{E}-\frac{(\vec{p}-\vec{l})^2}{2m_{\chi}}+i0^{+}} \frac{i}{E-\tilde{E}-\frac{(\vec{q}-\vec{l})^2}{2m_{\chi}}+i0^{+}}
    \times (\xi_{s_{1}^{\prime}}^{\dagger}\xi_{s_{1}})(\xi_{s^{\prime}_{2}}^{\dagger}\xi_{s_{2}})\ .
\end{align}
The cross box $\cu{A}_{\mathrm{cross\ box}}$ amplitude evaluates to $0$ since the integration over the energy $\tilde{E}$ is not pinched by its poles appearing in the integrand, and one can consequently deform the contour away making the integration zero. Thus, at leading power in velocity counting, the cross-box exchange vanishes, and does not contribute to the SE.

This completes the discussion for the scalar mediator case. We now proceed to the pseudoscalar case.

\subsection{Pseudoscalar mediators}
For a pseudoscalar mediator we set the UV Lagrangian for the DM self interactions as
\begin{align}\label{eq:ps_uv_l}
\mathcal{L}_{\rm PS}^{\rm UV} = \overline{\chi}(i\slashed{\partial}-m_{\chi})\chi + ig \overline{\chi}\gamma_5\chi \phi + \frac{1}{2}(\partial_\mu \phi)(\partial^{\mu} \phi)\ ,
\end{align}
where the mediator $\phi$ is now $P$-odd. We now wish to match this to an NREFT akin to Eq.~\eqref{eq:nreft_scalar} by careful matching from the full theory in Eq.~\eqref{eq:ps_uv_l}. We will show that by matching amplitudes from the UV theory, the relevant NREFT Lagrangian for pseudoscalar mediators is
\begin{align}
    \cu{L}_{\rm PS}^{\rm NREFT} &= \sum_{\vec{p}}\chi_{\vec{p}}(x)^{\dagger}\left(i\partial_t - \frac{\vec{p}^{\ 2}}{2m_{\chi}}\right)\chi_{\vec{p}}(x) + \sum_{\vec{q}} \phi_{s,\vec{q}}(x) \left[q_{0}^2-\vec{q}^{\ 2}\right] \phi_{s,\vec{q}}(x) + \frac{1}{2}\partial_{\mu}\phi_{\rm us}(x)\partial^{\mu}\phi_{\rm us}(x) \nonumber \\
    & +\ \sum_{\vec{p}, \vec{q}}\frac{g^2}{4m_{\chi}^2(\vec{p}-\vec{q})^2} \chi_{\vec{q}}^{\dagger}(\vec{q}-\vec{p})\cdot \vec{\sigma}~\chi_{\vec{p}}~\chi_{-\vec{q}}^{\dagger}~(\vec{q}-\vec{p})\cdot \vec{\sigma}\chi_{-\vec{p}}~-~\frac{g^2}{m_{\chi}} \sum_{\vec{p}, \vec{q}, \vec{p}^{\prime}, \vec{q}^{\prime}} \chi_{\vec{p}}^{\dagger} \chi_{\vec{p}^{\ \prime}} \phi_{s,\vec{q}}~\phi_{s,\vec{q}^{\ \prime}} \nonumber \\
    &+ g\sum_{\vec{p},\vec{q}} \frac{1}{m_{\chi}}\chi_{\vec{q}}^{\dagger} (\vec{p}-\vec{q})\cdot \vec{\sigma}~\chi_{\vec{p}}~\phi_{\rm{us}}(0,t)+\cdots\label{eq:nreft_ps}
\end{align}
at  leading power in $v$ counting.
\\ \\
Note that we use the same notation for the DM and mediator fields as in the scalar case. Since the kinetic terms do not depend on interaction details, one sees that the form of the kinetic terms in Eq.~\eqref{eq:nreft_ps} is the same as that of the scalar mediator case (Eq.~\eqref{eq:nreft_scalar}). Focusing thus on the interaction terms, we now illustrate their matching using scattering amplitudes of the DM particles. Consider the scattering process $\chi_{s_{1},~\vec{p}}~\chi_{s_{2},-\vec{p}}\rightarrow \chi_{s^{\prime}_{1},~\vec{q}}~\chi_{s^{\prime}_{2},-\vec{q}}$, where the first subscript denotes spin and the second denotes the label momenta. At tree level the relevant amplitude with potential modes is given in Eq.~\eqref{eq:tree_diagram_ps},
\\ \\
\input{tikz_figs/tree_amp_snippet}
\\ \\
which in the full theory takes the form
\begin{align}
    i\cu{A}_{\rm tree} = \lim_{v \rightarrow 0} \frac{g^2}{4m_{\chi}^2}~\overline{u}_{s_{1}^{\prime}}(q)\gamma_5 u_{s_{1}}(p)\frac{i}{(q-p)^2}\overline{u}_{s^{\prime}_{2}}(q^{\prime})\gamma_5 u_{s_{2}}(p^{\prime}) \label{eq:tree_amp}
\end{align}
In the NR limit, one can power expand the spinor bilinears appearing in Eq.~\eqref{eq:tree_amp} in the Weyl representation as follows
\begin{align}
    \overline{u}_{r}(q)\gamma_5 u_{s}(p) &= \begin{pmatrix}
        \xi_{r}^{\dagger}\sqrt{q\cdot \overline{\sigma}} & \xi_{r}^{\dagger}\sqrt{q\cdot \sigma}
    \end{pmatrix}
    \begin{pmatrix}
        -1 & 0 \\
        0 & 1
    \end{pmatrix}
    \begin{pmatrix}
        \sqrt{p\cdot \sigma} \xi_{s} \\
        \sqrt{p\cdot \overline{\sigma}} \xi_{s}
    \end{pmatrix}\ ,\\
    &=-\xi_{r}^{\dagger} \sqrt{q\cdot \overline{\sigma}} \sqrt{p\cdot \sigma} \xi_s + \xi_{r}^{\dagger} \sqrt{q\cdot \sigma} \sqrt{p\cdot \overline{\sigma}} \xi_s\ , \\
    &\approx m_{\chi}\left[-\xi_{r}^{\dagger}\left(1+\frac{\vec{q}-\vec{p}}{2m_{\chi}}\cdot \vec{\sigma}\right)\xi_{s}+\xi_{r}^{\dagger}\left(1-\frac{\vec{q}-\vec{p}}{2m_{\chi}}\cdot \vec{\sigma}\right)\xi_{s}\right]+\cdots\ ,\\
    &=-\underbrace{\xi_{r}^{\dagger}~(\vec{q}-\vec{p})\cdot \vec{\sigma}~\xi_{s}}_{\sim v} +\cdots\ .
\end{align}
giving us that
\begin{align}
    i\cu{A}_{\rm tree} &= \lim_{v \rightarrow 0} \frac{g^{2}}{4m_{\chi}^2}~\overline{u}_{s_{1}^{\prime}}(q)\gamma_5 u_{s_{1}}(p)\frac{i}{(q-p)^2}\overline{u}_{s^{\prime}_{2}}(q^{\prime})\gamma_5 u_{s_{2}}(p^{\prime}) \\
    \implies \cu{A}_{\rm tree} &= \frac{g^2}{4m_{\chi}^2} \underbrace{\frac{\xi_{s_{1}^{\prime}}^{\dagger}~(\vec{q}-\vec{p})\cdot \vec{\sigma}~\xi_{s_{1}}~~\xi_{s^{\prime}_{2}}^{\dagger}~(\vec{q}-\vec{p})\cdot \vec{\sigma}~\xi_{s_{2}}}{(\vec{q} -\vec{p})^{2}}}_{\sim v^0 } +\cdots \:\: .
    \label{eq:tree_ps_amp}
\end{align}
Thus, at leading power in $v$ counting, the amplitude for DM-DM scattering scales as $g^2 v^{0}$, making the pseudoscalar interaction velocity suppressed (relative to the pure scalar interaction) in the NR limit. Since the NREFT should reproduce the amplitude, one can simply read off the 4-pt fermion operator  arising from potential modes as
\begin{align}
    \cu{L}^{\rm NREFT}_{\rm PS} \supset \sum_{\vec{p}, \vec{q}, \mathrm{spins}}\frac{g^2}{4m_{\chi}^2(\vec{p}-\vec{q})^2}~\chi_{\vec{q}}^{\dagger}(\vec{q}-\vec{p})\cdot \vec{\sigma}~\chi_{\vec{p}}~\chi_{-\vec{q}}^{\dagger}~(\vec{q}-\vec{p})\cdot \vec{\sigma}\chi_{-\vec{p}} \sim v^{0}\ , \label{eq:pot_ps_term}
\end{align}
as duly indicated in the first term of the second line in Eq.~\eqref{eq:nreft_ps}. The interaction mediated by the potential modes suffices to describe elastic scattering of the DM in the NR limit where the scattering amplitude takes the form as shown in Eq.~\eqref{eq:ladder_ps}.

\input{tikz_figs/ladder_chain_ps}
Here, the potential mode mediated interaction (Eq.~\eqref{eq:pot_ps_term}) leads to the following one loop contribution to the scattering amplitude,

\begin{multline}
   i\cu{A}_{1-\rm loop} =
     \frac{1}{2}\left(\frac{g^2}{4m_{\chi}^2}\right)^2 \int \underbrace{\frac{d\tilde{E}}{2\pi}~\frac{d^3\vec{l}}{(2\pi)^3}}_{\sim v^5} \underbrace{\frac{\xi_{s_{1}^{\prime}}^{\dagger}~(\vec{q} -\vec{l})\cdot \vec{\sigma}~(\vec{l}-\vec{p})\cdot \vec{\sigma}~\xi_{s_{1}}}{(\vec{l}-\vec{p})^{2}}}_{\sim v^0 }\underbrace{\frac{i}{\left[E+\tilde{E}-\frac{\vec{l}^{\ 2}}{2m_{\chi}}+i0^+\right]}}_{\sim v^{-2}} \\ \times \underbrace{\frac{\xi_{s_{2}^{\prime}}^{\dagger}~(\vec{q} -\vec{l})\cdot \vec{\sigma}~(\vec{l}-\vec{p})\cdot \vec{\sigma}~\xi_{s_{2}}}{(\vec{q}-\vec{l})^{\ 2}}}_{\sim v^0 }\underbrace{\frac{i}{\left[E-\tilde{E}-\frac{\vec{l}^{\ 2}}{2m_{\chi}}+i0^+\right]}}_{\sim v^{-2}}\ .\label{eq:loop_amp_ps}
\end{multline}

Again, power counting the amplitudes (Eqs~\eqref{eq:loop_amp_ps} and \eqref{eq:tree_ps_amp}) in $v$ gives us
\begin{align}
    \cu{A}_{1-\rm loop}&\sim g^4 v\ , & \frac{\cu{A}_{1-\rm loop}}{\cu{A}_{\rm tree}}&\sim g^2 v \ .
\end{align}
Thus, we see that in the NR limit, the pseudoscalar mediator interactions lead to a suppression of the iteration of the tree-level potential, and produce no enhancement of the elastic DM interaction cross section in the $v\ll 1$ limit. Said differently, the NREFT analysis above allows us to explain using velocity power counting arguments, the lack of SE as seen in previous work~\cite{Agrawal:2020lea}. While we have set the mediator mass to be zero in the NREFT analysis thus far, the argument can be extended to a non-zero mediator mass as long as the mediator mass $m_\phi$ is much lower than the smallest energy scale in the NREFT loops (the kinetic energy of the DM fields). Thus if,
\begin{align}
    m_{\phi} &\ll m_{\chi} v^2\ ,
\end{align}
along with $g^{2}\sim v \ll 1$, one expects to find no SE in the DM scattering cross section mediated by pseudoscalar interactions (arising from renormalizable UV QFTs). We shall verify the no-SE observation using numerical approaches in Section~\ref{sec:bound_sq_well}.

To complete the derivation of the NREFT Lagrangian for pseudoscalars, we also demonstrate the origin of the soft mediator interaction terms (second term on second line in Eq.~\eqref{eq:nreft_ps}). In fact, we will find that unlike the scalar mediator in Eq.~\eqref{eq:nreft_scalar}, the soft interactions are not power suppressed in $v$ relative to the potential interactions. We proceed by computing the matching by evaluating the $\chi_{\vec{p},s_{1}} \phi_{\vec{q}} \rightarrow \chi_{\vec{p}^{\ \prime},s_{1}^{\prime}} \phi_{\vec{q}^{\ \prime}} $ amplitude $\cu{A}_{\rm s\chi s\chi}$ in  Eq.~\eqref{eq:comp_ps_amp}\ .
\input{tikz_figs/soft_2}
Taking the $v\ll 1$ limit of the amplitude in  Eq.~\eqref{eq:comp_ps_amp} yields
\begin{align}
    i\cu{A}_{\rm s\chi s\chi} &= \lim_{v \rightarrow 0}~\frac{i(-g)^2}{2m_{\chi}}~\left[\frac{\overline{u}_{s_{1}^{\prime}}(p^{\prime})\gamma_5 \left(\slashed{p}+\slashed{q}+m_{\chi}\right) \gamma_5 u_{s_{1}}(p)}{(q+p)^2 -m_{\chi}^2}+ \frac{\overline{u}_{s_{1}^{\prime}}(p^{\prime})\gamma_5 \left(\slashed{p}-\slashed{q}^{\prime}+m_{\chi}\right) \gamma_5 u_{s_{1}}(p)}{(p-q^{\prime})^2 -m_{\chi}^2}\right]\ ,\\
    &= \lim_{v \rightarrow 0}~\frac{i(-g)^2}{2m_{\chi}}~\left[\frac{\overline{u}_{s_{1}^{\prime}}(p^{\prime})\left(-\slashed{p}-\slashed{q}+m_{\chi}\right) u_{s_{1}}(p)}{2q\cdot p +q^{2}}+\frac{\overline{u}_{s_{1}^{\prime}}(p^{\prime})\left(-\slashed{p}+\slashed{q}^{\prime}+m_{\chi}\right) u_{s_{1}}(p)}{-2q^{\prime}\cdot p +q^{\prime\ 2}}\right]\ ,\\
    &= \lim_{v \rightarrow 0}~\frac{i(-g)^2}{2m_{\chi}}~\left[\frac{\overline{u}_{s_{1}^\prime}(p^{\prime})\left(-\slashed{q}\right) u_{s_{1}}(p)}{2m_{\chi} q^{0}}+\frac{\overline{u}_{s_{1}^{\prime}}(p^{\prime})\left(\slashed{q}^{\prime}\right) u_{s_{1}}(p)}{-2m_{\chi} q^{\prime\ 0}}\right]\ ,\\
    &=-i \underbrace{\frac{g^2}{m_{\chi}}\xi_{s_{1}^{\prime}}^{\dagger} \xi_{s_{1}}}_{\sim v^0 }\ +\ \cu{O}(v)\ . \label{eq:low_v_ps_soft}
\end{align}
As can readily be seen from the second term on the second line in Eq.~\eqref{eq:nreft_ps}, the soft mediator operator term correctly reproduces Eq.~\eqref{eq:low_v_ps_soft}. The power counting in $v$ of these soft interaction terms is $v^0$ and thus is the same as that of the interaction term mediated by potential modes (Eq.~\eqref{eq:pot_ps_term}).

As remarked earlier, the soft mediator interactions are needed for the consistency of the NREFT. In particular, they do contribute to the DM scattering at the one-loop level via diagrams such as those shown in Eq.~\eqref{eq:soft_loop}.
\\ \\
\input{tikz_figs/double_dash_bubble_snippet}
\\ \\
 Such soft mediator loop diagrams in NREFTs yield $\mathcal{O}(g^4)$ terms, which are $g^2$ corrections relative to the tree-level interaction potential between the DM particles~\cite{Luke:1999kz,Beneke:2019qaa} (see appendix~\ref{app:loop_pot}). The advantage of the velocity power counting in the NREFT is that we can systematically determine the $v$ counting of these loop corrected potentials, and assess if they can generate SE as well.

\subsection{Vector, axial vector and pseudo-Goldstone mediators}

For completeness, we also discuss the NREFT Lagrangians for vector, axial-vector mediator and pseudo-Goldstone interactions.

For vector/axial-vector interactions, the Lagrangian in the UV is taken to be
\begin{align}
    \cu{L}^{\rm UV} = \overline{\chi}(i\slashed{\partial}-m_{\chi})\chi -\frac{1}{4}F_{D\mu\nu}F_{D}^{\mu\nu} + g~(\overline{\chi}\gamma_{\mu}\Gamma\chi) A_{D}^{\mu} \label{eq:v_av}
\end{align}
where $A_{D}^{\mu}$ is a massless vector mediator, and $F_{D}$ is the associated field strength tensor. When the interaction is pure vector-like the matrix $\Gamma$ is taken to be $\Gamma_{V}\equiv\mathbbm{1}$, and the resultant NREFT for $\Gamma_{V}$ interactions is simply a dark copy of the NRQED/QCD Lagrangian, which is elaborated in detail in Ref.~\cite{Luke:1999kz}. For the purpose of assessing the presence of Sommerfeld enhancement, we need only the 4-pt DM interaction operator at leading power in $v$ counting, which goes as
\begin{align}
    \cu{L}_{\rm vector}^{\rm NREFT} \supset \frac{g^2}{(\vec{p}-\vec{q})^2}~\chi_{\vec{q}}^{\dagger}~\chi_{\vec{p}}~\chi_{-\vec{q}}^{\dagger}~\chi_{-\vec{p}}\ .
\end{align}
Thus, the form of the potential mode mediated interactions make the vector mediator case almost identical to that of the scalar mediator potential interactions. Thus, the NREFT $v$ power counting argument is identical for scalars and vectors, and corroborates the observation that vector-mediated DM interactions exhibit SE.

As discussed earlier, the interaction with a pure axial-vector ($\Gamma_{A}\equiv\gamma_5$) is subtle. A massless $A_\mu$ field (with $g$ finite) leads to an inconsistent UV theory, while making it a massive Proca field with mass $m_A$ necessitates that the UV theory as written is incomplete (because it couples to an anomalous current). In such scenarios the construction of the NREFT with a consistent velocity power counting necessitates a careful treatment of two small parameters $\epsilon_A \equiv m_{A}/m_{\chi}$ and $v$. We leave a detailed analysis of such a system to future work.

We now demonstrate that the NREFT also predicts no SE for pseudoscalar mediators that couple to the DM fields via derivative terms in the UV Lagrangian. These will be henceforth referred to as pseudo-Goldstone mediators, and the relevant interaction terms in the UV Lagrangian take the form,
\begin{align}
    \cu{L}_{\rm int} = \frac{g}{\Lambda}~ \overline{\chi}\gamma_\mu \gamma_5 \chi ~\partial^\mu \phi\ .
\end{align}
Such interactions have been considered in previous work~\cite{Bedaque:2009ri}. Consider the scattering process $\chi_{s_{1},~\vec{p}}~\chi_{s_{2},-\vec{p}}\rightarrow \chi_{s^{\prime}_{1},~\vec{q}}~\chi_{s^{\prime}_{2},-\vec{q}}$, where the first subscript denotes spin and the second denotes the label momenta. In the NR limit the amplitude takes the form,
\\ \\
\input{tikz_figs/tree_amp_pg}
\\ \\
\begin{align}
    i\cu{A}_{\rm tree} = \underbrace{-\frac{i}{
    (\vec{p}-\vec{q})^2}\frac{g^2}{\Lambda^2}\left[-\xi_{s_{1}^{\prime}}^{\dagger} \left(\vec{p}-\vec{q}\right)\cdot \vec{\sigma}~\xi_{s_{1}}\right] ~ \left[-\xi^{\dagger}_{s^{\prime}_{2}} \left(\vec{q}-\vec{p}\right)\cdot \vec{\sigma}~\xi_{s_{2}}\right]}_{v^{0}}\ +\ \mathcal{O}(v)
\end{align}
making it identical to that of the pseudoscalar case (Eq.~\eqref{eq:tree_ps_amp}) with the replacement $g^2/4 m_\chi^2 \rightarrow g^2/\Lambda^2$, i.e. the coupling is effectively multiplied by $2 m_\chi/\Lambda$. Thus, even for such derivative coupling to a pseudoscalar, the NREFT predicts that there should be no SE in the regime that $m_{\phi} \ll m_{\chi} v^2$, provided that the `effective' coupling $ g m_\chi/\Lambda \ll 1$ is perturbative.

A corollary of the velocity power counting in the NREFT is that if the interaction between the DM and scalar mediator has both scalar and pseudoscalar contributions,  i.e.~the Lagrangian has terms of the form $g_{\text{s}} \phi\overline{\chi}\chi + ig_{\text{ps}} \phi\overline{\chi}\gamma_5\chi$, then it follows from the previous sections that in the NREFT, the potential interaction is generically dominated by the parity-even scalar interactions (for $g_\text{s} \sim g_\text{ps}$),
\begin{align}
    \underbrace{\frac{g_{\text{s}}^2}{(\vec{p}-\vec{q})^2} \chi_{\vec{q}}^{\dagger}~\chi_{\vec{p}}~\chi_{-\vec{q}}^{\dagger}~\chi_{-\vec{p}}}_{\sim \frac{g_{\text{s}}^2}{v^2}}  \:\:
    \gg \:\:
    \underbrace{\frac{g_{\text{ps}}^2}{4m_{\chi}^2(\vec{p}-\vec{q})^2} \chi_{\vec{q}}^{\dagger}(\vec{q}-\vec{p})\cdot \vec{\sigma}~\chi_{\vec{p}}~\chi_{-\vec{q}}^{\dagger}~(\vec{q}-\vec{p})\cdot \vec{\sigma}\chi_{-\vec{p}}}_{\sim g_{\text{ps}}^2 v^{0}}\ .
\end{align}
This explains previous observations that the SE (and correspondingly the bound state formation rate) are primarily scalar-like (Coulombic) for a general interaction with the scalar mediator~\cite{Biondini:2021ccr}.

While we have demonstrated the NREFT power counting arguments explicitly for $\chi\chi$ scattering, the argument carries through for the $\chi \overline{\chi}$ scattering, as the $t$-channel diagrams dominate over the contact $s$-channel interactions in powers of $v$. Thus, the ladder diagrams in the pseudoscalar NREFT are velocity suppressed in both cases, and consequently we also do not expect the annihilation rate to be Sommerfeld-enhanced (since the ladder diagram interactions dress the contact annihilation). Similarly, given the non-enhancement in $v$ of the ladder diagrams for the pseudoscalar mediator, and coupled with the fact that the same diagrams enter the all-order sum to determine the bound-state poles in the spectrum, we expect a lack of weakly coupled bound states of DM in such systems. However, a more nuanced computation of the bound states (or their absence) would need an all-order matching computation~\cite{Luke:1996hj}, which we leave to future work.

The power of the NREFT is thus to systematize the NR potential in terms of velocity counting, and using successive matching computations of the potential to all orders in $v$, to compute the SE from each contribution of it (or the lack thereof).

Note that the NREFT arguments for predicting SE rely on the existence of perturbative matching computations from the full theory (QFT) to the NREFT at the scale $\sim m_{\chi}$. Thus for our arguments here to be valid, the NREFT coupling $g$ (or the effective coupling such as $gm_{\chi}/\Lambda$ in the pseudo-Goldstone mediator case from matching) should be $\leq 1$, in order to systematically truncate the NREFT Lagrangian. If the coupling (or effective coupling) $g \sim 1$, then such a truncation becomes invalid, and non-trivial SE from PS mediator exchange can arise, as we show in Sec.~\ref{sec:numerics}.

\section{Quantum mechanics framework for bound states and Sommerfeld-enhanced annihilation}\label{sec:bound_sq_well}

As discussed previously, we expect the standard QM approach to break down for singular potentials at short distances, requiring regularization.
In this section we will first show how to generalize the position-space matching approach employed in Ref.~\cite{Agrawal:2020lea} to obtain a criterion for bound states, which makes explicit the interaction between the long-range and UV/short-range physics. This approach will {\it not} rely on the tree-level matching used to compute the UV physics in Ref.~\cite{Agrawal:2020lea}; we will tacitly assume the scattering amplitude from UV physics can be determined to all orders. We will review a similar method to describe the Sommerfeld-enhanced annihilation introduced in Ref.~\cite{Parikh:2024mwa}.

Armed with these (exact) results, we will then discuss the sensitivity of the physics (bound-state formation and Sommerfeld-enhanced annihilation) to the UV regulator in both cases.\footnote{We will use the terms ``regulator-independent'' and ``cutoff-independent'' (and similarly ``regulator-dependent'' and ``cutoff-dependent'') interchangeably, and also identify this (in)dependence with the result being (in)dependent of the matching radius defining the boundary between UV and IR physics, since varying the matching radius effectively modifies the choice of regulator potential in the region between the original and shifted matching radii.}  For both cases, we will argue and  demonstrate numerically that knowing the low-energy scattering amplitude at leading order (or replacing the singular potential at short distances with a regular potential that matches the leading-order scattering amplitude) is not sufficient to obtain regulator-independent results. For our numerical examples, we will focus on the case of the renormalizable pseudoscalar coupling of Eq.~\eqref{eq:potential-pseudoscalar}, where the QFT Lagrangian can in principle be valid to arbitrarily high energies. We will presume the UV/relativistic physics relevant to scattering is fully described by this Lagrangian, and can thus be derived from first principles in the perturbative regime.

\subsection{Separating UV and IR physics}

The basic idea of Ref.~\cite{Agrawal:2020lea} is to separate the full scattering amplitude into a short-range contribution confined to some region $r < a$ where relativistic physics may be important, and a long-range contribution obtained by solving the Schr\"{o}dinger equation for $r > a$, with boundary conditions at the matching radius $r=a$ sourced by the short-distance term. The same basic approach can be applied to Sommerfeld-enhanced annihilation, where the relativistic short-range physics includes an absorptive term; the formalism was worked out in  Ref.~\cite{Parikh:2024mwa}. For a single-state system, where we set the long-range potential to zero within the matching radius $r=a$, and where the long-range potential does not mix partial waves, the Sommerfeld-enhanced inclusive annihilation cross section (for the $\ell^{\rm th}$ partial wave) can be written as \cite{Parikh:2024mwa}:
\begin{align}(\sigma_\text{ann} v_\text{rel})_\ell & = \frac{2\pi i c}{\mu} (2\ell+1) |\Sigma_\ell|^2 (f_\ell^\dagger - f_\ell + 2 i p f_\ell^\dagger f_\ell), \nonumber \\
\Sigma_\ell &  = \Sigma_{0,\ell}/(1 - (\bar{Z}_\ell + i p |\Sigma_{0,\ell}|^2) f_{\ell}).\label{eq:annihilation} \end{align}
Here $\bar{Z}_\ell$ and $\Sigma_{0,\ell}$ are (cutoff-dependent) coefficients that can be determined solely from the long-range potential outside $r=a$, and $f_\ell$ is the short-range scattering amplitude encoding all physics within $r=a$ (including scattering and, via a non-Hermitian contribution, annihilation). The cutoff dependence in $f_\ell$, $\bar{Z}_\ell$ and $\Sigma_{0,\ell}$ must cancel out by construction. The parameter $c$ is 2 for identical initial-state particles and 1 for distinguishable particles, and $\mu$ is the reduced mass.

While elastic scattering is not our focus in this work, the $S$-matrix can likewise be written (in this single-state case) as \cite{Parikh:2024mwa}:
\begin{align} S_\ell = S_{0,\ell} \left(1 + \frac{2 i p |\Sigma_{0,\ell}|^2}{f_\ell^{-1} - i p |\Sigma_{0,\ell}|^2 - \bar{Z}_\ell} \right), \end{align}
where $S_{0,\ell}$ is the $S$-matrix element with only the long-range/IR interactions.

We can likewise derive the analogous formalism for the presence of bound states (which is novel to our knowledge). Given a (real) 3-momentum $p$ (corresponding to the 3-momentum of one of the incoming particles in the center-of-momentum frame), and considering scattering between two angular momentum and spin configurations $(\ell,s)$ and $(\ell^\prime, s^\prime)$, let us define $w_{\ell s}^{(\ell^\prime s^\prime)}(p, r)$ to be the solution to the Schr\"odinger equation in the $(\ell, s)$ channel, with the following boundary conditions:
\begin{enumerate}
    \item For each $(\ell, s)$, $w_{\ell s}^{(\ell^\prime s^\prime)}(p, r)$ is purely outgoing for asymptotically large $r$. This  ensures the wavefunction will be exponentially decaying when we rotate $p \rightarrow i \rho$ (with $\rho > 0$).
    \item At some matching radius $r=a$, we can write:
\begin{align} w_{\ell s}^{(\ell^\prime s^\prime)}(p, r) \simeq \mathcal{B}_{\ell s}^{\ell^\prime s^\prime}(p) s_\ell(p r) + \delta_{\ell^\prime s^\prime, \ell s} (c_\ell(p r) + i s_\ell(p r)). \label{eq:bsmatching} \end{align}
Here $s_\ell(x) = x j_\ell(x)$ and $c_\ell(x) = -x y_\ell(x)$ are defined in terms of the spherical Bessel functions, and the $\simeq$ sign means that the values and first derivatives of the two sides agree at $r=a$.
\end{enumerate}
We typically want the matching radius to be large enough that for $r > a$, the NR potential description is valid; either the onset of relativistic corrections or the presence of other new physics beyond the EFT description may require such a cutoff at short distances. We can take a linear combination of such solutions to obtain a desired boundary condition at $r=a$, i.e.~$u_{\ell s}(p, r) = \sum_{\ell^\prime s^\prime} C^{(\ell^\prime s^\prime)}(p)  w_{\ell s}^{(\ell^\prime s^\prime)} (p, r) $ for some set of constants $C^{\ell^\prime s^\prime}(p)$.

In order to ensure the wavefunction has the correct physical behavior  at short distances (for example, if the short-distance physics can be described by a modified non-singular potential, this corresponds to picking out the regular solution), we can match it onto the short-range scattering amplitude $f_{\ell s}^{\ell^\prime s^\prime}(p)$ at $r=a$. Specifically, this is done by imposing the condition:
\begin{align} u_{\ell s}(p, r) \simeq \sum_{\ell^\prime s^\prime} \gamma^{\ell^\prime s^\prime}(p) (\delta_{\ell^\prime s^\prime, \ell s} s_\ell(p r) +  p f_{\ell s}^{\ell^\prime s^\prime}(p)  (c_\ell(p r) + i s_\ell(p r))  ). \end{align}
This condition imposes the relation between ingoing modes (purely in the $(\ell^\prime s^\prime)$ channel) and outgoing modes (in all $(\ell s)$ channels) required by the UV physics encoded in $f_{\ell s}^{\ell^\prime s^\prime}(p)$. Comparing coefficients of $c_\ell(p r)$ and $s_\ell(p r)$ at the matching radius $r=a$ yields the relations:
\begin{align} & \gamma^{\ell s}(p) = \sum_{\ell^\prime s^\prime} \mathcal{B}_{\ell s}^{\ell^\prime s^\prime}(p) C^{\ell^\prime s^\prime}(p), \quad
 C^{\ell s}(p)  = p \sum_{\ell^\prime s^\prime}  f_{\ell s}^{\ell^\prime s^\prime}(p) \gamma^{\ell^\prime s^\prime}(p) \nonumber \\
\Rightarrow & \,  C^{\ell s}(p) = p \sum_{\ell^\prime s^\prime} \sum_{\ell^{\prime \prime} s^{\prime \prime}} f_{\ell s}^{\ell^\prime s^\prime}(p) \mathcal{B}_{\ell^\prime s^\prime}^{\ell^{\prime \prime} s^{\prime\prime}}(p) C^{\ell^{\prime \prime} s^{\prime\prime} }(p).\end{align}

If we define a vector $\vec{C}$ with entries $C^{\ell s}(p)$ (here each choice of $\ell s$ labels one component), and matrices $f$ and $\mathcal{B}$ with entries $f_{\ell s}^{\ell^\prime s^\prime}(p)$ and $\mathcal{B}_{\ell s}^{\ell^\prime s^\prime}(p)$, then we can write this requirement in the form:
\begin{align} \vec{C} = p f \mathcal{B} \vec{C} \Rightarrow (p f \mathcal{B} - 1)\vec{C}=0,
\label{eq:multistatecond} \end{align}
Hence we see that in order to have a bound state, with both the short-distance and long-distance boundary conditions being satisfied, we must have $\text{det} (p f \mathcal{B} - 1)=0$ for some choice of $p=i\rho$ (with $\rho > 0$), ensuring that $p f \mathcal{B} - 1$ has at least one zero eigenvalue. The momentum associated with the bound state is $\rho$ where $p=i\rho$, and the coefficient vector $\vec{C}$ controls the eigenstate and must lie in the null space of $p f \mathcal{B} - 1$.

As a cross-check, in the case where the long-range potential induces no/negligible evolution in the wavefunction, then a wavefunction that is purely outgoing at large $r$ will also be purely outgoing at the matching radius $r=a$, and so this limit corresponds to $\mathcal{B} \rightarrow 0$. In this case, we see that (at least some elements of) $f$ must blow up, such that $f \mathcal{B}$ remains finite and non-zero in this limit, in order for a bound state to be possible; this corresponds to the familiar criterion that the scattering amplitude must have a (negative energy / imaginary momentum) pole for a bound state. However, for finite $\mathcal{B}$ and $f$, in principle both IR and UV contributions can work together to ensure the presence of a bound-state solution.

\subsection{When is tree-level matching sufficient?}

The methods reviewed/derived above allow us to characterize the Sommerfeld enhancement to annihilation and the presence of bound states in terms of the separate UV physics (encoded in $f$ or $f_\ell$) and IR physics (encoded in $\mathcal{B}$ for bound states, and $\bar{Z}_\ell$, $\Sigma_{0,\ell}$ for Sommerfeld enhancement). So long as the computation of the NR {\it potential} is sufficiently accurate (which may break down for the tree-level potential at strong coupling), we can compute the IR quantities including ladder diagrams at all orders, via the Schr\"{o}dinger equation.

However, if we want to compute the short-range/UV coefficients ($f$, $f_\ell$), they intrinsically originate from a regime where relativistic effects cannot be ignored, and so the calculation is more challenging. While physically the binding energies and Sommerfeld-enhanced annihilation cross sections should be independent of the matching radius and any choice of UV regulator, approximations or errors in $f$ and $f_\ell$ (which are generally matching-radius-dependent) can induce an apparent $a$-dependence or regulator dependence in physical predictions. One can also take the approach of simply treating $f$ and $f_\ell$ as phenomenological inputs; in this case, we want to understand how many parameters are needed to encode the short-distance physics.

This question already arises for elastic scattering. However, there is an important distinction between elastic scattering on one hand, and annihilation and bound state formation on the other hand: in the former case, it often suffices to compute the short-range amplitude $f$ in tree-level QFT / the first Born approximation; this was done in Ref.~\cite{Agrawal:2020lea,Parikh:2020ggm}, and shown to give cutoff-independent results. This procedure is equivalent to replacing the singular potential within $r=a$ with any regular potential that has the same first Born approximation for scattering in the $r < a$ region; because at leading order the UV scattering amplitude is well-approximated by the first Born approximation, any short-range regulating potential that shares the same first Born approximation for the scattering amplitude (over $0 < r < a$ and in the low-velocity limit) will give equivalent (and cutoff-independent) results for low-energy scattering, as discussed in e.g.~Ref.~\cite{Beane:2000wh}.

However, for bound states and Sommerfeld-enhanced annihilation, the processes of interest intrinsically enter $f$ at higher order (at least 1-loop to obtain the imaginary part that describes annihilation, and all orders for bound states), and so a tree-level matching will generally {\it not} be sufficient, or will only be sufficient in special circumstances (e.g. bound states that are almost entirely IR-supported where the short-range amplitude is perturbative). This difference will generally manifest itself as a cutoff/regulator dependence, even when restricting to the class of regulators that correctly match the tree-level behavior and so give cutoff/regulator-independent results for elastic scattering (as in Ref.~\cite{Agrawal:2020lea,Parikh:2020ggm}).

Let us first examine the case of Sommerfeld-enhanced annihilation. Here $f_\ell$ can be split into a pure-scattering term $f_{s,\ell}$ that has a tree-level contribution and persists in the absence of annihilation, and all other terms, which appear first at 1-loop order. Let us suppose there is some coupling $\alpha$ that governs the interactions, such that the tree-level amplitude is $\mathcal{O}(\alpha)$ and one-loop amplitudes are $\mathcal{O}(\alpha^2)$. Then at  $\mathcal{O}(\alpha)$, $f_{s,\ell}$ is Hermitian and its contribution to $f_\ell^\dagger - f_\ell$ cancels, so the leading-order term in the numerator of Eq.~\eqref{eq:annihilation} is $\mathcal{O}(\alpha^2)$. This corresponds to the tree-level (un-enhanced) annihilation cross section. The first perturbative correction to the annihilation cross section from the potential at $r<a$ (corresponding to the perturbative version of the Sommerfeld enhancement) thus enters at $\mathcal{O}(\alpha^3)$.

$f_{s,\ell}$ appears in both the numerator (squared) and the denominator of $\sigma_\text{ann} v_\text{rel}$. The matching prescription of Ref.~\cite{Parikh:2020ggm} corresponds to fixing $f_{s,\ell}$ to the result of applying the first Born approximation to the NR potential within $r < a$, or equivalently by matching the overall tree-level scattering cross section to the QFT calculation to fix $f_{s,\ell}$ (at tree level). We may replace the singular potential within $r=a$ with any alternative (regular) potential that has the same tree-level scattering amplitude $f_{s,\ell}$, and this will manifestly leave the $\mathcal{O}(\alpha)$ contributions to $f_{s,\ell}$ unchanged. However, this is not sufficient to fix the one-loop contributions, and these contribute to the numerator at the $\mathcal{O}(\alpha^3)$ level and to the denominator at the $\mathcal{O}(\alpha^2)$ level (corresponding to a $\mathcal{O}(\alpha^4)$ correction to the leading $\mathcal{O}(\alpha^2)$ term in the denominator). We work out an explicit example in appendix~\ref{app:coulomb}, where the long-range potential is the Coulomb potential, and within $r < a$ it is either unregulated or regulated with a spherical well. We see that even if the spherical well is chosen to match the short-range elastic scattering amplitude at tree level, the differences between the perturbative Sommerfeld enhancement in the two cases are of comparable size to the overall effect. (Another way to say this is that using the uncorrected annihilation amplitude amounts to ignoring the Sommerfeld enhancement from the region with $r < a$, and this contribution need not be negligible.)

Of course, when the short-range physics is in the perturbative regime, it is a viable solution to say that we should continue the matching to one-loop order, and require that $f_\ell$ (including annihilation and scattering) must match the QFT calculation at this order, effectively absorbing the calculation of short-range Sommerfeld enhancement into the QFT calculation (at least at leading order -- to go beyond the perturbative regime we would need a way to compute the full Sommerfeld enhancement from short distances, returning us to the problem that the matching was meant to solve). However, in this case it is more complicated to perform the position-space matching in the framework of quantum mechanics, since the 2nd Born approximation for a singular potential may diverge even if the 1st Born approximation is well-behaved (and it is easily checked that this is the case for the pseudoscalar-mediated potential), and the one-loop correction to the pseudoscalar-mediated potential yields terms that are singular even at the level of the 1st Born approximation (see appendix~\ref{app:loop_pot}).

An alternative approach  is to simply treat $f_\ell$ as a UV coefficient that must be measured. Provided $a \ll 1/p$ and the theory is weakly-coupled, we expect $f_\ell$ to be velocity-independent for $\ell=0$, and to behave as a contact interaction for higher $\ell$. Thus we may define the Sommerfeld enhancement as the ratio between the actual annihilation cross section and the annihilation cross section at some reference velocity, and consider all corrections from the potential within $r<a$ as higher-order corrections to the ``bare'' annihilation cross section, rather than contributors to the Sommerfeld enhancement. This provides a well-defined result for the IR Sommerfeld enhancement which should not depend on how the short-range physics is regulated, and will agree well with the standard Sommerfeld-enhancement calculation when there is a large Sommerfeld enhancement from IR physics and the higher-order corrections to the annihilation amplitude are negligible (as is typically the case for theories with small $\alpha$ but a long-range Yukawa potential). However, if the corrections are dominated by the short-range physics, then this redefined IR Sommerfeld enhancement may be negligible. This is similar to the ``wavefunction renormalization'' discussed in \cite{Bellazzini:2013foa}, where the enhancement is normalized to be 1 as $v\rightarrow 1$. However, we caution that for $v\sim\mathcal{O}(1)$, the momentum may be sufficiently large to probe the small-$r$ regulated region, and so the definition of the renormalized annihilation cross section may depend non-negligibly on the reference velocity.

The situation for bound states is closely related. Just as two short-range regulator potentials that have the same leading-order scattering amplitude may differ in their higher-order contributions, they may give rise to quantitatively different bound state spectra. It has been previously demonstrated that even in the regimes of perturbative couplings, IR bound states obtained in QM/NREFT may not correspond to existing bound states in the full QFT, and the correspondence failure arises from the need to perform an all-order matching both in the coupling and higher dimension operators to obtain the QM Hamiltonian or NREFT Lagrangian~\cite{Luke:1996hj}. In our framework, if we regulate the singular potential at short distances with a potential constrained via tree-level matching, we should only expect this to lead to a cutoff/regulator-independent spectrum of bound states when $f$ is small and perturbative (and well-approximated by the 1st Born approximation), and thus $\mathcal{B}$ must have very large components to meet the bound state condition. This corresponds to a bound state that is primarily IR-supported.

\subsection{Numerical tests with tree-level matching for a pseudoscalar mediator}

We can test the discussion above explicitly by numerically solving for the Sommerfeld enhancement to annihilation, in the case where we replace the singular potential with a regulating potential inside a matching radius $r=a$. We will focus on the $j=1$, $s=1$ sector where the $\ell=0,2$ states are coupled by spin-dependent interactions (from pseudoscalar, pseudo-Goldstone or axial vector exchanges), and specifically on the case of pseudoscalar exchange with a renormalizable coupling, using results summarized in Sec.~\ref{sec:potentials}. For this scenario, we expect the NR potential to be valid so long as the momentum scale is well below $m_\chi$ (i.e.~there is no intermediate cutoff scale $\Lambda < m_\chi$ above which the theory must be modified), and so we choose $a\sim 1/m_\chi$. We will focus on the weak-coupling region $\alpha \lesssim 1$, as at higher couplings we expect our analysis to become inaccurate due to large corrections to the potential.

We could similarly numerically solve for the presence of bound states (the Schr\"{o}dinger equation is identical). However, in a regulated (and hence non-singular) potential, the peaks in the Sommerfeld enhancement signal the entry of bound states into the spectrum (we have verified this by explicit calculation). Thus we can focus on the enhancement calculation to explore the apparent onset of bound states, and the degree to which this onset is cutoff/regulator-dependent.

\subsubsection{Constructing the regulator potentials}
\label{sec:regulator}

We can parameterize the possible contributions to the matrix potential (Eq.~\eqref{eq:potential-longRange}) from pseudoscalar, pseudo-Goldstone or axial vector exchange as:
\begin{align}
    V_C(r) & = 4\pi \kappa_1 \delta^{(3)}(\vec{r}) + \kappa_2 e^{-m_\phi r}/r, \quad V_T(r) = \kappa_3 (1 + m_\phi r + m_\phi^2 r^2/3) e^{-m_\phi r}/r^3,
    \label{eq:VC-VT-in-terms-of-kappas}
\end{align}
where $m_\phi$ is replaced by $m_A$ in the axial vector case.
We can read off the leading-order scattering amplitude in the Born approximation, for scattering within $r=a$, following Refs.~\cite{Agrawal:2020lea,Parikh:2020ggm}, as:
\begin{align}
p f_{\ell s}^{\ell^\prime s^\prime}(p) =- \frac{m_\chi}{p} \int^a_0 dr~s_{\ell^\prime}(p r) (V(r))_{\ell \ell^\prime} s_\ell(p r).
\end{align}

In this case, $s=1$ for both relevant states ($\ell=0$ and $\ell=2$), and the matrix elements are:\footnote{Note the 3D delta function is normalized as $\int d^3 r~\delta^{(3)}(\vec{r}) = 1$, which in terms of the 1D radial integral we write as  $\int r^2 dr~\delta^{(3)}(\vec{r}) = \frac{1}{4\pi}$.}
\begin{align} p f_{01}^{01} & = -\frac{m_\chi}{4p} \int^a_0 dr~ (4\pi \kappa_1 \delta^{(3)}(\vec{r}) + \kappa_2 e^{-m_\phi r}/r) \sin(p r)^2 \nonumber \\
p f_{01}^{21} & =  p f_{21}^{01} =  -\frac{\sqrt{8} m_\chi}{4p} \kappa_3 \int^a_0 dr~ (1 + m_\phi r + m_\phi^2 r^2/3) (e^{-m_\phi r}/r^3) \sin(p r) s_2(p r) \nonumber \\
p f_{21}^{21} & = -\frac{m_\chi}{4p} \int^a_0 dr~ (\kappa_1 \delta^{(3)}(\vec{r}) + \kappa_2 e^{-m_\phi r}/r - 2 \kappa_3  (1 + m_\phi r + m_\phi^2 r^2/3) (e^{-m_\phi r}/r^3)) s_2(p r)^2 .  \end{align}
If we choose the matching radius such that $a \ll 1/p$ and $a \ll 1/m_\phi$ (for example, for $a\approx 1/m_\chi$ we expect both these conditions to hold for NR scattering and a light mediator), then we can approximate $e^{-m_\phi r} \approx 1$ and $s_\ell(p r) \equiv p r j_\ell(p r) \approx \frac{1}{(2\ell+1)!!} (p r)^{\ell+1}$. In particular, $s_0(p r) \approx p r$, and $s_2(p r) \approx \frac{1}{15} (p r)^3$. With this assumption the matrix elements simplify to:
\begin{align} p f_{01}^{01} & = -\frac{m_\chi p}{4} \left[ \kappa_1 + \kappa_2 a^2/2 \right] \nonumber \\
p f_{01}^{21} & = p f_{21}^{01} =  -\frac{\sqrt{8} m_\chi p^3}{60} \kappa_3 \left[a^2/2 + m_\phi a^3/3 + m_\phi^2 a^4/12 \right] \approx  -\frac{\sqrt{2} m_\chi p^3}{60} \kappa_3 a^2  \nonumber \\
p f_{21}^{21} & = -\frac{m_\chi p^5}{4 \times 15^2}  \left[\kappa_2 a^6/6 - 2 \kappa_3 (a^4/4 + m_\phi a^5/5 + m_\phi^2 a^6/18)  \right] \approx -\frac{m_\chi p^5}{4 \times 15^2}  \left[\kappa_2 a^6/6 - \kappa_3 a^4/2  \right]     \end{align}
where we have used $m_\phi a \ll 1$. As expected from Refs.~\cite{Agrawal:2020lea,Parikh:2020ggm}, these scattering amplitudes are finite for finite coupling strengths.

Note that for a spherical well potential with value $V(r)=\overline{V}$ extending out to $r=a$, under the same approximations we would have:
\begin{align}
    p f_{01}^{01} & = -m_\chi p \overline{V} a^3/3, \nonumber \\
    p f_{21}^{01} & = -m_\chi p^3 \overline{V} a^5 /75, \\
    p f_{21}^{21} & = -m_\chi p^5 \overline{V} a^7 /(7\times 15^2).
\end{align}
Thus a spherical well potential that matches the pseudoscalar/axial-vector potential in terms of the low-momentum scattering amplitude has the structure (within $r=a$):
\begin{align}
    V(r) = \frac{1}{4}
    \begin{pmatrix}
        3 \left( \frac{\kappa_1}{a^3} + \frac{\kappa_2}{2 a}\right) & 5 \sqrt{2} \frac{\kappa_3}{a^3} \\ 5 \sqrt{2} \frac{\kappa_3}{a^3} & 7 \left( \frac{\kappa_2}{6 a} - \frac{\kappa_3}{2 a^3} \right)
    \end{pmatrix}
    \label{eq:ShortRangePotential-in-terms-of-kappas}
\end{align}

Similarly, if the regulator potential was written as $V(r) = \overline{V}/r$ within $r=a$ for some constant $\overline{V}$, the scattering amplitudes would take the form:
\begin{align} p f_{01}^{01} & = -m_\chi p \overline{V} a^2/2, \nonumber \\
p f_{21}^{01} & = -m_\chi p^3 \overline{V} a^4 /60, \\
p f_{21}^{21} & = -m_\chi p^5 \overline{V} a^6 /(6\times 15^2), \end{align}
so the potential that matches the low-momentum scattering amplitude for the full pseudoscalar/axial-vector potential would take the form:
\begin{align} V(r) & = \frac{1}{4 r} \begin{pmatrix}   \frac{2 \kappa_1}{a^2} + \kappa_2 & 4 \sqrt{2} \frac{\kappa_3}{a^2} \\4 \sqrt{2} \frac{\kappa_3}{a^2} & \kappa_2 - 3 \frac{\kappa_3}{a^2} \end{pmatrix}. \end{align}

As discussed previously, we do not expect this tree-level matching to be sufficient to match the bound states (and in fact it can be readily checked that the two possible regulator potentials described above do not have the same bound-state spectrum; an explicit example to this effect is given in appendix~\ref{app:coulomb}). However, we can still examine qualitatively where the first bound states would appear for either regulating potential, to give us a sense of where corrections to the tree-level short-range amplitude are likely to become large.

In a single-state 3D spherical well of depth $\overline{V}$, the condition for the first bound state to exist is $ \overline{V} > (\pi/2)^2/(m_\chi a^2)$ (this is a standard result found in many textbooks, e.g.~\cite{griffiths_introduction_2018}). Thus in order for a bound state to be supported by our more complicated multi-state potential, we would guess that we need at least one component of this effective short-distance potential matrix to satisfy the same relation. In terms of the $\kappa$ coefficients, this would correspond to requiring:
\begin{align} \kappa_1, \kappa_3 \gtrsim a/m_\chi, \, \text{or} \, \kappa_2 \, \gtrsim 1/(m_\chi a). \end{align}
Since we expect $a\sim m_\chi^{-1}$, the condition on $\kappa_2$ amounts to $\kappa_2 \gtrsim 1$. For the pseudoscalar (pseudo-Goldstone) potentials, the corresponding conditions on $\kappa_1$, $\kappa_3$ become $\alpha \gtrsim 1$ ($G \gtrsim 1$). For the axial vector case, the condition becomes $\alpha/m_A^2 \gtrsim 1/m_\chi^2$; this naively only seems to require quite a small coupling $\alpha \gtrsim (m_A/m_\chi)^2$, but this corresponds to a large Yukawa coupling in the axial vector scenario of Ref.~\cite{Agrawal:2020lea}.

For the case where the short-range potential is Coulomb-like, $V(r)=\overline{V}/r$, again, we would estimate that a bound-state can only exist in this case if the range of the potential $a$ and the strength of the potential $\overline{V}$ satisfy $\overline{V} m_\chi a \gtrsim 1$. So again this suggests we would require either $\kappa_2 \gtrsim 1/(m_\chi a)$ or $\kappa_1, \kappa_3 \gtrsim a/m_\chi \sim 1/m_\chi^2$. That is, forming a bound state with a constant potential with support only within $r\sim 1/m_\chi$ would require the theory to have large dimensionless effective couplings (which is consistent with the standard lore for $1/r$ potentials). For weaker couplings, the short-distance amplitude will be  small and perturbative, which supports the validity of the tree-level matching discussed here, but also means that bound states can only occur if they are primarily supported by the IR physics.
We will next explore this possibility numerically.

\subsubsection{Numerical procedure for Sommerfeld-enhanced annihilation}
\label{sec:numeric_approach}

To solve for the Sommerfeld enhancement to annihilation with a regulated short-distance potential, we need to solve the radial Schr\"odinger equation for the $\ell^{\rm th}$ partial wave (the reduced mass is $m_\chi/2$ and as previously $E$ is the single-particle energy)
\begin{align}
    -\frac{1}{m_\chi}\frac{\dd^2 u_{\ell}}{\dd r^2} + V_\text{eff}(r)\, u_{\ell}(r) = 2 E \, u_{\ell}(r)\:,\: V_\text{eff}(r) = V(r) + \frac{\ell(\ell+1)}{m_\chi r^2}\:,
\end{align}
where the potential $V(r)$ has the general form
\begin{align}
    V(r) = \Theta (a-r)\, V_S(r)  + \Theta (r-a)\,V_L(r)\:.
\end{align}
Here the subscripts $S, L$ denote the short and long-range parts of the potential, $\Theta$ is the Heaviside step function, and the boundary separating the long and the short-range parts is at $r = a$. The singular terms appear only in $V_L$ and since it is restricted to $r > a$, the Schr\"odinger equation is well posed.

Since the potential in Eq.~\eqref{eq:potential-pseudoscalar} has a non-zero $V_T(r)$ part,
it mixes $\ell=0$ and $\ell=2$ partial waves as described in Eq.~\eqref{eq:potential-longRange}. We will refer to the corresponding reduced radial wavefunctions as $u$ and $w$ respectively. Therefore one has to solve a matrix valued equation
\begin{align}
    &-\frac{1}{m_\chi}\Psi''(r) + V_\text{eff}(r) \Psi(r) = 2 E\, \Psi(r)\:,\:\:
    \Psi(r) = \begin{pmatrix}
        u(r) \\
        w(r)
    \end{pmatrix}
    \nonumber \\
    & V_\text{eff}(r) = \Theta (a-r)\, V_S(r)  + \Theta (r-a)\,V_L(r) + \begin{pmatrix}
        0 & 0 \\
        0 & \frac{6}{m_\chi r^2}
    \end{pmatrix}
\end{align}
where both $V_S, V_L$ are now $2\times2$ matrices and the last term in the second line comes from the centrifugal barrier.  The explicit expression for $V_L(r)$ is given in Eq.~\eqref{eq:potential-longRange}, and is repeated here for convenience:
\begin{align}
    V_L(r) = \frac14
    \begin{pmatrix}
        V_C(r) \:\: & \:\: \sqrt{8} V_T(r) \\
        \sqrt{8}V_T(r) \:\:  & \:\: V_C(r) - 2V_T(r)
    \end{pmatrix}\:,
\end{align}
with $V_C, V_T$ defined in Eq.~\eqref{eq:potential-pseudoscalar}.

To solve the equations, we also need to provide the boundary conditions. Regularity at the origin $r=0$ requires $u(r)\propto r$ and $w(r) \propto r^3$, for small $r$. In addition to this, we also want the large-$r$ solution to be of the form~\cite{Bedaque:2009ri}
\begin{align}
    \lim_{r\to\infty} u(r) = \frac{e^{i\delta_0}}{p}\sin(p r + \delta_0)\:,\:\:
    \lim_{r\to\infty} w(r) = \frac{\sqrt{2}e^{i\delta_2}}{p}\sin(p r -\pi + \delta_2)\:,
    \:\: p = \sqrt{2 m_\chi E} \:. \label{eq:large_r}
\end{align}
Here, we have dropped an overall multiplicative factor of $\sqrt{4\pi}$ in the wave function, which drops out in the computation of SE. At this point, the coefficients $A_u^S, A_w^S$ and the phase shifts $\delta_0, \delta_2$ are undetermined, and to find the right large $r$ solution we need to compute $\delta_0, \delta_2$. However one can construct the solution with the right $r\to\infty$ behavior by constructing two linearly independent solutions $(u_1,w_1)$ and $(u_2, w_2)$ that satisfy $u_1(0) = u_2(0) = w_1(0) = w_2(0) = 0$. For large $r$, they behave like
\begin{align}
    &u_1(r) \to A_u^{(1)} e^{-i p r} - B_u^{(1)} e^{ipr}\:,
    u_2(r) \to  A_u^{(2)} e^{-i p r} - B_u^{(2)} e^{ipr}\:,
    \nonumber \\
    &w_1(r) \to A_w^{(1)} e^{-i p r} - B_w^{(1)} e^{ipr}\:,
    w_2(r) \to  A_w^{(2)} e^{-i p r} - B_w^{(2)} e^{ipr}\:.
    \label{eq:linearly-independent-sol-large-r}
\end{align}
One can then show that the linear combination~\cite{Bedaque:2009ri}
\begin{align}\label{eq:correct-sol-from-linear-combination}
    &\begin{pmatrix}
        u_1 & u_2 \\
        w_1 & w_2
    \end{pmatrix} \cdot
    \begin{pmatrix}
        A_u^{(1)} & A_u^{(2)} \\
        A_w^{(1)} & A_w^{(2)}
    \end{pmatrix}^{-1} \cdot
    \begin{pmatrix}
        1 \\
        -\sqrt{2}
    \end{pmatrix} \times (i/2p)
\end{align}
has the appropriate large $r$ behavior in Eq.~\eqref{eq:large_r}.\footnote{E.g. when there is no potential, $w$ solutions decouple. Setting $A_u^{(1)} = B_u^{(1)} = i/(2p), A_u^{(2)} = B_u^{(2)} = 0, A_w^{(1)} = B_w^{(1)} = 0$,
and $A_w^{(2)}$ to any arbitrary value, the first component of the solution in Eq.~\eqref{eq:correct-sol-from-linear-combination} reduces to $u(r)=u_1(r)$, which from Eq.~\eqref{eq:linearly-independent-sol-large-r} is given by $u_1(r)=i/(2p) (-2i\sin(p r)) = \sin(pr)/p$ as expected.
}

The numerical procedure is therefore the following. We first solve for the regular (vector) solution to the short-range regulating potential within $r<a$,
evaluated at $r=a$, to determine the boundary conditions needed to solve from $r=a$ to some large $r = r_\infty$. We solve it twice, once with $A_u^S = 0, A_w^S = 1$ and once with $A_u^S = 1, A_w^S = 0$. These two provide the two linearly independent solutions satisfying the right boundary conditions at $r = 0$. We then solve for the coefficients $A_u^{(1)}, A_u^{(2)}, A_w^{(1)}, A_w^{(2)}$ using Eq.~\eqref{eq:linearly-independent-sol-large-r} as an equation at $r = r_\infty$. Finally we construct the right solution using Eq.~\eqref{eq:correct-sol-from-linear-combination}. The ($s$-wave) SE is then computed using
\begin{align}
    \text{SE} = \lim_{r\to 0}\left|\frac{u(r)}{u_0(r)}\right|^2
    = \lim_{r\to 0}\left|\frac{u(r)}{\sin (pr)/p}\right|^2
    \:,
    \label{eq:SE-def}
\end{align}
where in the last equality we have used the expression for the reduced wave function with no potential, $u_0 = \sin (pr)/p,\ p = \sqrt{2m_\chi E}$. Here we have assumed that annihilation from the $d$-wave channel can be ignored due to its greater suppression at low velocities (in the presence of a Coulomb potential, the additional suppression by powers of $v$ is replaced by powers of the coupling $\alpha$, but as a result the contribution to the annihilation rate is still suppressed).

For numerical convenience it is more useful to work with variables
\begin{align}
    x &= r\, m_\phi\ ,\ & \xi^2 &= \frac{2 E\,m_\chi}{m_\phi^2} = \frac{p^2}{m_\phi^2}\ ,\ &
    v(x) &= \frac{m_\chi V_\text{eff}(r)}{m_\phi^2}\ .\label{eq:dimensionless-variables}
\end{align}

The condition of being NR, i.e. $E \ll m_\chi$,
translates to $\xi \ll m_\chi/m_\phi$.

In terms of these dimensionless variables, the equation to be solved is
\begin{align}
    -\Psi''(x) + v(x) \Psi(x) = \xi^2\, \Psi(x)\:,
\end{align}
where
\begin{align}
    \Psi(x) &= \begin{pmatrix}
        u(x) \\
        w(x)
    \end{pmatrix}\:,\:\:
    v(x) = \Theta(x_0 - x)\,v_S(x)  + \Theta(x - x_0)\,v_L(x) +
            \begin{pmatrix}
                0 & 0 \\ 0 & 6/x^2
            \end{pmatrix}
   \nonumber \\
    v_L(x) &=
    \frac{\alpha}{12}\frac{m_\phi}{m_\chi}\, \frac{e^{-x}}{x^3}\,\begin{pmatrix}
                x^2 &
                \sqrt{8}  \left(3 + 3 x + x^2\right) \\
               \sqrt{8}  \left(3 + 3 x + x^2\right) &
               \:\:\:\: -x^2 - 6 x - 6
           \end{pmatrix}\:,
\end{align}
and we have chosen $v_L(x)$ to correspond to the case of pseudoscalar exchange (Eq.~\eqref{eq:potential-pseudoscalar}).

Note that the centrifugal term $6/x^2$ is present on both sides of the matching radius.
$x_0$ represents the matching radius, and by default we will choose $x_0 = m_\phi/m_\chi$, corresponding to $r=1/m_\chi$ (the Compton length of the DM particles); however, we will study the effect of varying this parameter by a $\mathcal{O}(1)$ factor to diagnose regulator/cutoff-dependence.

In terms of these dimensionless quantities, the regime of our earlier NREFT, $m_\phi \ll E$, becomes
\begin{align}
    \xi \gg \left(m_{\chi}/m_{\phi}\right)^{1/2}
    \:.
\end{align}
Since $m_\chi/m_\phi \gg 1$, one can have a scenario where $\xi$ is small enough to be in the NR regime, $\xi \ll m_\chi/m_\phi$, but large enough to violate the NREFT requirement.

To this point, nothing about this numerical approach has required specifying the short-range potential $v_S(x)$, and in the next section we will consider a form for $v_S(x)$ with more freedom, representing a broader range of possible UV physics. However, for our analysis of the renormalizable theory with pseudoscalar exchange, we model the short-distance potential as a spherical well and fix its normalization (as a function of the matching radius $a$) by matching the first Born approximation for the scattering amplitude (which should be equivalent by construction to matching the tree-level QFT calculation). Starting with Eq.~\eqref{eq:potential-pseudoscalar}, one can write $\kappa_1, \kappa_2, \kappa_3$ appearing in Eq.~\eqref{eq:VC-VT-in-terms-of-kappas} by a term by term comparison, and translate that to the short-range potential parametrized in terms of $\kappa_1, \kappa_2, \kappa_3$ in Eq.~\eqref{eq:ShortRangePotential-in-terms-of-kappas}. After doing this, the dimensionless potential $v_S(x)$ in the short distance region $x < x_0= a\,m_\phi$ is given as
\begin{align}
   v_S(x) =
   \frac{\alpha}{72} \frac{m_\phi}{m_\chi}
   \begin{pmatrix}
        -18/x_0^3 + 9/x_0 & 90\sqrt{2} /x_0^3 \\
        90\sqrt{2} /x_0^3 & 7/x_0 - 63/x_0^3
   \end{pmatrix}
   \:\:,\:\: x_0 = a\, m_\phi\:.
   \label{eq:ShortRangePotential-Goldstone-in-terms-of-G}
\end{align}
Recall that this result holds for both $\chi \chi$ and $\chi\bar{\chi}$ scattering (with the latter being relevant for annihilation), since the contact terms that are only present in the $\chi\bar{\chi}$ state vanish in the spin-triplet configuration.

\subsubsection{Numerical results}
\label{subsubsec:pseudoscalar-numerical-results}

\begin{figure}
    \centering
    \includegraphics[width=0.9\linewidth]{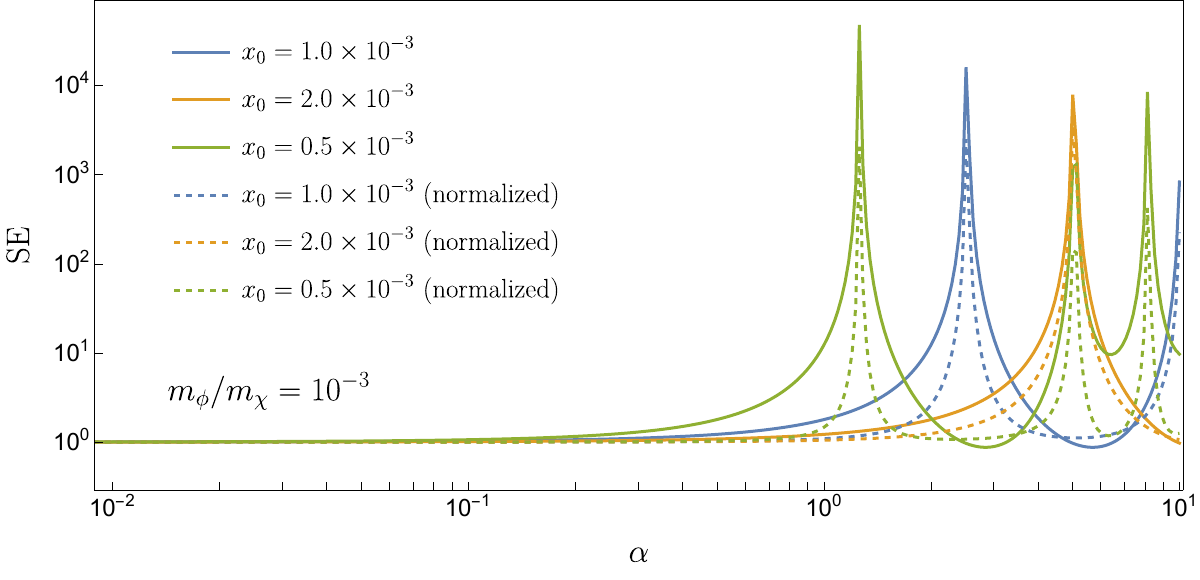}
    \caption{
    \small{SE as a function of $\alpha$ for the pseudoscalar mediator, varying the choice of matching scale $x_0$. The long-range and the short-range parts of the potential are fixed by $\alpha$. We have taken $m_\phi/m_\chi = 10^{-3}$. The solid lines show $\text{SE}(\xi = 1, \alpha)$ as a function of $\alpha$. The dashed lines show SE normalized by its value at $\xi = 500$, i.e. $\text{SE}(\xi = 1, \alpha)/\text{SE}(\xi = 500, \alpha)$, as a function of $\alpha$.}
    }
    \label{fig:SE-vs-alpha-pseudoscalar-ShortRange-fixed-by-LongRange}
\end{figure}

In Figure~\ref{fig:SE-vs-alpha-pseudoscalar-ShortRange-fixed-by-LongRange}, we show the numerically evaluated SE from a pseudoscalar exchange as a function of the coupling $\alpha$, for $m_\phi/m_\chi=10^{-3}$, and for a few values of the matching scale $x_0$. The solid lines show SE at $\xi = 1$, while the dashed lines show SE at $\xi = 1$ normalized by SE at $\xi = 500$, both evaluated as a function of $\alpha$.
We observe that for $x_0 = 10^{-3}$ there is a peak in the SE at $\alpha\sim 2.5$ which we expect to correspond to a zero-energy bound state entering the spectrum; as we have discussed, we expect this to correspond to our matching calculation breaking down, so the onset of this bound state should be cutoff-dependent. This is seen clearly by the movement of the location of the peaks in SE, as the parameter $x_0$ is varied in Fig.~\ref{fig:SE-vs-alpha-pseudoscalar-ShortRange-fixed-by-LongRange}. We observe that except at the cores of the peaks, the enhancement can be largely removed by rescaling by the enhancement at $\xi=500$ (i.e.~this apparent enhancement is not generally a low-velocity effect), but at the centers of the (cutoff-dependent) peaks we also observe an apparent enhancement at low velocities. We find no non-negligible inferred SE (especially when factoring out the high-velocity enhancement) for $\alpha$ well below the onset of the first peak. This is consistent with the absence of non-perturbative modifications to the scattering rate found in Ref.~\cite{Agrawal:2020lea} and our earlier NREFT analysis.

\begin{figure}
    \centering
    \includegraphics[width=0.6\linewidth]{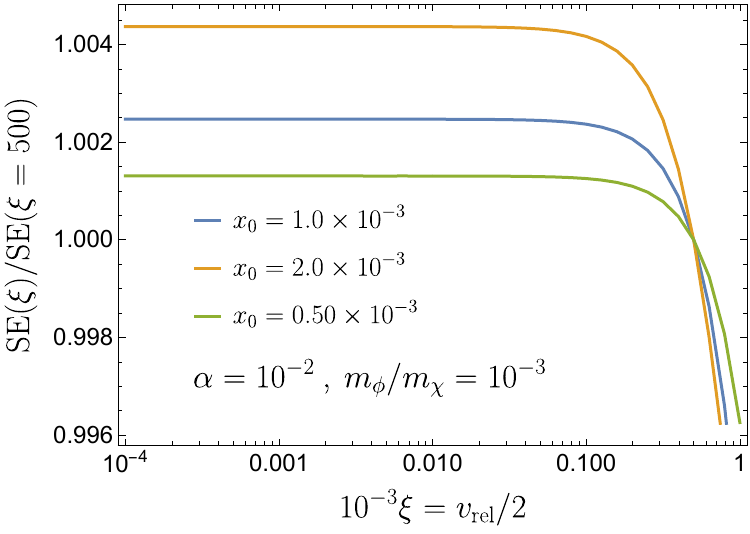}
    \:
    \caption{\small{SE as a function of $\xi$, normalized at $\xi = 500$, for three choices of matching scales $x_0$ and $\alpha = 10^{-2}$. For this perturbative value of the coupling $\alpha$, away from any peaks, we see there is no appreciable SE, as expected from NREFT.    }}
    \label{fig:SE-vs-alpha-on-peak-pseudoscalar-ShortRange-fixed-by-LongRange}
\end{figure}

Fig.~\ref{fig:SE-vs-alpha-on-peak-pseudoscalar-ShortRange-fixed-by-LongRange} shows SE as a function of $\xi$, normalized at its value at $\xi = 500$, for a generic perturbative value of $\alpha = 10^{-2}$, for three choices of the matching scale $x_0$. The choice of $\xi_0 = 500$ corresponds to $E_0/m_\chi = (\xi_0 m_\phi/m_\chi)^2/2 = 0.5^2/2$, corresponding to a single-particle velocity of $0.5 c$. For all three cases, no significant SE is seen and the deviation from 1 is cutoff-dependent at the $\mathcal{O}(1)$ level. We also note that close to the reference velocity, the SE is quite velocity-dependent, suggesting that this velocity is high enough to probe the UV/short-range potential and this is the reason for the cutoff dependence. If we had chosen a reference velocity corresponding instead to $\xi=100$ (factor of 5 smaller velocity), this would greatly mitigate the cutoff dependence, but in that case we would also find that there is no SE at lower velocities. This behavior implies that all the inferred SE in the baseline case is coming from regions that require velocities $\gtrsim 0.1 c$ to probe and cannot be separated cleanly from the relativistic physics, in contrast to the case of non-singular potentials where the enhancement dominantly arises from larger radii where the NR limit is clearly valid. This is consistent with the result from our NREFT calculation suggesting a lack of any low-velocity enhancement. Note that the condition for the NREFT analysis to hold is $\xi \gg (m_\chi/m_\phi)^{1/2} \approx 33$, but the QM analysis suggests that for small couplings there should be no enhancement even outside the range of validity of the EFT.

\begin{figure}
    \centering
    \includegraphics[width=0.7\linewidth]{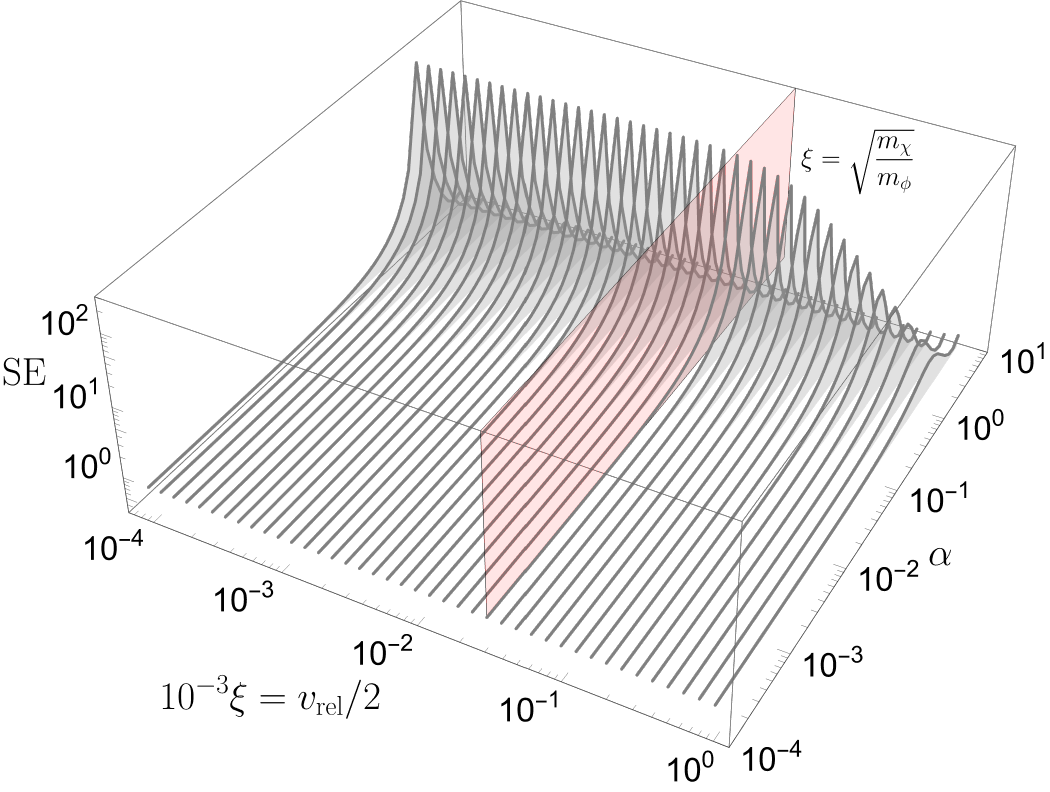}
    \caption{
    \small{SE for pseudoscalar mediator, as a function of $\alpha$ and $10^{-3} \xi =  v_\text{rel}/2$, where $v_\text{rel}$ is the relative velocity between the DM particles. The plane $\xi = \sqrt{m_\chi/m_\phi}$ is shown in light red. NREFT holds for $\xi \gg \sqrt{m_\chi/m_\phi}$ and $\alpha\lesssim 1$. For such values of $\xi, \alpha$ we do not see a substantial SE, as expected. We have taken $m_\phi/m_\chi = 10^{-3}.$}
    }
    \label{fig:SE-vs-G-vs-xi}
\end{figure}

Finally, Fig.~\ref{fig:SE-vs-G-vs-xi} shows the behavior of the SE as the coupling $\alpha$ and the velocity $\xi$ vary simultaneously, for the same fixed value of $m_\phi/m_\chi=10^{-3}$ and matching radius $x_0=10^{-3}$. The $\xi = (m_\chi/m_\phi)^{1/2}$ plane (corresponding to the boundary of validity of the NREFT approach) is also shown. We see that for perturbative couplings, there is no SE, as expected from the NREFT, across a broad range of $\xi$.

Thus we confirm that for weak couplings ($\alpha \ll 1$), there is no low-velocity enhancement associated with the pseudoscalar-mediated potential. Where we do see an apparent enhancement, for small couplings it is largely velocity-independent for NR velocities, and can be fully absorbed into the UV coefficient. For large couplings ($\alpha \gtrsim 0.75$), the calculation nominally gives large enhancements associated with the onset of a bound state, but we see that these enhancements (and thus the associated bound state) are strongly cutoff-dependent, and are also nearly velocity-independent (i.e.~already present at high velocity) except at the center of the peaks. For this reason, we view the large cutoff-dependent enhancements at high $\alpha$ as signaling a breakdown in the tree-level matching approach we have used to constrain the short-range regulator potential, rather than a physical effect (in this $\alpha \gtrsim 1$ regime there are also likely to be substantial corrections to the potential we have used at long distances).

\section{The pseudo-Goldstone case and the interplay of UV/IR physics}\label{sec:numerics}

We have established in the previous sections that both the NREFT and QM approaches predict negligible Sommerfeld enhancement from pure pseudoscalar exchange in the weak-coupling regime ($\alpha \ll 1$). We can see (directly from Eqs.~\eqref{eq:potential-pseudoscalar},\eqref{eq:potential-goldstone}) that the pseudo-Goldstone case predicts the same potential up to the replacement $\alpha \rightarrow 3 G =g^2 m_\chi^2/(\pi \Lambda^2)$. However, the pseudo-Goldstone theory has a cutoff at $p\sim \Lambda$, in addition to the breakdown of the NR limit at $p\sim m_\chi$. For $\Lambda \gtrsim m_\chi$ the two theories are essentially identical for our purposes and all our results for the pseudoscalar apply also to the pseudo-Goldstone case. However, for $\Lambda < m_\chi$, the situation is different in several ways: (1) the potential we have derived is only valid for $r \gtrsim 1/\Lambda$ (so we should set $a\sim 1/\Lambda$ instead of $a\sim 1/m_\chi$), (2) the region $r < 1/\Lambda$ cannot be well-described by the relativistic Lagrangian, instead we need to include the $\Lambda$-scale physics that has been integrated out to obtain the pseudo-Goldstone Lagrangian term, and (3) the effective strength of the potential is enhanced by a factor $(m_\chi/\Lambda)^2$, even as the underlying theory remains weakly-coupled. This is similar to the situation studied in Ref.~\cite{Ferrante:2025lbs} where a large ratio $m_\chi/\Lambda$ appears to give rise to Sommerfeld enhancement (albeit with dependence on the modeling of the UV physics) in an EFT that gives rise to a $1/r^3$ potential.

In this section we will thus focus on the pseudo-Goldstone case, allowing $\Lambda < m_\chi$, and parameterizing the short-distance physics ($r < 1/\Lambda$) with a spherical well. We will explore the interplay between the short-range and long-range physics with regard to bound state formation and Sommerfeld-enhanced annihilation in this setting.

We will follow the same analysis setup as in Sec.~\ref{sec:numeric_approach} (and work with the same dimensionless quantities defined in Eq.~\eqref{eq:dimensionless-variables}), with the exception that we modify the short-range dimensionless potential $v_S(x)$ to be a diagonal spherical well, with two free parameters $v_0$ and $\lambda$:
\begin{align}
    v_S(x) &=
    \begin{pmatrix}
        -v_0 & 0 \\
        0 & -v_0\lambda
    \end{pmatrix}\:.
\end{align}
The long-range potential in dimensionless quantities can be obtained using Eq.~\eqref{eq:potential-goldstone}, and is given as
\begin{align}
     v_L(x) &= G\frac{m_\phi}{4m_\chi}\,\frac{e^{-x}}{x^3}
     \begin{pmatrix}
        x^2 &
        \sqrt{8}  \left(x^2+ 3x + 3\right) \\
        \sqrt{8}  \left(x^2+ 3x + 3\right) &
        \:\:\:\: -x^2 - 6x - 6
    \end{pmatrix}\:.
\end{align}
Crucially, the matching scale $x_0 = m_\phi/\Lambda$ (for $\Lambda < m_\chi$) is different than the previously-considered case of a renormalizable pseudoscalar interaction. Note that our results in this section will apply to both $\chi \chi$ and $\bar{\chi} \chi$ scattering (because the only differences are encoded in contact operators, which in any case vanish for the spin-triplet case as discussed previously). This analysis builds on and is similar to earlier work in Ref.~\cite{Bedaque:2009ri}, but unlike that analysis, we aim to carefully test the separate effects of the short-distance and long-distance potential.

 Since the short-range part of the potential is simple, this fixes the wavefunctions in the $x < x_0$ region to be
\begin{align}
    u(x) &= A_u^S \, \sin (x p_u)\:,\:\: p_u = \sqrt{\xi^2 + v_0}\:,
    \nonumber \\
    w(x) &= A_w^S
    \frac{(x^2 p_w^2-3)\sin(x p_w)+ 3 x p_w \cos(x p_w)}{x^2 p_w^2}
    \:,\:\: p_w = \sqrt{\xi^2 + v_0 \lambda}
    \:.
    \label{eq:sol-small-r}
\end{align}

\subsection{Parameterizing the UV physics}

For calibration, note that the $s$-wave Sommerfeld enhancement from a spherical well is given by (e.g.~\cite{Arkani-Hamed:2008hhe}):
\begin{align}S = \frac{1}{\frac{\xi^2}{\xi^2 + v_0} \sin^2(x_0 \sqrt{\xi^2 + v_0}) + \cos^2(x_0 \sqrt{\xi^2 + v_0})} \label{eq:analyticSE} \end{align}
Thus in the limit of small $\xi^2 \ll v_0$, the enhancement is solely determined by $x_0\sqrt{v_0}$ (note that this is not the same scaling as the Born approximation for the scattering amplitude, which scales as $x_0^3 v_0$). Accordingly, if we parameterize the strength of the short-range potential well by $v_0 x_0^2$, the enhancement from the short-range spherical well alone should be independent of $x_0$ in the case of $\lambda=0$ (where there is no $d$-wave contribution).

Thus if we parameterize the cutoff by $x_0$ and the strength of the short-range potential (for purposes of the Sommerfeld enhancement) by $x_0^2 v_0$, the purely UV Sommerfeld enhancement will be a function of only $x_0^2 v_0$, whereas IR effects will depend  on $x_0$ (via their dependence on the cutoff $\Lambda$ which sets the coupling of the IR Lagrangian). Note that for $\lambda \ne 0$, the IR physics can provide non-negligible mixing with the $d$-wave component of the short-range potential, also inducing a dependence on $x_0$ and $x_0^2 v_0$ separately, but since this only happens in the case where there is a long-range potential, this does not violate the earlier statement that SE purely from the short-distance spherical well is expected to depend solely on $x_0^2 v_0$.

More broadly, we could ask how many parameters are needed in general to capture the full UV physics, for purposes of computing observables such as the Sommerfeld-enhanced annihilation rate or the elastic scattering cross section. We have argued that it is always possible to write the annihilation cross section, the full scattering amplitude, and the condition for bound-state formation, in terms of IR physics + the UV scattering amplitude, which sets the boundary conditions at the matching radius; the UV scattering amplitude is a complex number (or matrix of numbers in the case of coupled channels) so corresponds to two real parameters (multiplied by the number of independent elements of the channel matrix). However, in principle these parameters are  not just numbers but functions of momentum. Thus to parameterize the UV physics, we need to consider the possible momentum dependence of the UV scattering amplitude (at least at small momenta).

One might na\"ively expect that the UV scattering amplitude would become essentially momentum-independent for sufficiently low momenta (relative to the cutoff scale), allowing it to be fully described by a matrix of complex numbers. This matrix could be determined by measuring, for example, the elastic scattering cross sections between different channels, and the annihilation rates associated with different initial conditions, at some fixed reference momentum. The physical cross sections at other momenta should then be cutoff-independent.

However, this expectation of momentum-independence can be violated in special circumstances, notably where there is a new low-energy scale associated with the UV physics. For a spherical-well regulator, for example, this occurs when the well supports a near-zero-energy bound state, leading to resonances in the associated Sommerfeld enhancement and a velocity-dependent enhancement at momenta well below the inverse radius of the well. Nonetheless, these resonances feature universal behavior that can again be used to simplify the description of the UV scattering amplitude \cite{Braaten:2013tza}.

In the simplified case of a single-channel $s$-wave interaction and in proximity to a resonance, the non-relativistic scattering amplitude can be parameterized in the  form \cite{Braaten:2013tza}:
\begin{align} f(p) \approx \left(-\gamma - i p \right)^{-1},\end{align}
where $\gamma$ is a complex constant. This essentially amounts to keeping the first two terms in the effective range expansion \cite{Bethe:1949yr}. Within this parameterization, the inelastic (annihilation) cross section is given by \cite{Braaten:2013tza}:
\begin{align} \sigma_\text{ann} \approx \frac{1}{p} \frac{4\pi \text{Im}\gamma}{|\gamma + i p|^2}. \end{align}
If we can choose a reference momentum $p_\text{ref}$ such that $|\gamma| \ll p_\text{ref} \lesssim \Lambda$ (with the 2nd condition enforcing that the EFT is valid), then it follows that:
\begin{align} \frac{\sigma_\text{ann} v_\text{rel}}{\sigma_\text{ann} v_\text{rel}|_{p=p_\text{ref}}} \approx \frac{p_\text{ref}^2}{|\gamma + i p|^2} =   p_\text{ref}^2 |f(p)|^2. \end{align}
The alternate case with $|\gamma| \gtrsim \Lambda$ recovers the case of a momentum-independent scattering amplitude discussed above.

Thus in this case, the relevant cross sections can again be described in terms of a momentum-independent complex number $\gamma$; the Sommerfeld-enhanced annihilation rate as a function of momentum can also be inferred from measuring the Sommerfeld-enhanced annihilation cross section at a single reference momentum, and then the scattering phase shift. This latter approach is similar to that employed in Ref.~\cite{Bellazzini:2013foa}, which parameterizes the Sommerfeld enhancement in terms of the IR parameters, a real UV coefficient that controls the phase shift (at a given velocity), and the Sommerfeld enhancement evaluated as $v_\text{rel} \rightarrow 1$ (in our case we would need to choose the reference momentum within the range of validity of the EFT).\footnote{When checking the numerical experiments shown in Fig.~1 of Ref.~\cite{Bellazzini:2013foa}, we have found that the provided non-zero values of the UV parameter denoted $A/B$ require tuning the UV spherical well very close to a resonance, so we expect  this near-resonance description to be fairly accurate for these examples. We thank Brando Bellazzini for enlightening discussions on this point.} As we have argued earlier, it is not generally true that specifying the low-energy phase shift is sufficient to specify the Sommerfeld enhancement (from either the UV potential alone, or the combined IR and UV potentials), but we see that adding information on the Sommerfeld enhancement at a specific momentum can be sufficient.

For our numerical calculations in this work, we will not focus on obtaining cutoff-independent results in terms of physically-measured quantities. Instead, we will treat the cutoff and UV parameters as inferred from some unknown UV theory, and the effect of scanning over those parameters as describing the range of possible UV completions.

\subsection{Conditions for a large enhancement}

Our expectation is that for a substantial enhancement from the UV well, we will need $x_0^2 v_0 \gtrsim 1$. On the other hand, for any enhancement from the IR physics, a necessary (but not sufficient) condition is that the exponential drop in the potential associated with non-zero $m_\phi$ must lie outside the cutoff radius, i.e. $m_\phi \lesssim \Lambda \Rightarrow x_0 \lesssim 1$. In appendix~\ref{app:altdimlesscoords} we give an argument that a substantial enhancement from the IR physics in a region where the EFT is valid also requires:
\begin{align} \sqrt{2 E / m_\chi } \lesssim \Lambda/m_\chi \lesssim g^2/(12\pi),\label{eq:validity} \end{align}
which in terms of our present dimensionless parameters translates into:
\begin{align} \xi \lesssim \Lambda/m_\phi, \quad (m_\phi/m_\chi) \xi \lesssim \Lambda/m_\chi \lesssim g^2/(12 \pi)\label{eq:highSEcond} \end{align}
In terms of the coupling $G$, these conditions impose $G = \frac{g^2 m_\chi^2}{3\pi \Lambda^2} \gtrsim\frac{g^2 }{3\pi (g^2/12\pi)^2} = \frac{48\pi}{g^2}$, and thus obtaining a large Sommerfeld enhancement in a weakly-coupled theory requires $G \gg 1$. We  discuss possible UV completions and whether they can give rise to this hierarchy in appendix~\ref{app:UV-completion-Goldstone-interaction}. We will test these intuitions numerically
in Sec.~\ref{sec:Goldstonenumerics}.

\subsection{Channels without coupling between different-$\ell$ modes}

Before moving on to the numerical results, let us briefly discuss the enhancement in the $\ell=s=0$ state, where the potential (for $r > 1/\Lambda$) is $V(r) = -(3/4) V_C(r) = -(3/4)G (m_\phi/m_\chi)^2 e^{-m_\phi r}/r$, as discussed in Sec.~\ref{sec:litreview}. In that section we argued there would be no significant SE from this potential for $G \lesssim 1$, but as we are now considering $\Lambda \ll m_\chi$ and hence $G \gtrsim 1$, we need to reconsider this scenario. We see that in this case the long-range potential is simply an attractive Yukawa potential with effective coupling $g^2 m_\phi^2/(4 \pi \Lambda^2)$, and range $1/m_\phi$. Our expectation is that a significant Sommerfeld enhancement from the IR Yukawa potential will require the standard conditions:
\begin{equation*}
    v_\text{rel} \lesssim \frac{g^2}{4\pi} \left( \frac{m_\phi}{\Lambda}\right)^2, \quad \frac{m_\phi}{m_\chi} \lesssim \frac{g^2}{4\pi} \left( \frac{m_\phi}{\Lambda}\right)^2.
\end{equation*}
Furthermore, we expect that the short-range cutoff at $r \approx 1/\Lambda$ will not significantly affect the IR Sommerfeld enhancement if this cutoff is small compared to the Bohr radius, i.e.:
\begin{equation*} \Lambda \gtrsim m_\chi \frac{g^2}{4\pi} \left( \frac{m_\phi}{\Lambda}\right)^2   \end{equation*}
So putting these together we have:
\begin{equation} v_\text{rel}, \frac{m_\phi}{m_\chi} \lesssim \frac{g^2}{4\pi} \left(\frac{m_\phi}{\Lambda}\right)^2 \lesssim \Lambda/m_\chi \end{equation}
We see that in the $m_\phi \rightarrow 0$ limit we expect no IR enhancement in this case (as the effective coupling of the long-range potential goes to zero), in contrast to the coupled-channel case. However, there could be an intermediate range of $m_\phi$  where we expect a substantial enhancement (at least for sufficiently low velocities), corresponding to:
\begin{align} \frac{4\pi}{g^2} \frac{\Lambda}{m_\chi} \lesssim \frac{m_\phi}{\Lambda} \lesssim  \sqrt{\frac{\Lambda}{m_\chi} \frac{4\pi}{g^2}} \lesssim 1\end{align}
In order for this range to exist, we see we need $\frac{g^2}{4\pi} \gtrsim \Lambda/m_\chi$, similar to our earlier estimate for IR SE in the coupled-channel ($j=1,s=1, \ell=0,2$) case.

In the case with $\ell=s=1$, $j=0$, there is an attractive singular potential and the $1/r^3$ component is of the form $-\frac{g^2}{\pi \Lambda^2} e^{-m_\phi r}/r^3$. The centrifugal term in the potential has the form $\frac{2}{m_\chi r^2}$ (since $\ell=1$ for this mode), so in order for the attractive potential to dominate over the centrifugal term at the cutoff, we require (assuming $\Lambda < m_\chi$ so $\Lambda$ sets the cutoff radius) $2 \Lambda^2/m_\chi < \frac{g^2}{\pi\Lambda^2} \Lambda^3 e^{-m_\phi/ \Lambda}$. Satisfying this condition, which we expect to be a prerequisite for a significant enhancement, will typically require both $m_\phi \ll \Lambda$ and $\Lambda/m_\chi \lesssim \frac{g^2}{2\pi}$.  This condition is parametrically and conceptually very similar  to Eq.~\eqref{eq:validity}, but slightly weaker (less restrictive) due to the numerical prefactor.

\subsection{Validity of the EFT}

Finally, let us comment briefly on the validity of the pseudo-Goldstone Lagrangian term, which arises from an EFT where physics at the scale $\Lambda$ has been integrated out. We will generally assume that the pseudo-Goldstone Lagrangian is valid so long as $m_\phi < \Lambda$ (which in any case we expect to be required for a substantial IR enhancement) and separately $m_\chi v < \Lambda$, with the latter condition being based on the momentum running through the mediator exchanges. The $\chi$ (or $\bar{\chi}$) fermions are treated largely as static sources, retaining only the small dynamical fluctuations with momenta $\lesssim m_{\chi}v$, allowing us to consider the regime $\Lambda < m_\chi$, when combined with $m_\chi v < \Lambda$. However, a full description of the entire scattering of the $\chi$ $(\overline{\chi})$ fermions (including modes with momenta up to $\sim m_{\chi}$) will require knowledge of the theory in the range between $\Lambda$ and $m_\chi$, including potentially four-fermion interactions from particles that have been integrated out at the $\Lambda$ scale. Such interactions would contribute to our UV/short-range potential.

These validity conditions are similar to those employed in Ref.~\cite{Ferrante:2025lbs}. The potential terms in the Hamiltonian for $r > 1/\Lambda$ have a maximum value of order $G \Lambda^3/m_\chi^2 \sim g^2 \Lambda$, which remains smaller than $\Lambda$ provided $g^2 \lesssim 1$. However, the {\it momentum} naively associated with those potential terms is $p \sim \sqrt{m_\chi E} \sim g \sqrt{m_\chi \Lambda}$, which exceeds $\Lambda$ if $g^2 \gtrsim \Lambda/m_\chi$, which is parametrically the same as our condition for a large Sommerfeld enhancement. To the degree that this is a concern, the same scaling will arise for any potential term of the form $g^2/(\Lambda^2 r^3)$. We will proceed under the assumption that we need only compare the asymptotic momentum to $\Lambda$, rather than the momentum associated with the potential at the cutoff scale.

\subsection{Numerical results for the coupled-channel case}
\label{sec:Goldstonenumerics}
We now use the numerical procedure discussed earlier to solve for SE numerically, as a function of various parameters. There are two qualitative differences in this case compared to the pseudoscalar case in sec.~\ref{subsubsec:pseudoscalar-numerical-results}. First, the effective coupling $G=g^2 (m_\chi/\Lambda)^2/3\pi$ can be larger than one even for small couplings $g$, in the case where $\Lambda \lesssim m_\chi$ (see appendix~\ref{app:UV-completion-Goldstone-interaction} for possible UV completions). Second, the cutoff, being set by $\Lambda$, must be adjusted as $G$ varies. As before, in what follows, we will set $m_\phi/m_\chi = 10^{-3}$.

\begin{figure}
    \centering
    \includegraphics[width=0.95\linewidth]{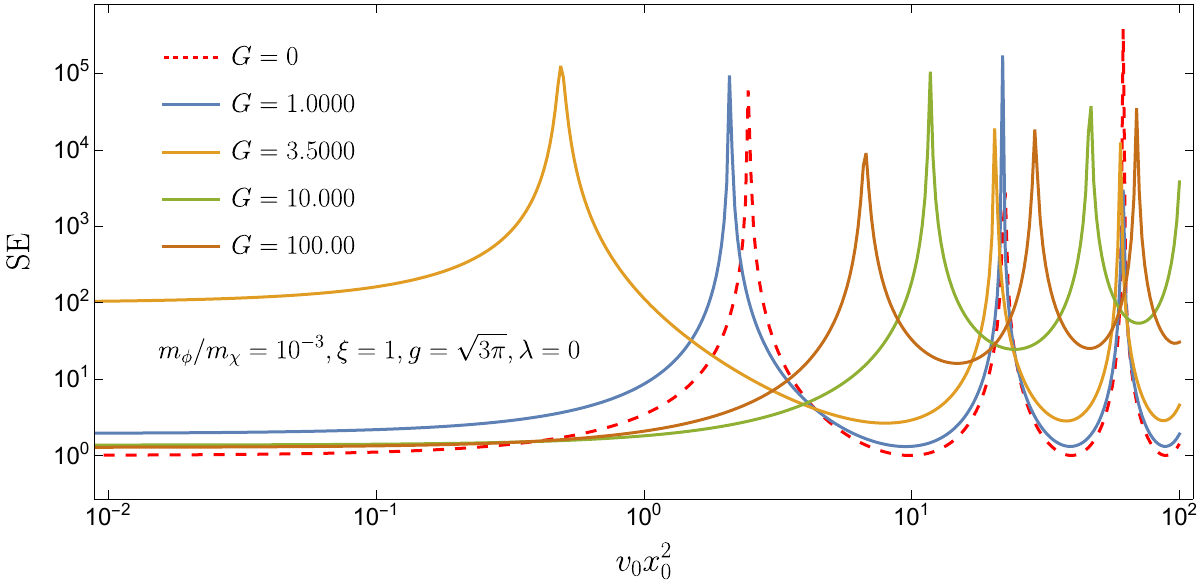}
    \caption{
    \small{SE for pseudo-Goldstone mediator, as a function of $v_0 x_0^2$, for several values of $G=(m_\chi/\Lambda)^2$ (note we have chosen $g=\sqrt{3\pi})$. The choices for other parameters are indicated on the plot. We have chosen the cutoff $x_0 = \text{max}(m_\phi/m_\chi, m_\phi/\Lambda)$, as discussed in the text.}}
    \label{fig:SE-vary-params}
\end{figure}

We would like to understand the interplay between the long-range and the short-range potentials in giving a large SE. Figure~\ref{fig:SE-vary-params} shows SE as a function of $v_0 x_0^2$ (which as discussed above parametrizes the enhancement from purely short-range physics), for $ m_\phi/m_\chi = 10^{-3}$ and $\xi = 1$. We see that there is SE at $G = 0$ (red dashed line) i.e. when there is no long-range potential; we have confirmed this SE agrees with the analytic result of Eq.~\eqref{eq:analyticSE}.
This is therefore a purely short-range effect sourced by the spherical well in the $\ell=0$ channel. Once we include the long-range potential by making $G$ non-zero (holding $g$ fixed but varying $\Lambda/m_\chi$), there is an interplay between the long- and short- range physics, and the location and the height of the peaks change (blue, orange, green, and brown curves). We see that the main effect of turning on the long-range potential is to shift the peaks with respect to $v_0 x_0^2$, although it also increases the enhancement between the peaks.

\begin{figure}
    \centering
    \includegraphics[width=0.95\linewidth]{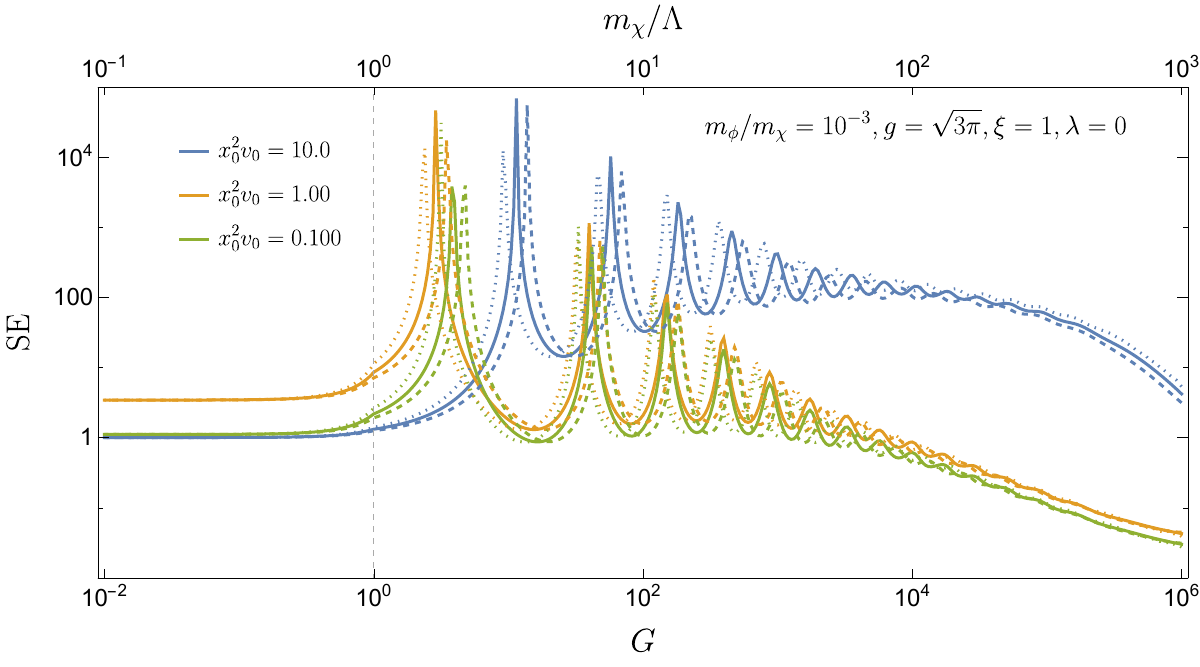}
    \caption{
    \small{SE for pseudo-Goldstone mediator, as a function of $G=(m_\chi/\Lambda)^2$ for some fixed values of $v_0x_0^2$. The values of other parameters are indicated on the plot. The range of $G$ translates to a range of $m_\chi/\Lambda$ for fixed $g=\sqrt{3\pi}$, which is shown on the top horizontal axis. The solid/dashed/dotted lines are with the cutoff $x_0/\text{max}(m_\phi/m_\chi, m_\phi/\Lambda) = 1, 1.1, 0.9$ respectively.
    }}
    \label{fig:SE-Goldstone-vs-G}
\end{figure}

\begin{figure}
    \centering
    \includegraphics[width=0.49\linewidth]{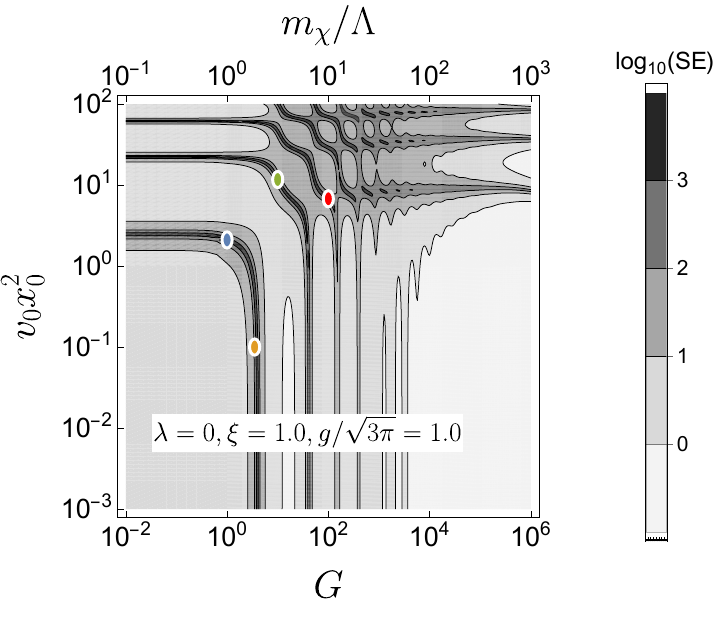}
    \includegraphics[width=0.49\linewidth]{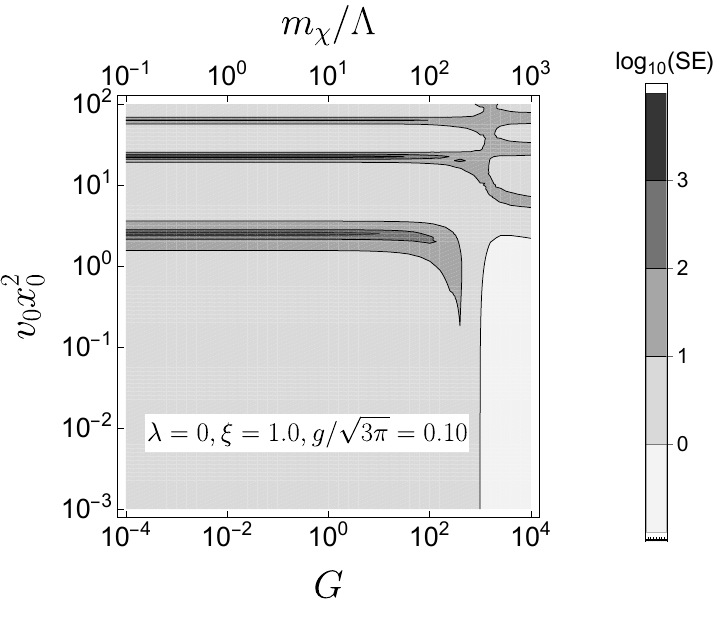}
    \includegraphics[width=0.49\linewidth]{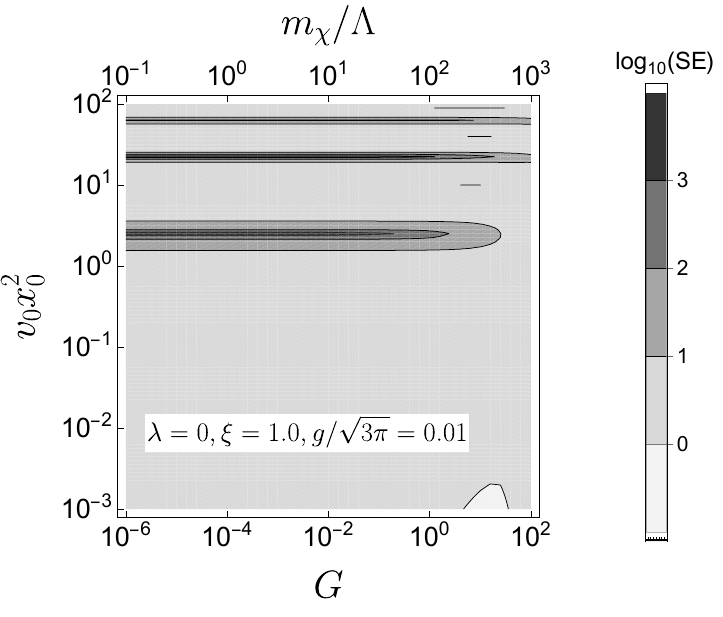}
    \caption{
    \small{Contours of $\log_{10}$SE for the pseudo-Goldstone mediator, as a function of $G$ and $v_0x_0^2$, for $\lambda = 0, \xi = 1$ and $g/\sqrt{3\pi} = 1$ (top left), $0.1$ (top right) and $0.01$ (bottom). For fixed $g$, the range of $G$ corresponds to a range of $m_\chi/\Lambda$ which is shown in the top horizontal axis of each plot. The colored blobs in the top left plot correspond to the curves shown in Fig.~\ref{fig:SE-v-dependence-Goldstone}. See Fig~\ref{fig:SE_massless} for the case with $m_\phi=0$. }
    }
    \label{fig:SE-Goldstone-vs-G-vs-v0}
\end{figure}

\begin{figure}
    \centering
    \includegraphics[width=0.95\linewidth]{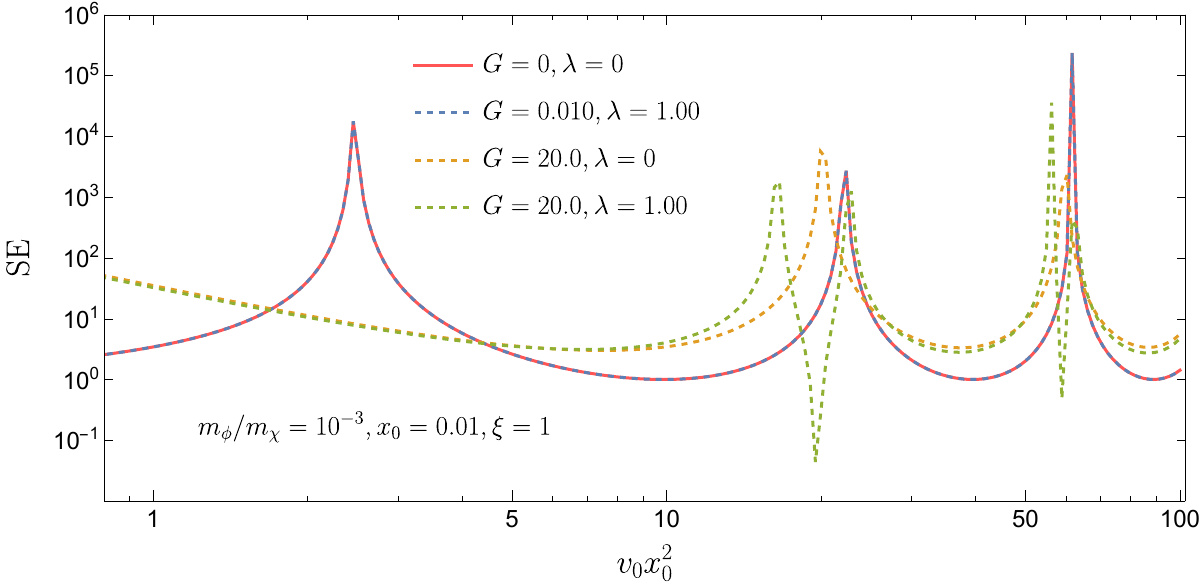}
    \caption{
    \small{
    SE for pseudo-Goldstone mediator as a function of $v_0x_0^2$, for several values of $G, \lambda$. Here we have taken $m_\phi/m_\chi = 10^{-3}, \Lambda/m_\chi = 10^{-1}$ and the cutoff $x_0 = \max(m_\phi/m_\chi, m_\phi/\Lambda) = 10^{-2}$. Different values of $G$ correspond to different values of $g$. We see that a non-zero $\lambda$ affects SE only at larger values of $G$.}
    }
    \label{fig:SE-lambda-variation}
\end{figure}
\begin{figure}
    \centering
    \includegraphics[width=0.6\linewidth]{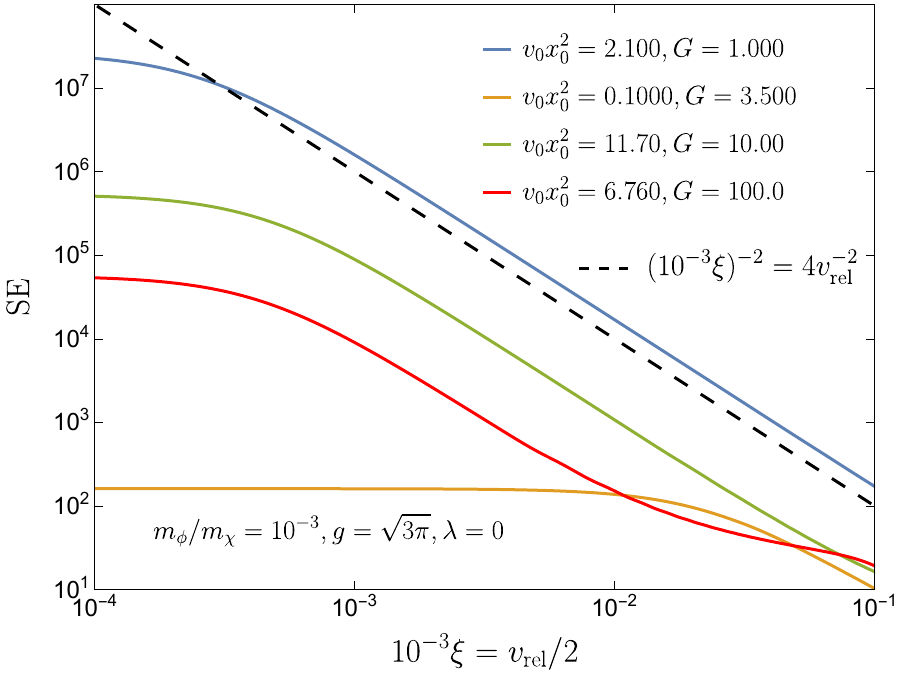}
    \caption{
    \small{SE as a function of relative velocity $v_\text{rel}$, for a few values of $v_0x_0^2, G$, which demonstrate the behavior in proximity to different peaks (see corresponding points in Fig.~\ref{fig:SE-vary-params}). We have also indicated the corresponding points in the $(G, v_0x_0^2)$ plane in Fig.~\ref{fig:SE-Goldstone-vs-G-vs-v0} with corresponding colored markers. We have taken $m_\phi/m_\chi = 10^{-3}$ and $\lambda = 0$. }
    }
    \label{fig:SE-v-dependence-Goldstone}
\end{figure}

In Fig.~\ref{fig:SE-Goldstone-vs-G} we show the complementary behavior of the SE when we vary $m_\chi/\Lambda$ for several choices of the fixed short-range potential, for the moment choosing $\lambda=0$ (i.e.~the short-range potential has no $d$-wave component). We also show the impact of varying the cutoff radius around the baseline value, although in this case we do not expect the results to be cutoff-independent (varying the cutoff radius modifies the range over which the potential takes its long-range form, with no compensating effect). We see that for $\Lambda \gtrsim m_\chi$ the enhancement is effectively $G$-independent, but for $m_\chi/\Lambda \gtrsim 1$ (which corresponds to $G \gtrsim 1$ for the illustrative choice of $g=\sqrt{3\pi}$ here), we see large SE effects including resonant peaks even when $x_0^2 v_0 \ll 1$.  We observe modest quantitative changes in the peak positions and normalizations when the cutoff radius is varied around its baseline value. It is interesting to note that at sufficiently large $G$ the SE no longer exhibits peaks (at least at this velocity scale), and actually becomes a suppression unless $x_0^2 v_0$ is sufficiently large.

To show the interplay between the long-range and the short-range potentials more clearly, in Fig.~\ref{fig:SE-Goldstone-vs-G-vs-v0} we show contour plots of SE as a function of $v_0 x_0^2$ and $G$, for different choices of the parameter $g$. For $g=\sqrt{3\pi}$, we see that for $G\lesssim 1$, the SE is essentially $G$-independent and fixed solely by $v_0 x_0^2$; on the other hand, for $v_0 x_0^2 \lesssim 1$, the SE is essentially independent of that parameter and is controlled by $G$ (including resonance peaks for $1 \lesssim G \lesssim 10^{3}$). In the region where both parameters are $\lesssim 1$ there is no enhancement, but where both are $\gtrsim 1$, we see interesting behavior where the peaks vary non-trivially with both parameters. The onset of the $G$-dependent behavior at $m_\chi/\Lambda \sim 1$ is broadly consistent with our estimate in Eq.~\eqref{eq:highSEcond} (we would have predicted a requirement of $m_\chi/\Lambda \gtrsim 4$, which is indeed roughly where the first purely $G$-dependent peak appears).

For smaller $g=0.1\sqrt{3\pi}$, depicted in the second panel, we see similar behavior, but with a less rich structure of peaks at high $m_\chi/\Lambda$. This is to be expected as our expectation is the onset of significant SE induced by the long-range potential occurs at $m_\chi/\Lambda \sim 12\pi/g^2 \approx 400$ for this case; this agrees quite well with the observed numerical results. Consequently, there is only a limited range of $m_\chi/\Lambda$ between this threshold and where the EFT loses validity at $m_\chi/\Lambda = 10^3$ (at least for this choice of velocity). Going to even smaller $g=0.01\sqrt{3\pi}$, we expect there to be no significant enhancement except that induced by the short-range spherical well within the range of validity of the EFT, and indeed we see the only peaks are essentially $G$-independent and are controlled by $v_0 x_0^2$. At these small couplings, any possibility of substantial enhancement from the long-range potential would require both the velocity and $m_\phi$ to be much smaller (together with a large hierarchy $m_\chi/\Lambda$). We have also checked that the massless mediator limit produces similar behaviour in the SE, as shown in the appendix Fig~\ref{fig:SE_massless}.

In Fig.~\ref{fig:SE-lambda-variation} we illustrate the impact of turning on the short-distance $d$-wave potential, i.e.~setting $\lambda=1$ vs $\lambda=0$. We see that as expected, when the long-range potential is negligible (zero or small $G$), modifying $\lambda$ has no effect, because the short-distance potential does not couple the $s$-wave and $d$-wave states and the Sommerfeld enhancement is read off solely from the $s$-wave component. However, when the long-range interaction is strong, we observe dips in the SE for $\lambda=1$ that are not present for $\lambda=0$, as well as modest shifts in the peak positions with respect to $v_0 x_0^2$; we attribute this to destructive interference induced by the mixing between channels.

Finally, in Fig.~\ref{fig:SE-v-dependence-Goldstone} we illustrate the velocity dependence of the enhancement on the peaks, for four locations in the $G-v_0 x_0^2$ plane (for fixed $g = \sqrt{3 \pi}$). We see that all four locations exhibit the standard $s$-wave resonance behavior where the enhancement is proportional to $v_\text{rel}^{-2}$, for at least some velocities, and saturates at low velocities, although some of the chosen points correspond to maxima with respect to $G$ (orange), some with respect to $x_0^2 v_0$ (blue), and some with respect to both $G$ and $x_0^2 v_0$ (green and dark orange). The dark orange line shows some non-trivial velocity structure which may be associated with its position in the region of $G-v_0 x_0^2$ space where both short-range and long-range effects are important, but the green curve does not show the same behavior and so it does not seem to be ubiquitous in this region.

\subsection{Analytic approximate results and bound states}

So far the results obtained for SE were computed numerically, since there are no closed form solutions for the long-range part of the potentials involved, and also because the equations are coupled. However, it is possible to solve the equations approximately and obtain an analytical approximation for the SE. The advantage of having an analytical expression for SE is that one can easily continue $\xi$ to the complex plane and look at the pole structure in the amplitude for the wavefunction, and therefore establish a connection between peaks in SE and bound states close to threshold.

The first thing to observe is that different components in $v_L(x)$ have different fall-off behavior. Recall that the full potential also involves the centrifugal barrier term in the $\ell=2$ channel. Denoting the $\ell=0$ and $\ell=2$ channels by $u$ and $w$ respectively, we immediately observe that for small $x$,
\begin{align}
        (v_L)_{uu} \sim e^{-x}/x\:,\:\: (v_L)_{uw} \sim  (v_L)_{ww} \sim e^{-x}/x^3\:.
\end{align}
This suggests that we can approximate the long-range potential by a potential involving only delta functions:
\begin{align}
    v_L(x) \to
    \delta (x - x_{23})
    \begin{pmatrix}
        0 & \Delta_M \\
        \Delta_M & \Delta_T
    \end{pmatrix}
    +
    \delta (x - x_{34})
    \begin{pmatrix}
        \Delta_C & 0 \\
        0 & 0
    \end{pmatrix}\:.
\end{align}
Here, $\Delta_M$ models the mixing that $(v_L)_{uw}$ generates, $\Delta_T$ models propagation in the $\ell = 2$ channel and $\Delta_C$ models the long-range propagation in the $\ell = 0$ channel. We have placed the delta-functions for terms with similar $x$ dependence at the same location.
The advantage of this approximation is that one can solve for the wavefunction in between the regions bounded by the delta functions and match them using the discontinuity generated by the delta functions.

The Schr\"odinger equation for the approximated potential can be solved easily. We effectively have separated the full range $0< x < \infty$ into four regions: $i)$ the short-range region $x < x_{12} = x_0$ as before, $ii)$ the intermediate region $x_0 = x_{12} < x < x_{23}$, $iii)$ $x_{23} < x < x_{34}$, and $iv)$ $x > x_{34}$. In each of the four regions, the solution to the Schr\"odinger equation has an analytical form. The forms in region $i)$ and region $iv)$ are further constrained by the boundary conditions. Matching the wavefunction and derivatives across the delta functions at $x = x_{23}$ and at $x = x_{34}$ fixes the solution uniquely. We define SE as before by comparing the $\ell=0$ channel solution in the $x < x_0$ region evaluated at origin to the no-potential case (see Eq.~\eqref{eq:SE-def}). The explicit steps of the computation, and the resulting approximate expression for SE are given in appendix~\ref{app:Analytical-Approximation-for-SE}.

\begin{figure}[h!]
    \centering
    \includegraphics[width = 0.95\linewidth]{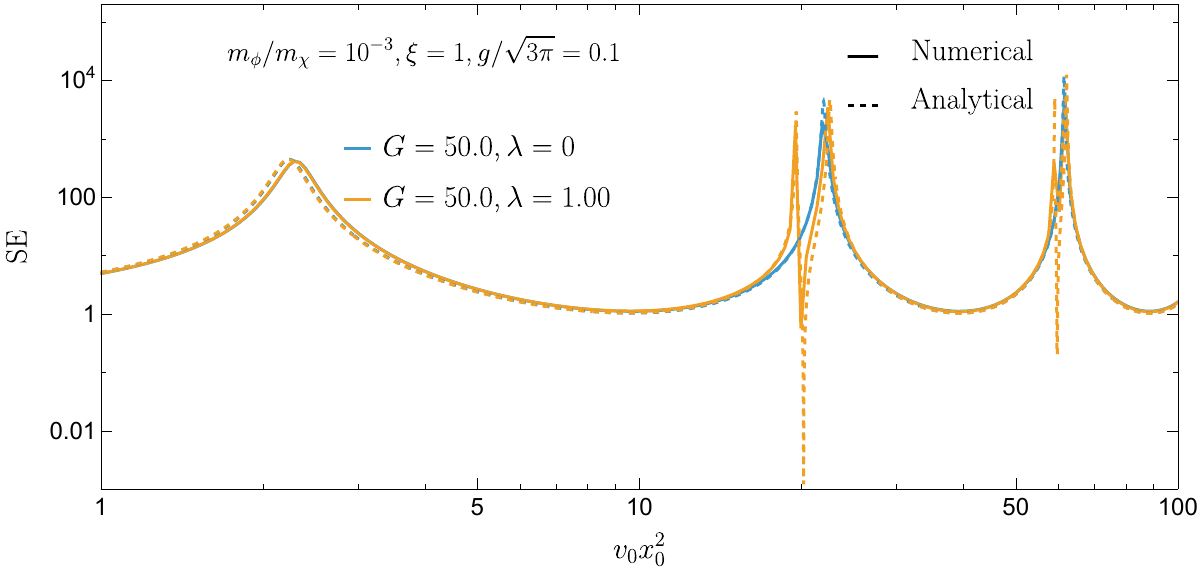}
    \includegraphics[width = 0.95\linewidth]{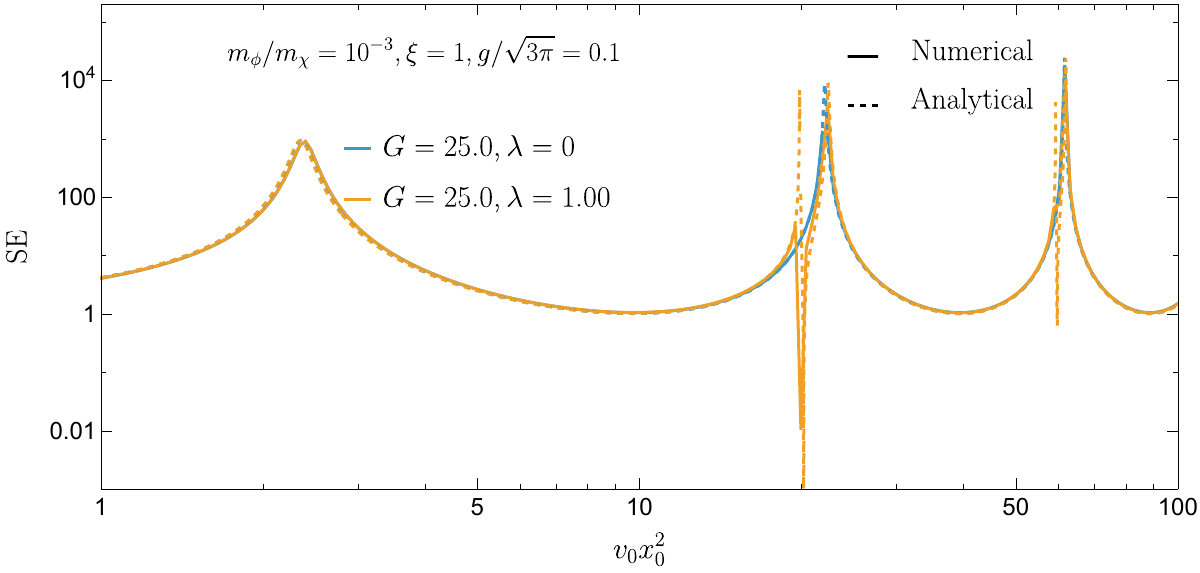}
    \caption{
    \small{A comparison of SE as a function of $v_0x_0^2$, computed analytically (in dashed) and numerically (in solid), for a few values of $G, \lambda$. The top panel shows $G = 50, \lambda = 0, 1$ while the bottom panel shows $G=25, \lambda = 0, 1$. The analytical result is normalized to match the numerical result at small $v_0x_0^2$. We have taken $x_{23}/x_0 = 1.1$, $\Delta_M = 10$ and $\Delta_T = 0$ for both the plots, while $\Delta_C$ and $x_{34}$ are fixed using Eq.~\eqref{eq:DeltaC-definition} and Eq.~\eqref{eq:x34-estimate} respectively.}
    }
    \label{fig:SE-numerical-vs-analytical}
\end{figure}

Figure~\ref{fig:SE-numerical-vs-analytical} shows the numerical and analytical results for some choice of parameters. We see that the agreement is very good and captures the peak and the dip locations well, thus validating the approximation. We found that setting $\Delta_T = 0$ still gave a good agreement between numerical and analytical results, so we made that choice to simplify the expressions to follow. After this choice, there are four parameters that need to be specified: $x_{23}, x_{34}, \Delta_M$ and $\Delta_C$. We adjusted $x_{23}$ and $\Delta_M$ by hand to get a good match to the numerical result. Variation in $x_{23}$ moves the location of the first peak (both at $\lambda = 0, 1$), while variation in $\Delta_M$ moves all the peaks and the extra feature seen for $\lambda = 1$. We found that $x_{23}/x_0 = 1.1$, $\Delta_M = 10$ gave a good fit to the numerical results for both the plots in Fig.~\ref{fig:SE-numerical-vs-analytical}. We were able to find an analytical expression for $\Delta_C$ and $x_{34}$ which worked well, removing the need to adjust them by hand (see Eq.~\eqref{eq:DeltaC-definition} in appendix~\ref{app:Analytical-Approximation-for-SE} and discussions nearby). We have also normalized the SE to match the numerical result away from the peaks, at small $v_0 x_0^2$.

Having computed the SE analytically allows us to immediately look at the bound states and identify how the bound states move and come close to threshold at the special values of parameters where a peak in SE is seen. To look at the location of bound states, recall that we need to look at the poles of $A_u^{(1)}$ in the complex $\xi$ plane, where $A_u^{(1)}$ is the coefficient of the $\ell=0$ wavefunction in the $r<a$ part (see appendix~\ref{app:Analytical-Approximation-for-SE} for details). The general expectation is that when there is a SE, there is a corresponding bound state at a purely positive imaginary value of $p$ or equivalently $\xi$. Since we already have an analytical expression for $A_u^{(1)}$ we can do this easily.

\begin{figure}[h!]
    \centering
    \includegraphics[width=0.99\linewidth]{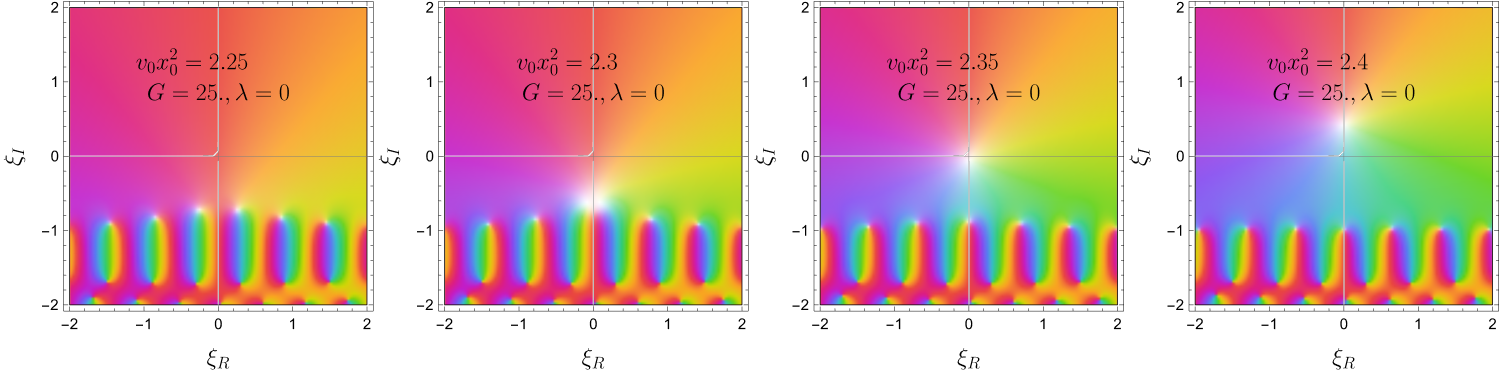}
    \\
    \includegraphics[width=0.99\linewidth]{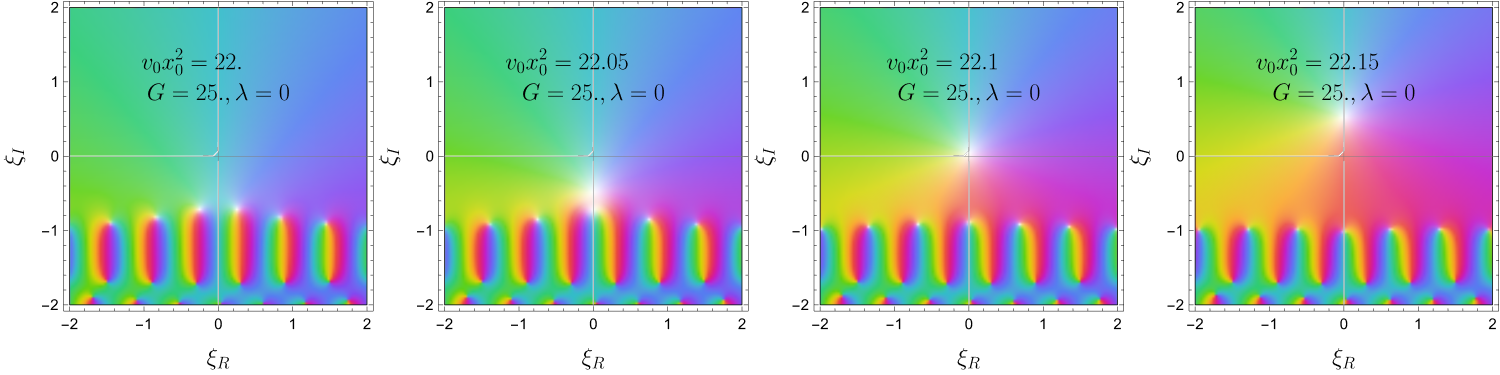}
    \\
    \includegraphics[width=0.99\linewidth]{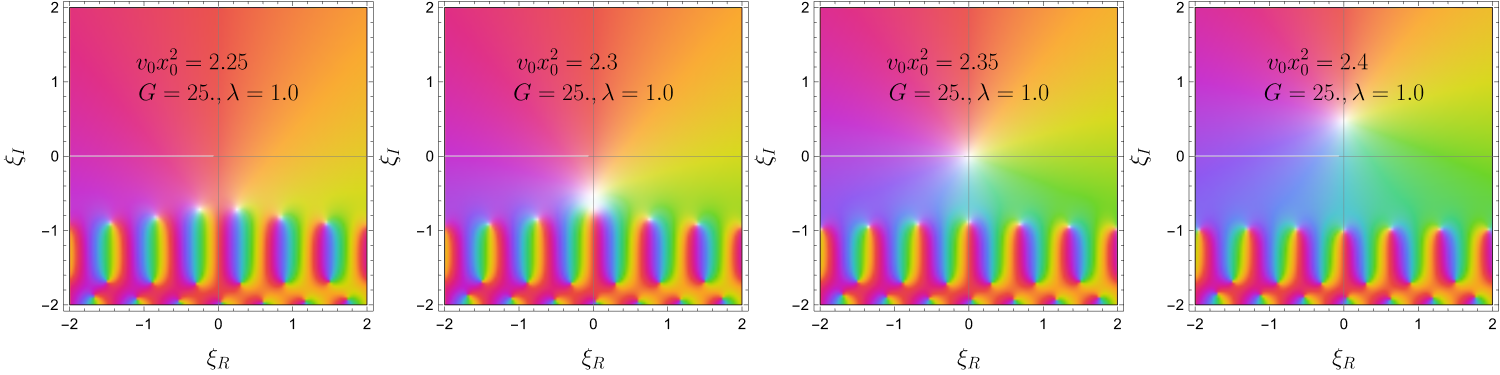}
    \\
    \includegraphics[width=0.99\linewidth]{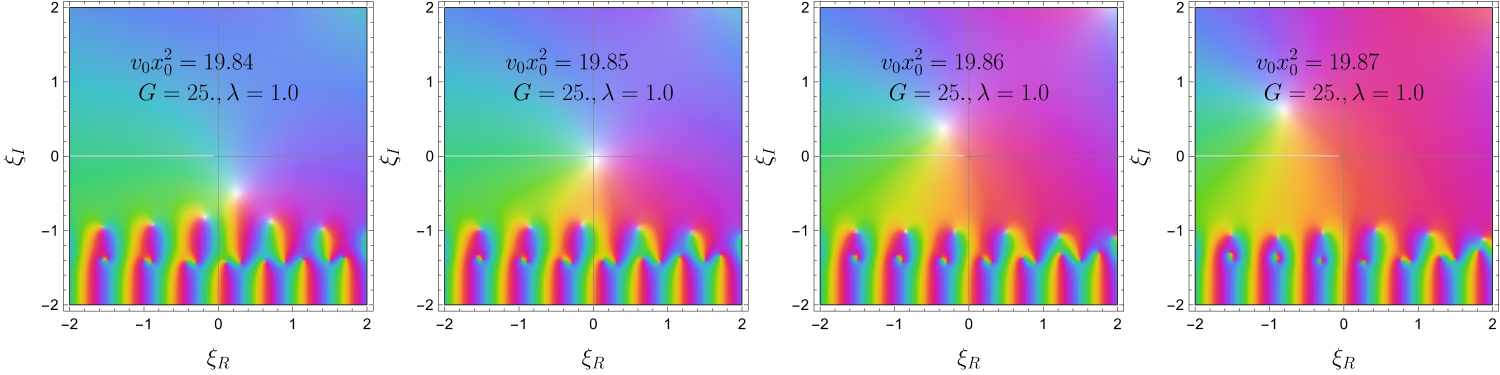}
    \caption{
    \small{$A_u^{(1)}$ (see appendix \ref{app:Analytical-Approximation-for-SE}), in the complex $\xi = \xi_R + i \xi_I$ plane. The poles are seen as white spots. Top two rows are for $\lambda = 0$ and bottom two rows are for $\lambda = 1$. For these plots we have taken $G = 25$ and $m_\phi/m_\chi = 10^{-3}$. We see that for both $\lambda = 0, 1$, close to where a peak is seen in the SE (see corresponding plots in Fig.~\ref{fig:SE-vary-params}), a corresponding state exists close to threshold, for a slightly positive imaginary value of $\xi$.
    }
    }
    \label{fig:BoundState-vs-alpha}
\end{figure}

Figure~\ref{fig:BoundState-vs-alpha} shows $A_u^{(1)}$ in the complex $\xi$ plane, as $v_0x_0^2$ is scanned near a peak in SE for a given value of $G, \lambda$. The poles are seen as localized white spots in the plots. The top two rows show $\lambda = 0$, and the bottom two rows show $\lambda = 1$. We have taken $ G = 25$ and $m_\phi/m_\chi = 10^{-3}$ for all the plots. At values of $v_0$ where one of the poles is on the imaginary axis at slightly positive values, there is a corresponding peak in SE (see Fig.~\ref{fig:SE-vary-params}). These results also demonstrate that in the vicinity of parameters where there is a peak in the SE, there are states which are better called resonances because they are not at purely imaginary values of $\xi$.

We would like to point out that the periodic structure of the amplitude seen in the complex plane for negative imaginary values of $\xi$ comes purely from the effect of the long-range potential. As a first step, if we set $G=0$, one can explicitly check that it goes away. One can understand this periodic feature analytically by calculating the amplitude as an expansion in the IR parameters $\Delta_M, \Delta_C$ to linear order. Appendix~\ref{app:Analytical-Approximation-for-SE} discusses this in detail. We would also like to note that for $\lambda = 1$, one of the poles moves in a different way than the rest (e.g. the last row in Fig.~\ref{fig:BoundState-vs-alpha}), and this corresponds to a peak in SE where a sharp enhancement and reduction is seen (peak near $v_0x_0^2 = 19.8$ in Fig.~\ref{fig:SE-numerical-vs-analytical}).

In conclusion, we are able to demonstrate that the peaks in the SE as a function of $v_0x_0^2$, which parametrizes the short distance physics, are in one to one correspondence with shallow bound states that appear on the slightly positive and purely imaginary values of $\xi$, as per expectations. In the presence of a long-range potential, the location of the peaks gets shifted and there are new structures in the complex $\xi$ plane that emerge. It would be interesting to further explore these features.

\section{Conclusions}\label{sec:con}

In this work, we have established possible origins of Sommerfeld enhancement in singular potentials using several approaches. First, we demonstrated that the issue of analyzing SE in singular potentials arising from ill-posed boundary conditions for the Schr\"odinger equation in coordinate space $(r\rightarrow 0)$, can be circumvented via velocity power counting of scattering amplitudes in NREFTs. We have established that for perturbative coupling strengths, long-range singular potentials arising from parity-odd mediator exchange in renormalizable theories furnish no low-velocity enhancement or SE in the scattering or annihilation cross sections, consistent with previous observations for elastic scattering in the framework of NR quantum mechanics with an explicit short-distance regulator~\cite{Agrawal:2020lea,Parikh:2020ggm}.

We have shown how to write down the criterion for bound-state formation when the position-space singularity of the short-range potential is removed by matching to an appropriate short-distance scattering amplitude. We have explored modeling this short-distance amplitude by an appropriate non-singular potential, in particular demanding that it reproduce the scattering amplitude from the long-range potential using the first Born approximation. However, we have shown that while this matching condition is sufficient for the elastic-scattering problem studied in \cite{Agrawal:2020lea}, both the Sommerfeld enhancement to annihilation and the bound-state spectrum will require a higher-order matching (which cannot be easily computed from the singular potential via the Born approximation). As a consequence, with this matching procedure, the inferred Sommerfeld enhancement is generically cutoff-dependent; however, we have demonstrated that for any plausible cutoff the enhancement remains small for weakly-coupled exchange of a fundamental pseudoscalar. This statement breaks down when the coupling $\alpha$ becomes $\mathcal{O}(1)$, when we do find an apparent substantial SE and the onset of a bound state; however, the coupling at which the bound state first forms is quite cutoff-dependent and we view the onset of large SE as more likely to indicate a breakdown of this treatment (perturbative matching + use of the tree-level potential) rather than a physical effect.

However, in the presence of non-renormalizable interactions, and in particular pseudo-Goldstone exchange, it is possible to have a large effective coupling inherited from a large ratio $\frac{m_{\chi}}{\Lambda}$ between the DM mass $m_\chi$ and the cutoff scale of the EFT $\Lambda$, provided the momentum exchange remains small relative to $\Lambda$. We have explored the interplay between the UV physics at $r \lesssim 1/\Lambda$ and the IR potential associated with pseudo-Goldstone exchange, and have argued that in this case a large Sommerfeld enhancement from the IR potential  is achievable if $\Lambda/m_\chi \lesssim g^2/(12 \pi)$, where $g$ is a perturbative dimensionless coupling constant. We have numerically demonstrated the transition between the regimes where the UV and IR physics dominate the SE and have demonstrated the presence of Sommerfeld resonances (associated with near-threshold bound states) in both regimes. To make this association sharp, we used an analytical approximation by replacing the long-range potential by appropriately placed delta function terms and calibrated it by comparison with numerical results. By looking at the amplitude of the $\ell = 0$ wavefunction in the complex $\xi$ plane, we then established a one-to-one connection between the SE peaks and the bound states close to threshold (seen as poles at slightly positive imaginary value of the rescaled momentum $\xi$). Similar claims of SE have been made in the context of $1/r^3$ potentials in EFTs \cite{Ferrante:2025lbs}; we have demonstrated that the non-trivial coupling between different partial-wave channels (and the absence of a $1/r^3$ coupling in the pure $s$-wave channel) relevant to pseudo-Goldstone exchange does not prevent a large enhancement in the presence of a large $m_\chi/\Lambda$ hierarchy.

We thus reconcile existing results of SE in pseudoscalar-mediated singular potentials in previous literature; the lack of SE for perturbative long-range interaction strengths~\cite{Agrawal:2020lea}, and the presence of SE in the context of an EFT that gives a strong effective coupling in the low-energy theory~\cite{Bedaque:2009ri,Bellazzini:2013foa,Ferrante:2025lbs}. Note however that our results do differ in detail from those of \cite{Bedaque:2009ri,Bellazzini:2013foa}; in particular we impose the position-space cutoff at $r\sim 1/\Lambda$ rather than $r\sim 1/m_\chi$ when $\Lambda < m_\chi$, since the potential derived from the EFT should not be expected to be valid at position scales smaller than $1/\Lambda$.

There exist several interesting future directions to our work. We have worked out the velocity power counting for massless mediators in DM systems, and it would be interesting to generalize to NREFTs with massive mediators. The ratio of the mediator mass to the DM mass, $\epsilon_\phi = m_\phi/m_{\chi}$, serves as an additional dial in the NREFT.  Investigating the regimes $\epsilon_\phi \sim v$ and $\epsilon_\phi \gg v$ seem very natural, although based on our QM results we do not expect these regimes to allow for SE or bound-state formation in a weakly-coupled renormalizable theory. The regime with $\epsilon_\phi \sim v$ is pertinent for axial vector mediator NREFTs, as their UV completions generally require a non-zero $\epsilon_\phi$. Moreover, the momentum space analysis in NREFTs  may allow for a direct, albeit iterative solution of the Lippmann Schwinger equation to analytically obtain SE~\cite{Beneke:2013jia,Binder:2026fwe}.
Another natural direction is to systematically improve the position space regulation of singular potentials in this work, beyond the first Born approximation. In particular, a next step is computing and combining the first Born approximation from the long-range one-loop potential  with the second Born approximation of the tree-level potential. Both of these terms are divergent for pseudoscalar exchange, so some regularization would likely be required, and would shed light on the higher order matching and regularization, both in this example and in cases where there is a substantial Sommerfeld enhancement from the leading-order potential but higher-order corrections are singular. Such higher order computations are necessary to systematically reduce the matching radius dependence in the regulated potentials in position space, and serve to probe the regulator independence of perturbative SE obtained directly in position space.

We have identified a regime where we expect a substantial SE and bound state formation from a pseudoscalar mediator, when the low-energy behavior is well described by an effective pseudo-Goldstone term with cutoff $\Lambda$, and there is a large hierarchy $\Lambda \ll m_\chi$. It would be interesting to explore SE in UV-complete models whose low-energy EFT matches onto the desired behavior, which would also allow us to carefully study the matching at $r\sim 1/\Lambda$. One possibility is a QCD-like dark sector where the DM is a baryon and the pions play the role of the pseudo-Goldstone bosons (see App.~\ref{app:UV-completion-Goldstone-interaction}); alternatively, axion-like particles (which have been studied extensively both as DM candidates and as possible mediators) could play the role of the pseudo-Goldstone mediator. It would also be interesting to explore the phenomenology of strongly-coupled dark sectors, which may allow for large SE and bound-state formation even in the absence of the hierarchy $\Lambda \ll m_\chi$. In both of these cases, there may also be large effects on elastic scattering, as relevant for self-interacting dark matter.

\acknowledgments

This work was  performed in part at the Aspen Center for Physics, which is supported by National Science Foundation grant PHY-2210452; RKM and TRS thank the Center for their hospitality during the completion of this work. This work was also performed in part at the Erwin-Schrödinger International Institute for Mathematics and Physics at the University of Vienna during the program ``New Paradigms for Harnessing Quantum Field Theory at Collider'' (2026); AB thanks the centre for their hospitality during the completion of the work.
RKM acknowledges support from the Gravity Spacetime and Particle Physics (GRASP) initiative at Harvard
University.
TRS' work is supported by the Simons Foundation (Grant Number 929255, T.R.S) and by the U.S. Department of Energy, Office of Science, Office of High Energy Physics of U.S. Department of Energy under grant contract Number DE-SC0012567. During the course of this work, TRS was supported in part by a Guggenheim Fellowship; the Edward, Frances, and Shirley B.~Daniels Fellowship of the Harvard Radcliffe Institute; and the Bershadsky Distinguished Fellowship of the Harvard Physics Department. We would like to thank Prateek Agrawal, Brando Bellazzini, Simone Biondini, Matt Reece, and Bingrong Yu for useful discussions and valuable comments on the manuscript.

\appendix
\section{UV completions of pseudo-Goldstone interaction}
\label{app:UV-completion-Goldstone-interaction}
We present some possible UV completions to the dimension-5 pseudo-Goldstone interaction Lagrangian
\begin{align}
    \mathcal{L}_\text{int} = \frac{g}{\Lambda}\bar{\chi}\gamma^\mu\gamma^5\chi\,\partial_\mu\phi\:.
\end{align}
In our results for SE coming from the long-range part of non-relativistic potential we have used the fact that $g(m_\chi/\Lambda)$ can be parametrically large. We would like to view the possible UV completions of the pseudo-Goldstone interaction with this in mind, i.e. which ones can accomplish this without requiring any other parameters to be unnaturally large.

\subsection{A weakly coupled UV completion without a parametric enhancement}
Consider the UV Lagrangian to be
\begin{align}
    \mathcal{L}_\text{UV} = |\partial_\mu S|^2 - V(S^\dagger S)+ i\bar{\chi}\slashed{\partial}\chi - \left(y S \bar{\chi}_L\chi_R + \text{h.c.}\right)\:,
\end{align}
where $S$ is a scalar with a Yukawa interaction with $\chi$. There is an approximate $U(1)$ global symmetry under which $S$ and $\bar{\chi}_L\chi_R$ are oppositely charged. The potential gives $S$ a VEV, which breaks the approximate $U(1)$ symmetry. In the non-linearly realized phase, using the parametrization $S = \frac{(f+s)}{\sqrt{2}} e^{i\phi/f}$ and doing a chiral rotation gives the pseudo-Goldstone interaction with the identification $m_\chi  = y f/\sqrt{2}, \Lambda/g = 2f$
which gives
\begin{align}
    \frac{g \, m_\chi}{\Lambda} = \frac{y}{2\sqrt{2}}\:.
\end{align}
Hence we see that for realistic values of the Yukawa coupling, $y\lesssim 4\pi$, $g(m_\chi/\Lambda) \lesssim 4\pi/(2\sqrt{2}) \approx 4.4 $ can be only modestly large.

\subsection{A strongly coupled UV completion with a parametric enhancement}
Consider a QCD-like theory which confines at scale $\Lambda_c$, and has three flavors of quarks: two light quarks $q$ with equal or comparable mass $m \ll \Lambda_c$ and a heavy quark $Q$ with mass $M \gg \Lambda_c$. The approximate $SU(2)_L\times SU(2)_R$ symmetry is realized non-linearly, giving a pion $\pi$ with mass $m_\pi^2 \sim m \Lambda_c$. Consider the spin-$1/2$ baryon $QQq$~\cite{Hu:2005gf,Brambilla:2005yk} with mass $m_{QQq} \approx 2M$. Identifying $\chi$ with $QQq$ and $\phi$ with $\pi$, the chiral Lagrangian has a term of the form we are interested in.\footnote{We assume that the vector heavy-flavor number carried by $Q$ is exact, so that the lightest state in the sector with two units of heavy-flavor charge is stable and can be a DM candidate. The UV completion also contains an isospin partner and a nearby spin$-3/2$ baryon. These states must be included as additional coupled channels if their splittings are comparable to the non-relativistic energy, but their presence does not suppress the leading pion coupling.} This identification gives~\cite{Hu:2005gf,Qiu:2020omj}
\begin{align}
    \frac{g \, m_\chi}{\Lambda} = \frac{g_A m_{QQq}}{2f_\pi} \approx \frac{4 \pi g_A M}{\Lambda_c} \:,
\end{align}
where $g_A$ is the axial coupling of the pion to the baryon current in the chiral Lagrangian, $f_\pi = \Lambda_c/(4\pi)$ is the pion decay constant and we used $m_{QQq} \approx 2M$ in the last line. Due to the assumed hierarchy $M \gg \Lambda_c$, we see that $g(m_\chi/\Lambda)$ can be made large without making $g$ large.
The hierarchy $m \ll \Lambda_c \ll M$ is technically natural. The limit $m\to0$ restores the light-flavor chiral symmetry, while $M$ can be an independent vector-like mass and does not break the symmetry protecting the pion. If the quark masses instead arise from a common Higgs expectation value, the hierarchy corresponds to hierarchical Yukawa couplings, but no Yukawa coupling is required to be parametrically large.

\section{An analytical approximation for Sommerfeld enhancement}
\label{app:Analytical-Approximation-for-SE}

The equation to be solved is
\begin{align}
    -\frac{\dd^2\Psi}{\dd x^2} + v(x) \Psi(x) = \xi^2 \Psi (x)\:,
\end{align}
where $\Psi(x) = \left(u(x), w(x)\right)^T$ and the potential $v(x)$ is given as (using $x_{12} = x_0$)
\begin{align}
    v(x) &= \Theta(x_0-x)
    \begin{pmatrix}
    -v_0 & 0 \\
    0 & -v_0 \lambda
    \end{pmatrix}
    +
    \begin{pmatrix}
    0 & 0 \\
    0 & 6/x^2
    \end{pmatrix}
    \nonumber
    \\
    &\qquad
    +\delta(x-x_{23})\begin{pmatrix}
    0 & \Delta_M \\
    \Delta_M & \Delta_T
    \end{pmatrix}
    +
    \delta(x-x_{34})\begin{pmatrix}
    \Delta_C & 0 \\
    0 & 0
    \end{pmatrix}\:,
\end{align}
where the last line is the approximation used in replacing $v_L$ by a sum of delta function terms. There are five parameters that need to be specified: $x_{23}, x_{34}, \Delta_M, \Delta_T$ and $\Delta_C$. As discussed in the main text, we set $\Delta_T = 0$ and adjust $x_{23}$ and $\Delta_M$ by hand to match the numerical and analytical results. For $\Delta_C$, we estimate it by matching the matrix element in the first Born approximation, using the $\ell=0$ wavefunction for $\xi\to 0$.\footnote{While we use the unperturbed wavefunction, if we were to use the wavefunction matched to the one in the square well region, the results do not change.} The free wavefunction in the $\ell=0$ channel scales as $x$, for $\xi\to0$. We therefore estimate
\begin{align}
    \int_{x_0}^\infty \dd x \, x^2 \delta(x-x_{34})\Delta_C = \int_{x_0}^\infty \dd x\, \frac{G}{4}\frac{m_\phi}{m_\chi}\frac{e^{-x}}{x} x^2\:,
\end{align}
which gives
\begin{align}
    \Delta_C = \frac{G}{4}\frac{m_\phi}{m_\chi}\frac{1+x_0}{x_{34}^2} e^{-x_0}\:.
    \label{eq:DeltaC-definition}
\end{align}
To estimate $x_{34}$, we take the product of two copies of the $\ell = 0$ free-particle wavefunctions and $(v_L)_{uu}$, which gives a distribution proportional to $x e^{-x}$; we calculate the expectation value of $x$ under this distribution, which gives
\begin{align}
    x_{34} = \left(\int_{x_0}^\infty \dd x\, xe^{-x}\right)^{-1} \int_{x_0}^\infty \dd x \, x\, xe^{-x}  = 2 +\mathcal{O}(x_0)\:.
    \label{eq:x34-estimate}
\end{align}
We therefore use $x_{34} = 2$ in the analytical approximations.

We split the full region $0 < x < \infty$ into four regions in which the wavefunctions can be solved for analytically. Requiring the appropriate boundary conditions at $x = 0$ and for $x\to\infty$ (which restrict the leftmost and the rightmost regions respectively, but not the ones in the middle), the wavefunctions in the four regions are given as
\begin{align}
    &\textbf{1: } 0\leq x < x_{12} = x_0\:,\:\:
    \nonumber \\
    &u_1(x) = A_u^{(1)} \sin (x \xi_u )\:,\:\:
    w_1(x) = A_w^{(1)} \frac{3 x \xi_w \cos (x \xi_w ) + (x^2 \xi_w^2 - 3)\sin (x \xi_w)}{x^2\xi_w^2}\:,
    \label{eq:u_1}
    \\
    &\xi_u^2 = \xi^2+v_0\:,\:\: \xi_w^2 = \xi^2 +v_0\lambda\:,
    \nonumber \\[1em]
    &\textbf{2: } x_0 = x_{12} \leq x < x_{23}\:,
    \nonumber \\
    &u_2(x) = A_u^{(2)} \sin (x \xi ) + B_u^{(2)} \cos (x \xi )\:, \\
    &w_2(x) = A_w^{(2)} \left(\frac{3\sin(x\xi)}{x^2\xi^2} - \frac{3\cos(x\xi)}{x\xi}-\sin(x\xi)\right)
    + B_w^{(2)} \left(-\frac{3\cos(x\xi)}{x^2\xi^2} - \frac{3\sin(x\xi)}{x\xi} +\cos(x\xi) \right)
    \:, \\[1em]
    &\textbf{3: } x_{23} \leq x < x_{34}\:,
    \nonumber \\
    &u_3(x) = A_u^{(3)} \sin (x \xi ) + B_u^{(3)} \cos (x \xi )\:, \\
    &w_3(x) = A_w^{(3)} \left(\frac{3\sin(x\xi)}{x^2\xi^2} - \frac{3\cos(x\xi)}{x\xi}-\sin(x\xi)\right)
    + B_w^{(3)} \left(-\frac{3\cos(x\xi)}{x^2\xi^2} - \frac{3\sin(x\xi)}{x\xi} +\cos(x\xi) \right)
    \:, \\[1em]
    &\textbf{4: } x_{34}\leq x < \infty\:,
    \nonumber \\
    &u_4(x) = \frac{1}{2i\xi}\left(A_u^{(4)} e^{i x \xi} - e^{-i x \xi}\right)\:,
    \label{eq:u_3}\:\:
    \\
    &w_4(x) = -\frac{\sqrt{2}}{2i\xi}\left(A_w^{(4)} e^{i x \xi}\left(1+\frac{3i}{x\xi}-\frac{3}{x^2\xi^2}\right) - e^{-i x \xi}\left(1-\frac{3i}{x\xi}-\frac{3}{x^2\xi^2}\right)\right)\:.
\end{align}
Here we have dropped an overall multiplicative factor of $\sqrt{4\pi}$ in the wave function, which drops out in the computation of SE. At this point we have 12 complex constants $A_u^{(1)}, A_w^{(1)}$, $A_u^{(2)}, B_u^{(2)}, A_w^{(2)}, B_w^{(2)}, A_u^{(3)}, B_u^{(3)}, A_w^{(3)}, B_w^{(3)}, A_u^{(4)}, A_w^{(4)}$ to be solved for. These are solved by the condition that the wavefunction has to satisfy at the three boundaries $x = x_{12} = x_0$, $x = x_{23}$ and $x = x_{34}$. Around $x = x_{12}$, we use the standard requirement that the wavefunction and its derivative are equal:
\begin{align}
    &u_2(x_{12}) = u_1(x_{12})\:,\:\:
    w_2(x_{12}) = w_1(x_{12})\:,\:\:
    u_2'(x_{12}) = u_1'(x_{12})\:,\:\:
    w_2'(x_{12}) = w_1'(x_{12})\:.
\end{align}
Integrating the Schr\"odinger equation around $x = x_{23}$ requires the wavefunctions to satisfy
\begin{align}
    &u_3(x_{23}) = u_2(x_{23})\:,
    \nonumber \\
    &w_3(x_{23}) = w_2(x_{23})\:,
    \nonumber \\
    &u_3'(x_{23}) = u_2'(x_{23})  + \Delta_M \, w_2(x_{23})\:,
    \nonumber \\
    &w_3'(x_{23}) = w_2'(x_{23}) + \Delta_M \, u_2(x_{23}) \:.
\end{align}
Similarly, integrating the Schr\"odinger equation around $x = x_{34}$ requires the wavefunctions to satisfy
\begin{align}
    &u_4(x_{34}) = u_3(x_{34})\:,
    \nonumber \\
    &w_4(x_{34}) = w_3(x_{34})\:,
    \nonumber \\
    &u_4'(x_{34}) = u_3'(x_{34})  + \Delta_C \, u_3(x_{34})\:,
    \nonumber \\
    &w_4'(x_{34}) = w_3'(x_{34}) \:.
\end{align}
Together, these 12 conditions provide 12 linear equations for the 12 variables, and one can solve for the coefficients explicitly. The exact expressions are long and do not provide any further insight, so we will not write them here.

Having solved for the coefficients, the SE is straightforward to obtain, and is given by
\begin{align}
    \text{SE} = \lim_{x\to0} \:  \left|\frac{A_u^{(1)}\sin(x\sqrt{\xi^2 + v_0})}{\sin(x\xi)/\xi}\right|^2\:
    = \left|A_u^{(1)}\right|^2\left(\xi^2 + v_0\right),
\end{align}
where we have used the free reduced wave function to be $u_0(x) = \sin (x\xi)/\xi$, obtained by setting $A_u^{(4)} = 1$ in the expression for $u_4(x)$ in Eq.~\eqref{eq:u_3} and matching to the expression for $u_1(x)$ in Eq.~\eqref{eq:u_1}. We have also taken the $x\to0$ limit in the second equality.

In practice, the approximation is somewhat crude and captures the features in the SE only qualitatively. Further, it does not get the overall normalization correct. To address this, we redefine $\text{SE}$ to be
\begin{align}
    \text{SE}
    = \mathcal{N}\left|A_u^{(1)}\right|^2\left(\xi^2 + v_0\right)\:,
\end{align}
and fix $\mathcal{N}$ to match the numerical and analytical results at one point. We choose to match them at $v_0x_0^2 = 10^{-2}$. The plots in Fig.~\ref{fig:SE-numerical-vs-analytical} are shown after this rescaling.

As seen in Fig.~\ref{fig:BoundState-vs-alpha}, there are periodic features in the plot of $A_u^{(1)}$ for negative imaginary values of $\xi$, and these are periodic along the real $\xi$ direction. We would like to understand if this is coming from the long-range or the short-range part of the potential. To investigate this, we can calculate the amplitude $A_u^{(1)}$ as an expansion in $\Delta_M, \Delta_C$ (which parametrize the long-range potential)
\begin{align}
    A_u^{(1)} &= A_{u,0}^{(1)} + \Delta_M A_{u,M}^{(1)} + \Delta_C A_{u,C}^{(1)} + \cdots \:,
\end{align}
where $A_{u,0}^{(1)}, A_{u,M}^{(1)}, A_{u,C}^{(1)}$ are calculable analytically ($A_{u,0}^{(1)}$ is the pure square well result). One can plot these quantities over the complex plane in $\xi$ and see that in a range of $\xi$, while $A_{u,0}^{(1)}$ does not have a periodic structure, $A_{u,C}^{(1)}$ does.

The expression for $A_{u,C}^{(1)}$ can be simplified and allows understanding this structure analytically. For notational convenience we define
\begin{align}
    q = \sqrt{\xi^2 + v_0}\:,\:\: Q = q\cos(q x_0) -i \xi \sin(q x_0)\:,
\end{align}
in terms of which we get
\begin{align}
    A_{u,C}^{(1)}
    &=
    -\frac{e^{i\xi(x_{34}-2x_0)}}{Q^2}
    \left(e^{i\xi(x_{34}-x_0)} \sin (q x_0) + \frac{Q}{\xi}\sin(\xi(x_{34}-x_0))\right)
\end{align}
Using $x_0 \ll x_{34}$, we can immediately see that the exponential term $e^{i\xi x_{34}}$, upon the substitution $\xi = \xi_R + i \xi_I$ becomes $e^{i\xi_R x_{34}} e^{-\xi_I x_{34}}$. As a result, the amplitude is larger for $\xi_I < 0$, and the oscillations along $\xi_R$ axis appear with frequency $1/x_{34}$ which is order 1. Since this effect comes from the first order term in the expansion of the amplitude, this is a purely IR effect.

\section{A different scaling of coordinates}
\label{app:altdimlesscoords}

\begin{figure}
    \centering
\includegraphics[scale=0.42]{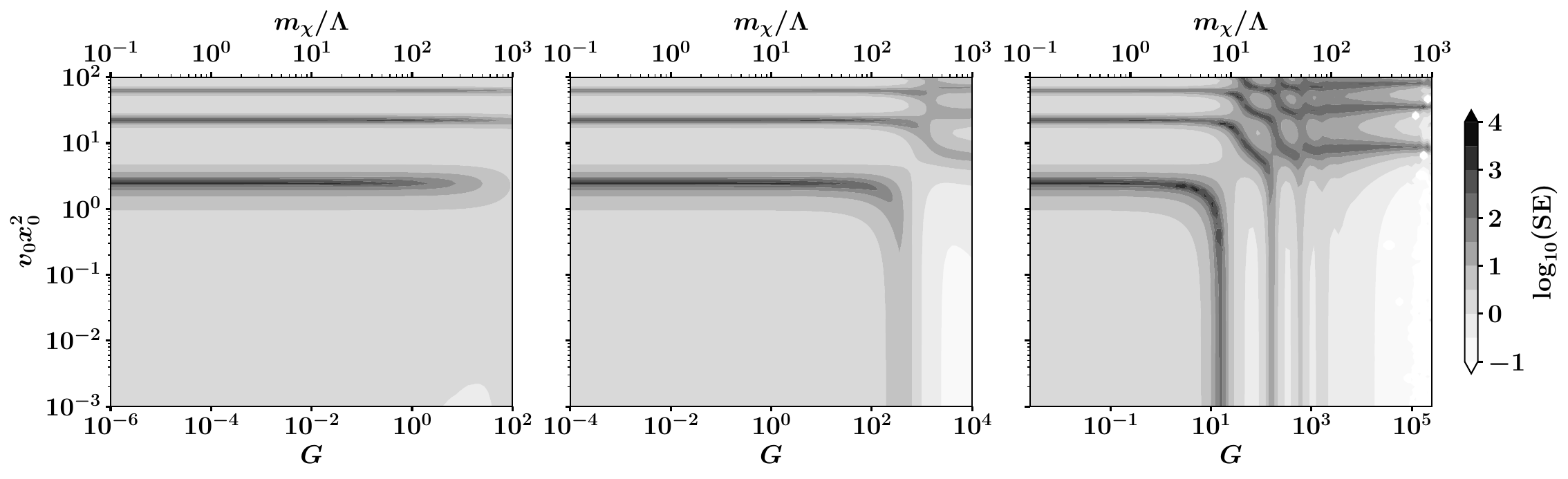}
    \caption{SE for a massless mediator $\epsilon_\phi=0$, at relative velocity $v_{\rm rel}=10^{-3}$, $\lambda=0$ for the pseudo-Goldstone mediator. The coupling $g$ increases from the left to right $g/\sqrt{3\pi}=0.01,\ 0.1,\  0.5$, and we vary the well depth $v_0 x_0^2$ and $m_\chi/\Lambda$. We see that the features of the SE in the massless mediator limit are similar to that when $m_{\phi} \ll m_{\chi}$, as shown in  Fig~\ref{fig:SE-Goldstone-vs-G-vs-v0}.}
    \label{fig:SE_massless}
\end{figure}

Here we provide an alternate rescaling of the Schr\"odinger equation that is amenable to the limit of the mediator mass going to zero. This is more in line with the scaling choice used in previous literature, and we have verified that the physical results such as the SE and bound state existence are the same as those computed using the scaling choice in the main text (see Fig.~\ref{fig:SE_massless}).
\noindent
The Schr\"odinger equation to be solved is
\begin{align}
    &-\frac{1}{m_\chi}\Psi''(r) + V_\text{eff}(r) \Psi(r) = 2 E\, \Psi(r)\:,\:\:
    \Psi(r) = \begin{pmatrix}
        u(r) \\
        w(r)
    \end{pmatrix} \:,
    \nonumber \\
    & V_\text{eff}(r) = \Theta (a-r)\, V_S(r)  + \Theta (r-a)\,V_L(r) + \begin{pmatrix}
        0 & 0 \\
        0 & \frac{6}{m_\chi r^2}
    \end{pmatrix}
    \:.
\end{align}
Setting
\begin{align}
    x = m_\chi r\:,\:\: \epsilon_{\phi} = \frac{m_\phi}{m_\chi}\:,\:\: \xi^2 = \frac{2 E}{m_\chi}\:, v(x)= \frac{V_{\rm eff}(r)}{m_\chi},
\end{align}
the equation becomes
\begin{align}
    &-\Psi''(x) + v(x) \Psi(x) = \xi^2 \Psi(x)\:,\:\:
\end{align}
where
\begin{align}
    v(x)
    = \Theta (x_0-x) v_S(x)
    + \Theta (x-x_0) v_L(x)
    +
    \begin{pmatrix}
        0 & 0 \\
        0  & 6/x^2
    \end{pmatrix}\:,
\end{align}
where the boundary between the long-range and the short-range part of the potential is at $x = x_0$ (i.e. $r = x_0/m_\chi$). In the formalism of Sec.~\ref{sec:numerics}, the potentials are given as
\begin{align}
    v_S(x)
    &=
    \begin{pmatrix}
        -v_0 & 0 \\
        0  & -\lambda v_0
    \end{pmatrix}\:,
    \nonumber \\
    v_L(x) &= \frac14 G \frac{e^{-\epsilon_{\phi} x}}{x}
    \begin{pmatrix}
        \epsilon_{\phi}^2 &
        \sqrt{8}  \left(\epsilon_{\phi}^2 + 3\epsilon_{\phi}/x + 3/x^2\right) \\
        \sqrt{8}  \left(\epsilon_{\phi}^2 + 3\epsilon_{\phi}/x + 3/x^2\right) &
        \:\:\:\: -\left(\epsilon_{\phi}^2 + 6\epsilon_{\phi}/x + 6/x^2\right)
    \end{pmatrix}\:.
\end{align}
In the massless mediator limit, $\epsilon_{\phi} = 0$, the potential reduces to
\begin{align}
v_L(x) &= \frac{G}{4x}
    \begin{pmatrix}
        0 &
        3\sqrt{8}/x^2  \\
        3\sqrt{8}/x^2 &
        \:\:\:\: - 6/x^2
    \end{pmatrix}\:.
\end{align}

Alternatively, we could instead rescale the radial coordinate by $\Lambda$, ensuring that the cutoff is at a roughly fixed coordinate position for $\Lambda < m_\chi$. In this case we would define:
\begin{align}
    x = \Lambda r\:,\:\: \epsilon_{\phi} = \frac{m_\phi}{\Lambda}\:,\:\: \xi^2 = \frac{2 E m_\chi}{\Lambda^2}\:, v(x) = \frac{m_\chi V_{\rm eff}(r)}{\Lambda^2}
\end{align}
and the equation becomes
\begin{align}
    &-\Psi''(x) + v(x) \Psi(x) = \xi^2 \Psi(x)\:,\:\:
\end{align}
where
\begin{align}
    v(x)
    = \Theta (x_0-x) v_S(x)
    + \Theta (x-x_0) v_L(x)
    +
    \begin{pmatrix}
        0 & 0 \\
        0  & 6/x^2
    \end{pmatrix}\:,
\end{align}
where the boundary between the long-range and the short-range part of the potential is at $x = x_0$ (i.e. $r = x_0/\Lambda$). In the formalism of Sec.~\ref{sec:numerics}, the potentials are given as
\begin{align}
    v_S(x)
    &=
    \begin{pmatrix}
        -v_0 & 0 \\
        0  & -\lambda v_0
    \end{pmatrix}\:,
    \:v_L(x) = \frac{g^2}{12 \pi } \frac{m_\chi}{\Lambda} \frac{e^{-\epsilon_{\phi} x}}{x}
    \begin{pmatrix}
        \epsilon_{\phi}^2 &
        \sqrt{8}  \left( \epsilon_{\phi}^2 + \frac{3\epsilon_\phi}{x} + \frac{3}{x^2}\right) \\
        \sqrt{8}  \left(\epsilon_{\phi}^2 + \frac{3\epsilon_\phi}{x} + \frac{3}{x^2}\right) &
        \:\:\:\: -\left(\epsilon_{\phi}^2 + \frac{6\epsilon_\phi}{x} + \frac{6}{x^2}\right)
    \end{pmatrix}\:.
\end{align}
In the massless mediator limit, $\epsilon_{\phi} = 0$, the potential reduces to
\begin{align}
v_L(x) &= \frac{g^2}{12 \pi} \frac{m_\chi}{\Lambda} \frac{1}{x^3}
    \begin{pmatrix}
        0 &
        3\sqrt{8}  \\
        3\sqrt{8} &
        \:\:\:\: - 6
    \end{pmatrix}\:.
\end{align}
We thus see that in general the long-range behavior can be described by the three parameters $\xi$ (given by $(v_\text{rel}/2) (m_\chi/\Lambda)$), $\epsilon_\phi$ (describing the ratio between the range of the Yukawa potential and the cutoff scale) and an effective coupling $(g^2/(12 \pi)) (m_\chi/\Lambda)$. We know from the results of Sec.~\ref{sec:bound_sq_well} that we only expect a long-range enhancement when the effective coupling is $\gtrsim 1$, which in this case amounts to requiring $\Lambda/m_\chi \lesssim g^2/(12\pi)$ -- that is, for weak couplings we need a large hierarchy between the cutoff scale $\Lambda$ and the mass $m_\chi$. We also expect that we will need $\epsilon_\phi \lesssim 1$ for the potential to have any significant effect (else it is already exponentially suppressed at the short-distance cutoff radius).

We can also estimate the requirement for the zero-$m_\phi$ potential to have large effects by noting that replacing $1/x^3 \rightarrow 1/x$ corresponds to making the potential strictly stronger at all $x$ for $x > 1$ (i.e.~outside the cutoff radius). For a $V(r)=-\alpha/r$ potential, we know that a necessary condition for a large Sommerfeld enhancement is $v_\text{rel}/2 \lesssim \alpha$. Rescaling to dimensionless coordinates $x=m_\chi r$ in that case, we would obtain the following dimensionless Schr\"{o}dinger equation:
\begin{align}
    &-\Psi''(x) - (\alpha/x) \Psi(x) = (v_\text{rel}/2)^2 \Psi(x)\:,\:\:
\end{align}
Thus the condition that $v_\text{rel}/2 \lesssim \alpha$ would translate in our case to the condition
\begin{align}
(g^2/(12\pi)) (m_\chi/\Lambda) \gtrsim \xi \ ,
\end{align}
up to the $\mathcal{O}(1)$ factors in the matrix potential, which in turn suggests we will need $v_\text{rel}/2 \lesssim g^2/(12\pi)$ to expect an appreciable enhancement (since if this condition is not satisfied, our potential would be everywhere weaker than one which does not support a large enhancement). The actual condition on the velocity may be stronger, since the $1/x^3$ potential is weaker than $1/x$ outside the cutoff. This inequality also ensures that the $v_L(x)$ potential term will be enhanced relative to the centrifugal repulsive term at $x=1$, although the centrifugal term may still dominate for sufficiently high $\ell$. In particular, the inequality corresponds exactly to the requirement that the lower diagonal component of $v_L(x)$ exceeds the $6/x^2$ centrifugal contribution to this component at $x=1$. \footnote{We thank Brando Bellazzini for this observation.}

Furthermore, validity of the EFT requires that $m_\chi v_\text{rel}/2 \lesssim \Lambda$, so we expect that in the regime where both the EFT is valid and a substantial enhancement can be expected, we will have:
\begin{align} \sqrt{2 E/m_\chi} = v_\text{rel}/2 \lesssim \Lambda/m_\chi \lesssim g^2/(12\pi) \end{align}

\section{Single channel singular potentials}\label{app:p_wave_sing_chan}

In this appendix, we work out the case of the $\ell=1$ single-channel attractive singular potential, as alluded to in Sec.~\ref{sec:potentials}. We focus on the pseudoscalar mediator, and find that for the quantum numbers $\ket{j,m_{j},\ell,s}=\ket{0,0,1,1}$, the associated potential is $V(r)=(1/4) V_C(r) - V_T(r)$ which is attractive and singular. Our arguments for the lack of SE in the  weakly coupled regime ($\alpha\lesssim 1$) also apply here (both via the NREFT power counting and the matching to a short-range potential via the first order Born approximation), and we demonstrate the numerical evidence for this at the parameter point $m_{\phi}=0$, i.e.~the massless mediator limit, where the potential has the longest possible range. Numerically, we solve the Schr\"odinger equation,
\begin{align}
    -u^{\prime\prime}(x) + v(x) u(x) = \xi^2 u(x)\ ,
\end{align}
where we have scaled to dimensionless quantities defined as
\begin{align}
   x = m_\chi r\:,\:\: \epsilon_{\phi} = \frac{m_\phi}{m_\chi}\:,\:\: \xi^2 = \frac{2 E}{m_\chi}\:, v(x)= \frac{V_{\rm eff}(r)}{m_\chi},
\end{align}
with the effective potential for the $p$ wave here being
\begin{align}
    V_\text{eff}(r) = \Theta (a-r)\, V_S(r)  + \Theta (r-a)\,V_L(r) + \frac{2}{m_\chi r^2}\ .
\end{align}
As done previously, the short-range potential is fixed by matching the first Born approximation for the free scattering states to that from a spherical well. In dimensionless units, we obtain
\begin{align}
    v(x) &= -\frac{5\alpha}{2(m_{\chi}a)^3}\Theta (m_{\chi}a-x) - \frac{\alpha}{4x^3}e^{-\epsilon_{\phi}x}(2+x\epsilon_{\phi})^2 \Theta (x-m_{\chi}a) + \frac{2}{x^2}\ , \label{eq:p_wav_sing_pot}
\end{align}
where $a$ denotes the matching radius or the radius of the short-range spherical well. For typical non-relativistic velocities and to ensure the applicability of QM outside the matching radius, the canonical choice is $a=m_{\chi}^{-1}$, i.e. $x_0=m_{\chi} a = 1$.

The SE for $p$-wave states ($\ell=1$) is given by
\begin{align}
    \mathrm{SE} = \lim_{x\rightarrow 0}~\bigg\vert \dv{}{x}\left(\frac{u(x)}{x}\right) \bigg\vert^2 \bigg\vert \dv{}{x}\left(\frac{u_0(x)}{x}\right) \bigg\vert^{-2}\ ,
\end{align}
where $u_0(x)$ is the reduced wavefunction when $v(x)=2/x^2$, corresponding to free propagation in the absence of a potential. The large-$x$ boundary conditions have to be specified in addition to the regularity condition at the origin, $u(x) \underset{x\rightarrow 0}{\propto}x^{2}$, i.e. $u(x)$ has to be the appropriate combination of the purely incoming and outgoing waves (at zero interaction potential) at large $x$. Practically, we implement that by dividing out by the Wronskian of the numerical solution $u(x)$ computed with respect to the purely-outgoing solution $\tilde{u}(x)=c_{1}(\xi x) + i s_{1}(\xi x)$, for both $u(x)$ and $u_{0}(x)$ evaluated at some large $x_{\rm{max}}$. Here $c_{\ell}(x) = -x y_{\ell}(x)$, and $s_{\ell}(x)=xj_{\ell}(x)$ are respectively irregular and regular solutions to the free Schr\"odinger equation for the $\ell^{\rm th}$ partial wave.

\begin{figure}[!htbp]
    \centering
    \begin{subfigure}[t]{0.6\textwidth}
        \centering
        \includegraphics[scale=0.7]{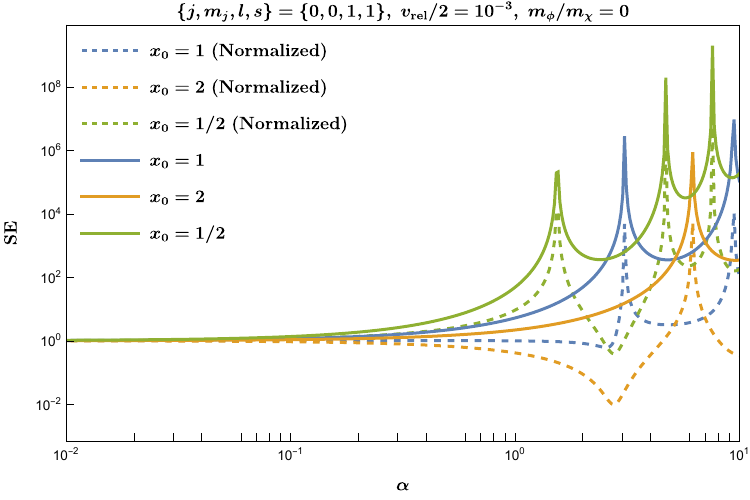}
    \end{subfigure}

    \begin{subfigure}[t]{0.6\textwidth}
        \centering
        \includegraphics[scale=0.7]{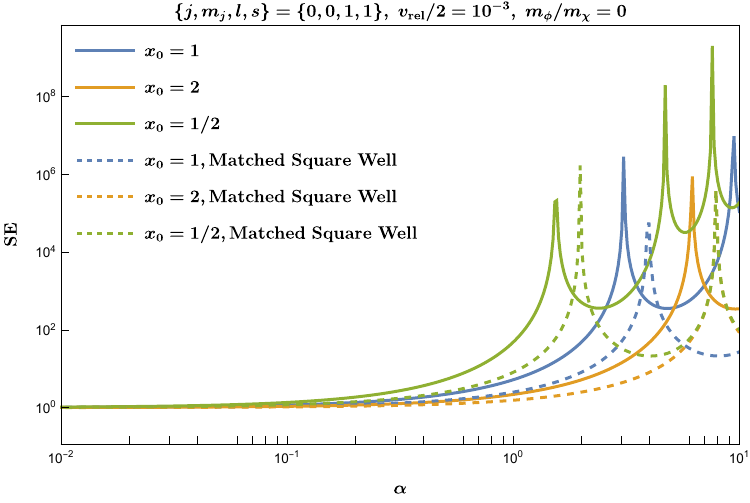}
    \end{subfigure}

    \begin{subfigure}[t]{0.6\textwidth}
        \centering
        \includegraphics[scale=0.7]{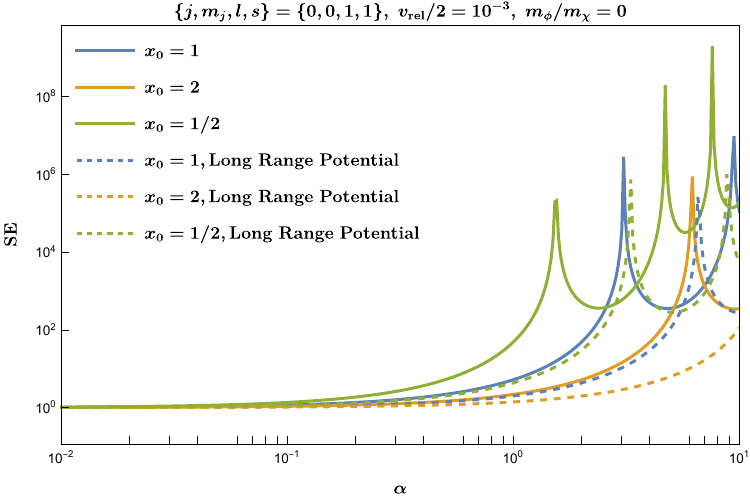}
    \end{subfigure}
    \caption{SE as a function of coupling $\alpha$ for pseudoscalar mediators for the state with quantum numbers $\ket{j,m_j, \ell, s}=\ket{0,0,1,1}$. We plot the SE as a function of the cutoff at $x_0=1,2,1/2$ respectively, for a massless mediator. The curves have been computed by numerically solving the Schr\"odinger equation from $x_{\rm min}=10^{-4}$ to $x_{\rm max}=10^{2}$, at $\xi = v_{\rm rel}/2 = 10^{-3}$. Solid curves show the overall SE and are the same across all panels. In the first row, the dashed curves show the SE normalized to the corresponding SE at $\xi_{\rm ref} = 0.5$. In the second row, the dashed curves show the SE arising purely from the short distance square well, and in the third row, the dashed curves show the SE arising purely from the long-range IR potential alone. All three panels demonstrate that the SE in the system is negligible for perturbative couplings $\alpha \lesssim 1$.}
    \label{fig:p_wave_SE}
\end{figure}

Implementing the procedure above, we plot the SE at $\epsilon_\phi =0$ for the potential in Eq.~\eqref{eq:p_wav_sing_pot} in Fig.~\ref{fig:p_wave_SE}. In the first row, we plot the SE at the parameter point $\xi = v_{\rm rel}/2 =10^{-3}$ as a function of coupling strength $\alpha$ for the cutoff choices of $x_0 =1,2, 1/2$ respectively. For reference, we also plot the normalized version of the SE, where we divide out each solid curve point-by-point for a reference SE computed at $\xi_{\rm ref}=0.5$. As is evident in the plot, there is no significant SE for $\alpha\lesssim 1$ (for comparison, SE at $\alpha=1$ is $\sim 50$ ($x_0=1/2$), $\sim 5$ ($x_0=1$) and $\sim 2$ ($x_0=2$)). This is in line with our NREFT predictions and matching arguments, as given in the main text. As in the coupled channel analysis, we do find that normalizing by the value of the SE at a high reference velocity reduces the cutoff dependence at small couplings, but retains the resonance structure ($\alpha \sim 3.08$ at $x_0=1$ is the first peak).

In the second row, we show the SE obtained purely from the short-range potential $V_{S}/m_{\chi}$, for each cutoff choice, with the dashed lines. Finally, in the third row, we plot with dashed lines the SE  purely from the long-range potential $V_{L}/m_{\chi}$, where we find that over the range of $\alpha \lesssim 1$, the SE purely from the IR potential is small ($\sim 5$ for the maximum of the three dashed curves at $\alpha=1$). We observe that the short-range physics is set here by matching to the long-range physics, and that consequently their effects are similar, in that the onset of a large SE occurs at a similar $\alpha$ and to a similar degree when either the long-range or short-range physics is set to zero.

Thus, much like the coupled-channel analysis, in this example of a single-channel singular potential,  we find empirically that the SE is not significant for pseudoscalar mediators with weak couplings.

\section{Loop potentials} \label{app:loop_pot}

In this appendix, we illustrate how soft loops generate the loop-level interaction potential in NREFT. For instance, for the pseudoscalar NREFT, the soft loop diagram in Eq.~\eqref{eq:soft_loop} generates the loop level amplitude
\begin{align}
    i\cu{A}_{\mathrm{1-loop\ potential}} &= -\lim_{v \rightarrow 0} \left(\frac{-ig^2}{m_{\chi}}\right)^2 \int \underbrace{\frac{d^d l}{(2\pi)^d}}_{\sim v^{4}}\ \underbrace{\frac{1}{(l_0^2-\vec{l}^{\ 2}) \left[(l_{0})^2-(\vec{l}+\vec{q}-\vec{p})^2\right]}}_{\sim v^{-4}}
    \xi_{s_{1}^{\prime}}^{\dagger}
    \xi_{s_{1}^{\phantom{\prime}}}^{\phantom{\dagger}}
    ~\xi_{s_{2}^{\prime}}^{\dagger}
    \xi^{\phantom{\dagger}}_{s_{2}^{\phantom{\prime}}} \\
    &=
    \frac{i g^4}{16\pi^2 m_{\chi}^2}\left[\frac{1}{\varepsilon} + 2 + \ln(\frac{4\pi e^{\gamma_E} \mu^2}{(\vec{q}-\vec{p})^2}) \right] \xi_{s_{1}^{\prime}}^{\dagger}
    \xi_{s_{1}^{\phantom{\prime}}}^{\phantom{\dagger}}
    ~\xi_{s_{2}^{\prime}}^{\dagger}
    \xi^{\phantom{\dagger}}_{s_{2}^{\phantom{\prime}}}
\end{align}
where the divergent pieces can be subtracted off using counterterms in the NREFT. Note however that the finite terms are at-most $\ln{v}$ enhanced, making the $v$ counting of the one loop potential marginal, and one can show that its iteration in momentum space (via solving the Lippmann Schwinger equation) will also not generate SE. In position space, the loop level potential is given by
\begin{align}
    V_{\mathrm{1-loop\ potential}} &= -\frac{g^4}{16\pi^2 m_{\chi}^2}\int~\frac{d^3\vec{q}}{(2\pi)^3}~e^{-i\vec{q}\cdot \vec{r}} \left[2+\ln(\frac{\mu^2}{\vec{q}^{\ 2}})\right]\\
    &=-\frac{g^4}{32\pi^3 m_{\chi}^2r^3} - \frac{g^4}{8\pi^2m_{\chi}^2} \delta^{(3)}(\vec{r})
\end{align}
which is also singular. This is a spin independent term~\cite{Ferrer:1998ue} and from its matrix elements in the basis of free scattering states $s_{\ell}(r)$, it is manifest that the first Born approximation for $\ell=0$ states is rendered divergent.

\section{Example: Coulomb potential regulated by spherical well}\label{app:coulomb}

To demonstrate the origin of cutoff/regulator-dependence in the Sommerfeld enhancement, let us examine the simpler case of a regular Coulomb potential, replaced at $r < a$ by a spherical well such that the tree-level scattering amplitude remains invariant. For simplicity, we will focus on the $\ell=0$ $s$-wave case.

Let us work in dimensionless coordinates $x=\alpha m_\chi r$, where $\alpha$ is the coupling for the Coulomb potential. If the Schr\"{o}dinger equation is initially:
\begin{align} - \frac{1}{m_\chi} \frac{d^2}{dr^2} u(r) + V(r) u(r) = (p^2/m_\chi) u(r),\end{align}
in these units it becomes:
\begin{align} - \frac{d^2}{dx^2} u(x) + \frac{V(x)}{\alpha^2 m_\chi} u(x) = \epsilon_v^2 u(x), \quad \epsilon_v = p/(\alpha m_\chi).\end{align}
We choose the potential as:
\begin{align} \overline{V}(x) \equiv \frac{V(x)}{\alpha^2 m_\chi} = - \begin{cases}1/x, & x > \alpha a m_\chi \\ \overline{V}_0, & x < \alpha a m_\chi\end{cases}\end{align}

If we require that the first Born approximation matches between the Coulomb potential and the spherical well with radius $a$, we have \begin{align} \int^{\alpha a m_\chi}_0 \overline{V}_0 x^2 dx =\int^{\alpha a m_\chi}_0 (1/x) x^2 dx \Rightarrow (\alpha a m_\chi)^2/2 = \overline{V}_0 (\alpha a m_\chi)^3/3, \label{eq:app_d_matching} \end{align}
so this condition (1st-order matching) imposes $\overline{V}_0=(3/2)/(\alpha a m_\chi)$.

Now the regular solution within $x<\alpha a m_\chi$ (corresponding to $r<a$) has the form $u(x) =\sigma \sin(\kappa x)$, where $\kappa=\sqrt{\epsilon_v^2 + \overline{V}_0}$ and $\sigma$ is a constant. Thus at $x=\alpha a m_\chi$ we have $u'(\alpha a m_\chi) = \kappa \sigma \cos(\kappa \alpha a m_\chi)$, $u(\alpha a m_\chi) = \sigma\sin(\kappa a \alpha m_\chi)$. Note that the usual Sommerfeld enhancement factor (to the amplitude) is determined from the $x\rightarrow 0$ limit as $\sigma \kappa/\epsilon_v = \sigma \sqrt{1 + \overline{V}_0/\epsilon_v^2}$. The naive Sommerfeld enhancement to the cross section thus takes the form $S=|\sigma|^2 (1 + \overline{V}_0/\epsilon_v^2)$.

We want to match this short-range solution onto the Coulomb solution with the canonically normalized incoming wave, which can be written for the $s$-wave case as:
\begin{align} u(x) & = \frac{-i}{2} e^{-\pi/(4 \epsilon_v)} W(-i/(2\epsilon_v), 1/2, 2 i \epsilon_v x) +  C \left[M\left(-\frac{i}{2 \epsilon_v},\frac{1}{2},2 i \epsilon_v x \right) \right. \nonumber \\
& \left. + e^{-\pi/(2 \epsilon_v)} \frac{1}{\Gamma(1 - i/(2\epsilon_v))} W(-i/(2 \epsilon_v),1/2,2  i \epsilon_v x)  \right].\label{eq:coulomb} \end{align}
Here $M$ and $W$ are the Whittaker functions.

If we take the limit as $a\rightarrow 0$, the boundary condition reduces to the usual condition of regularity. This imposes the requirement that the $W$ terms must cancel (since they are irregular as $x \rightarrow 0$), leading to the condition $C = \frac{1}{2} i e^{\pi/(4\epsilon_v)} \Gamma(1 - i/(2\epsilon_v))$, and hence (from the asymptotic behavior of the Whittaker $M$ function) $u(x) \rightarrow - e^{\pi/(4\epsilon_v)} \Gamma(1 - i/(2\epsilon_v)) (\epsilon_v x)$ as $r\rightarrow 0$. This leads to a Sommerfeld factor of $S=e^{\pi/(2\epsilon_v)} \frac{\pi}{2\epsilon_v} \frac{1}{\sinh(\pi/(2 \epsilon_v))} = \frac{\pi}{\epsilon_v} \frac{1}{1 - e^{-\pi/\epsilon_v}}$, which is the standard result for the Coulomb potential.

More generally, for non-zero $a$, we first solve for $C$ from:
\begin{align} \kappa \cot(\kappa a \alpha m_\chi) = \frac{u'(\alpha a m_\chi)}{u(\alpha a m_\chi)}, \end{align}
where $u(x)$ is evaluated as in Eq.~\eqref{eq:coulomb}. We can then substitute this result into Eq.~\eqref{eq:coulomb}, evaluate $u(\alpha a m_\chi)$, and set it equal to $\sigma \sin(\kappa \alpha a m_\chi)$ to obtain $\sigma$. The full resulting expression is rather complicated, but if we expand to lowest order in $\alpha a m_\chi$ (which we expect to be $\mathcal{O}(\alpha)$ for typical cutoff scales), we obtain:
\begin{align} \sigma \sqrt{1 + \overline{V}_0/\epsilon_v^2} \approx -e^{\pi/(4\epsilon_v)}  \Gamma\left(1 - \frac{i}{2 \epsilon_v}\right) (1 - \alpha a m_\chi /4) + \mathcal{O}((\alpha a m_\chi)^2). \end{align}
Consequently, the Sommerfeld enhancement takes the (cutoff-dependent) form:
\begin{align} S \approx S_0 (1 - \alpha a m_\chi/2), \quad S_0 = \frac{\pi}{\epsilon_v} \frac{1}{1 - e^{-\pi/\epsilon_v}}, \end{align}
where $S_0$ is the standard Coulomb Sommerfeld enhancement corresponding to the $a\rightarrow 0$ limit.\footnote{Here we have used the standard identity $\left|\Gamma\left(1 - \frac{i}{2 \epsilon_v}\right)\right|^2 = \frac{\pi}{2\epsilon_v \sinh(\pi/(2 \epsilon_v))}$.}

We observe that the modification is cutoff-dependent but not velocity-dependent in this small-$a$ limit; this is to be expected so long as $a$ corresponds to the smallest length scale in the problem (e.g. $a \ll 1/(m_\chi v), 1/(m_\chi \alpha)$). This corresponds to the statement that the physics at $r < a$ is not expected to be resolved provided the wavelength associated with the wavefunction is larger than $a$ (which for $a \sim 1/m_\chi$ simply corresponds to $\alpha \ll 1$, $v\ll 1$, i.e.~that the theory is both NR and weakly coupled). This suggests that a cutoff-independent Sommerfeld enhancement may still be defined by simply rolling the $(1 - \alpha a m_\chi/2)$ factor into the hard annihilation cross section.

If this redefinition of the annihilation cross section is not performed, the cutoff-dependent contribution to the enhancement is always $\mathcal{O}(\alpha)$ for $a\sim 1/m_\chi$. This means it is subdominant to any large Sommerfeld enhancement from the Coulomb potential, but if $S_0$ is also perturbative (i.e. of the form $1 + \mathcal{O}(\alpha)$), then the cutoff-dependent correction to the annihilation cross section can be of the same order as the $\mathcal{O}(\alpha)$ terms in $S_0$. Thus the cutoff dependence occurs at leading order in the corrections to the annihilation cross section in this case.

One might also ask whether the apparent  cutoff dependence is an artifact of the fact that we have used the conventional approach of reading off the Sommerfeld enhancement from the wavefunction as $x\rightarrow 0$; a number of recent papers have pointed out that for the enhancement to be unitary in general, corrections to the usual procedure are required \cite{Blum:2016nrz,Parikh:2024mwa,Flores:2024sfy,Flores:2025uoh,Flores:2026yay,Watanabe:2025kgw,Cimring:2026jzn}. However, those corrections are proportional to the unenhanced annihilation rate and so cannot fully compensate the annihilation-independent but cutoff-dependent effects we have identified here. We have checked that when we use the full corrected expression (Eq.~\eqref{eq:annihilation}), in the limit of a small annihilation rate, we obtain the same (cutoff/regulator-dependent) results as by reading off the Sommerfeld enhancement from the $x\rightarrow 0$ behavior. In this picture, the main source of the discrepancy in Sommerfeld enhancement between the original potential and the square-well-regulated potential is that the annihilation part of the short-range amplitude $f_\ell$ receives a Sommerfeld enhancement from the potential {\it within} $r < a$ (i.e. a loop-level correction to the tree-level annihilation cross section corresponding to potential exchange), and this enhancement differs at the $\mathcal{O}(1)$ level between the spherical well and Coulomb cases. It is quite natural to consider this apparent enhancement as part of the high-scale annihilation amplitude (since it also arises from short-range physics / high momentum scales), and where this short-range enhancement is small and perturbative, it could in principle be computed directly from loop diagrams in the underlying perturbative QFT.

We can similarly examine the bound states in this setup. We can write the exponentially decaying Coulombic wavefunction in the form:
\begin{align} u(x) = C_2 W(1/(2 \epsilon_E),1/2,2 \epsilon_E x),  \end{align}
where $\epsilon_E = \sqrt{E_b/(\alpha^2 m_\chi)}$ ($E_b$ denoting the binding energy), and match this at $x=\alpha m_\chi a$ to the solution $u(x) = \sin(\kappa^\prime x)$ where now $\kappa^\prime = \sqrt{\overline{V}_0 - \epsilon_E^2}$ (note the resulting solution will still need to be normalized). We can thus require $C_2 = \sin(\kappa^\prime \alpha m_\chi a)/ W(1/(2\epsilon_E), 1/2, 2 \epsilon_E \alpha m_\chi a)$, and the condition for a bound state is that $\kappa^\prime \cos(\kappa^\prime \alpha m_\chi a) = \left. C_2 \frac{d}{dx} W(1/(2 \epsilon_E),1/2,2 \epsilon_E x) \right|_{x=\alpha m_\chi a}$, or equivalently:
\begin{align} \kappa^\prime \cot(\kappa^\prime \alpha m_\chi a) = \left. \frac{d}{dx} \ln W(1/(2 \epsilon_E),1/2,2 \epsilon_E x) \right|_{x=\alpha m_\chi a}.\end{align}

Alternatively, using the framework described in the main text, we can solve for the dimensionless coefficient $A$ encoding the long-distance physics as in (the rescaled dimensionless version of) Eq.~\eqref{eq:bsmatching}, with the prescription $p=i (\alpha m_\chi) \epsilon_E$:
\begin{align} u(\alpha a m_\chi) = i A \sinh(\epsilon_E \alpha a m_\chi) + e^{-\epsilon_E \alpha a m_\chi}, \nonumber \\
u'(\alpha a m_\chi) \equiv i A \epsilon_E \cosh(\epsilon_E \alpha a m_\chi) - \epsilon_E e^{-\epsilon_E \alpha m_\chi a}.\end{align}
Solving these equations yields the expression:
\begin{align} A & = -\frac{i e^{-\alpha a m_\chi \epsilon_E} \text{csch}(\alpha a m_\chi \epsilon_E) }{W\left(\frac{1}{2 \epsilon_E},\frac{1}{2},2 \alpha a m_\chi \epsilon_E\right) (2 \alpha a m_\chi \epsilon_E^2 (\text{coth}(\alpha a m_\chi \epsilon_E) - 1) +1) + 2 \epsilon_E W\left(1+\frac{1}{2 \epsilon_E},\frac{1}{2},2 \alpha a m_\chi \epsilon_E\right)} \nonumber \\
& \times \left[(4 \alpha a m_\chi \epsilon_E^2-1) W\left( \frac{1}{2 \epsilon_E},\frac{1}{2},2 \alpha a m_\chi \epsilon_E\right) -2 \epsilon_E W\left(1+\frac{1}{2 \epsilon_E},\frac{1}{2},2 a \alpha m_\chi \epsilon_E\right)\right]. \end{align}
Now the short-range scattering amplitude for a finite spherical well of radius $a$ and dimensionless depth $\overline{V}_0=(3/2)/(\alpha a m_\chi)$ (to match the Coulomb potential at leading order as described below Eq.~\eqref{eq:app_d_matching}), with momentum $p = i(\alpha m_\chi) \epsilon_E$, is given at leading order by $p f(p) = (m_\chi/p) \alpha (p a)^2/2$. Here $f$ has units of $1/$mass, so to convert to dimensionless units we consider:
\begin{align}\alpha m_\chi f =  \alpha m_\chi^2 \alpha a^2/2 = (\alpha a m_\chi)^2/2. \end{align}

Thus the condition $pf\mathcal{B} =1$ becomes, to leading order in the short-range physics and all orders in the long-range physics:
\begin{align} 1 = (i \epsilon_E) (\alpha m_\chi) ((\alpha m_\chi a)^2/2)/(\alpha m_\chi) A
= (i A) \epsilon_E ((\alpha m_\chi a)^2/2) . \end{align}

Recall that our expectation is that the bound state will be dominated by the long-range physics in this case, with only a small perturbative contribution from the short-range physics (this is the case where the short-range physics is expected to be well-approximated by the tree-level amplitude). If we expand $A$ in the small-$a$ limit, we obtain:
\begin{align} A \epsilon_E = i \left[-2 \epsilon_E +  \ln(2 \epsilon_E \alpha a m_\chi) + \psi^{(0)}(-1/(2 \epsilon_E)) + 2 \gamma_E  \right] + \mathcal{O}(\alpha a m_\chi), \label{eq:Aapprox} \end{align}
where $\psi^{(0)}$ is the digamma function and $\gamma_E$ is the Euler-Mascheroni constant.

We observe that in order to obtain a bound state, we require $A \epsilon_E = -2 i/(\alpha a m_\chi)^2$, i.e. with magnitude much greater than 1. We see from Eq.~\eqref{eq:Aapprox} that this is only possible close to the pole in $\psi^{(0)}(-1/(2 \epsilon_E))$, which corresponds to the bound state in the pure Coulomb solution. Of course, if the short- and long-distance calculation were both done at all orders and the short-distance regulator matched the true UV theory, we would expect to exactly recover the usual Coulomb bound states independent of $a$, but because we are approximating the short-distance physics with a modified regulator, we expect $a$-dependent corrections to the bound state spectrum (and in practice there will also be $a$-dependent corrections sourced by our truncation of $A$ to the lowest order in $a$). Taking the pole term to dominate $A \epsilon_E$, the bound state condition becomes:
\begin{align} \psi^{(0)}(-1/(2\epsilon_E)) \approx - \frac{2 }{(\alpha m_\chi a)^2 }  \Rightarrow \epsilon_E \approx \frac{(1/2)}{n -  \frac{(\alpha m_\chi a)^2 }{2 }} \end{align}
Here we have approximated $\psi^{(0)}(-1/(2\epsilon_E)) \approx \frac{1}{-n + 1/(2\epsilon_E) } $ close to a pole, where $n$ is a positive integer.

This result suffices to demonstrate the convergence to the usual Coulombic bound states in the $a\rightarrow 0$ limit. However, when we checked this result numerically, we found that it did not accurately capture the behavior in the case where $a$ is small but non-zero, in particular the shift in the pole position was not correct (the shift using the full expressions was much smaller than we would infer from the approximate result above). This appears to be because of large $a$-independent factors associated with the pole which enhance the higher-order-in-$a$ terms in the expansion of $A$; we have checked explicitly that the leading-order-in-$a$ expansion is inaccurate (in particular it can have the wrong sign) for $\epsilon_E$ sufficiently close to the pole corresponding to the Coulombic bound state.

An alternative approach is to study the behavior in the vicinity of the Coulomb pole. Suppose we take $\epsilon_E=1/2$, corresponding to the $n=1$ pole (in the pure-Coulomb case). Then $A \rightarrow -\frac{2i}{e^{\alpha a m_\chi} (\alpha a m_\chi -1) +1} \approx-\frac{4i}{(\alpha a m_\chi)^2}$, with the latter approximate equality holding in the small-$a$ limit. We note that the divergence at the pole apparent in Eq.~\eqref{eq:Aapprox} is regulated by the non-zero value of $a$, and for small $a$ and this value of $\epsilon_E$, we find
\begin{align} i A \epsilon_E (\alpha m_\chi a)^2/2 \rightarrow -\frac{4i^2}{(\alpha a m_\chi)^2} (1/2)   ((\alpha m_\chi a)^2/2) = 1,\end{align}
which is the condition required for a bound state. We note that this only works because of the normalization of the spherical well; if we had taken a different value for $\overline{V}_0$ (corresponding to a different tree-level matching), it would directly modify $f$ and hence force $\epsilon_E A$ to a different value in order to satisfy the bound state condition. Thus at leading order the tree-level matching based on the scattering amplitude is working well, as expected.

Going beyond leading order, we can see that in order to satisfy the bound-state condition {\it exactly} at the same value of $\epsilon_E$, as $a$ varies, we would need $\alpha m_\chi f = 1/(i A \epsilon_E) = e^{\alpha a m_\chi}(\alpha a m_\chi-1) + 1 \approx \frac{(a\alpha m_\chi)^2}{2} + \frac{(a\alpha m_\chi)^3}{3} + \frac{(a\alpha m_\chi)^4}{8} + \cdots$. Computing the all-orders scattering amplitude for the Coulomb potential restricted to $r < a$, we obtain $\alpha m_\chi f \approx (\alpha a m_\chi)^2/2 + (\alpha a m_\chi)^3/3 + (11 - 12 \epsilon_E + 4 \epsilon_E^2) (\alpha a m_\chi)^4/48 + \cdots $. Interestingly, at the third order we see that it is not sufficient to take the $\epsilon_E\rightarrow 0$ limit (which is sufficient for the first two terms); we match the bound-state condition by evaluating the amplitude at the bound-state momentum $\epsilon_E=1/2$. For the spherical well regulator, however, the expansion already deviates at next-to-leading order; we find $\alpha m_\chi f \approx (\alpha a m_\chi)^2/2 + 3 (\alpha a m_\chi)^3/10$. Thus beyond leading order we expect that there will be a $a$-dependent shift in the binding energy associated with use of the spherical well regulator.

\end{fmffile}
\bibliographystyle{JHEP}
\bibliography{ref}
\end{document}

%% file: tikz_figs/pot_1.tex
\makeatletter\@ifundefined{DefineMPColorFromDvips}{\input{tikz_figs/dvips_mp_colors}}{}\makeatother
\DefineMPColorsFromDvipsList{mpRed/BrickRed,mpBlue/RoyalBlue,mpBlack/Black,mpOliveGreen/OliveGreen}%
\colorlet{BlobOliveFill}{OliveGreen!20!white}%
\colorlet{BlobOliveEdge}{OliveGreen!45!white}%
\DefineMPColorFromDvips{mpBlobFill}{BlobOliveFill}%
\DefineMPColorFromDvips{mpBlobEdge}{BlobOliveEdge}%
\colorlet{colvtx}{OliveGreen!85!black}%
\providecommand{\NreftQlabel}{\mbox{$l^\mu\!\sim\!(m_{\chi}v^2,m_{\chi}v)$}}%
\newcommand{\VblobLabel}{\mbox{$\bm{V}_{\rm{tree}}$}}%
\begin{align}
	\begin{gathered}
		\lim_{v\rightarrow 0}\ \qquad
		\parbox[c]{30mm}{\centering\resizebox{30mm}{!}{%
				\begin{fmfgraph*}(30,20)
					\fmfstraight
					\fmfleft{i1,i2}
					\fmfright{o1,o2}
					\fmf{fermion,f=mpRed}{i1,v1,o1}
					\fmf{fermion,f=mpRed}{i2,v2,o2}
					\fmffreeze				    \fmf{dashes,f=mpOliveGreen,l.s=right,l.d=6.5pt,label=\NreftQlabel, label.pos=right}{v1,v2}
				\end{fmfgraph*}%
		}}%
		\qquad\qquad\qquad  = \qquad\qquad
		\parbox[c]{30mm}{\centering\resizebox{30mm}{!}{%
				\begin{fmfgraph*}(30,20)
					\fmfstraight
					\fmfleft{i1,i2}
					\fmfright{o1,o2}
					\fmf{phantom}{i1,v,o1}
					\fmf{phantom}{i2,v,o2}
					\fmffreeze
					\fmf{fermion,f=mpRed}{i1,v}
					\fmf{fermion,f=mpRed}{i2,v}
					\fmf{fermion,f=mpRed}{v,o1}
					\fmf{fermion,f=mpRed}{v,o2}
					\fmfv{decor.shape=circle,decor.filled=empty,decor.size=27,
						f=mpBlobEdge,b=mpBlobFill,l=\color{colvtx}\VblobLabel,l.a=0,l.d=0}{v}
				\end{fmfgraph*}%
		}}%
		\sim
		\frac{g^2}{v^2}\ .
	\end{gathered} \label{eq:pot_diag}
\end{align}

%% file: tikz_figs/dvips_mp_colors.tex
%
%
%
\makeatletter
\newcommand{\MPrgb@dvips}{}
\newcommand{\DefineMPColorFromDvips}[2]{%
  \convertcolorspec{named}{#2}{rgb}\MPrgb@dvips
  \edef\@temp{\noexpand\fmfcmd{color #1; #1 := (\MPrgb@dvips);}}%
  \@temp
}
\newcommand{\DefineMPColorsFromDvipsList}[1]{%
  \@for\next:=#1\do{%
    \expandafter\DefineMPColorFromDvips@pair\next\@@
  }%
}
\def\DefineMPColorFromDvips@pair#1/#2\@@{%
  \DefineMPColorFromDvips{#1}{#2}%
}
\makeatother

%% file: tikz_figs/soft_1.tex
\makeatletter
\@ifundefined{DefineMPColorFromDvips}{%
  \newcommand{\MPrgb@dvips}{}%
  \newcommand{\DefineMPColorFromDvips}[2]{%
    \convertcolorspec{named}{#2}{rgb}\MPrgb@dvips
    \edef\@temp{\noexpand\fmfcmd{color #1; #1 := (\MPrgb@dvips);}}%
    \@temp
  }%
  \newcommand{\DefineMPColorsFromDvipsList}[1]{%
    \@for\next:=#1\do{%
      \expandafter\DefineMPColorFromDvips@pair\next\@@
    }%
  }%
  \def\DefineMPColorFromDvips@pair#1/#2\@@{%
    \DefineMPColorFromDvips{#1}{#2}%
  }%
}{}%
\makeatother
\DefineMPColorsFromDvipsList{mpRed/BrickRed,mpOliveGreen/OliveGreen,mpBlack/Black}%
%
\providecommand{\LabEll}{\mbox{$l^\mu\!\sim\!m_{\chi}(v,v)$}}%
\providecommand{\LabEllp}{\mbox{$l'{}^\mu\!\sim\!m_{\chi}(v,v)$}}%
%
\begin{equation} \label{eq:comp_scalar_nreft}
\scalebox{0.73}{$\displaystyle
	\begin{gathered}
		\lim_{v\rightarrow 0}\qquad 
		\parbox[c]{42mm}{\centering\resizebox{42mm}{!}{%
				\begin{fmfgraph*}(54,18)
					\fmfstraight
					\fmftop{tL,tR}
					\fmfbottom{bL,bR}
					\fmf{phantom,tension=1}{bL,v1,v2,bR}
					\fmffreeze
					\fmf{phantom,tension=1}{tL,v1}
					\fmf{phantom,tension=1}{v2,tR}
					\fmffreeze
					\fmf{fermion,f=mpRed,tension=1.5}{bL,v1}
					\fmf{fermion,f=mpRed,tension=0.22,label=\LabEll,label.side=bottom}{v1,v2}
					\fmf{fermion,f=mpRed,tension=1.5}{v2,bR}
					\fmf{dashes,f=mpOliveGreen,tension=1}{v1,tL}
					\fmf{dashes,f=mpOliveGreen,tension=1}{v2,tR}
					\fmflabel{$q$}{tL}
					\fmflabel{$q'$}{tR}
				\end{fmfgraph*}%
		}}%
		\qquad\;+\;\qquad
		\parbox[c]{42mm}{\centering\resizebox{42mm}{!}{%
				\begin{fmfgraph*}(54,18)
					\fmfstraight
					\fmftop{tL,tR}
					\fmfbottom{bL,bR}
					\fmf{phantom,tension=1}{bL,v1,v2,bR}
					\fmffreeze
					\fmf{phantom,tension=1}{tL,v2}
					\fmf{phantom,tension=1}{v1,tR}
					\fmffreeze
					\fmf{fermion,f=mpRed,tension=1.5}{bL,v1}
					\fmf{fermion,f=mpRed,tension=0.22,label=\LabEllp,label.side=bottom}{v1,v2}
					\fmf{fermion,f=mpRed,tension=1.5}{v2,bR}
					\fmf{dashes,f=mpOliveGreen,tension=1}{v2,tL}
					\fmf{dashes,f=mpOliveGreen,tension=1}{v1,tR}
					\fmflabel{$q$}{tL}
					\fmflabel{$q'$}{tR}
				\end{fmfgraph*}%
		}}%
		\,
		\qquad \;=\; \qquad
		\parbox[c]{24mm}{\centering\resizebox{24mm}{!}{%
				\begin{fmfgraph*}(26,22)
					\fmfstraight
					\fmftop{tL,tR}
					\fmfbottom{bL,bR}
					\fmf{phantom}{tL,v,tR}
					\fmf{phantom}{bL,v,bR}
					\fmffreeze
					\fmf{fermion,f=mpRed}{bL,v}
					\fmf{fermion,f=mpRed}{v,bR}
					\fmf{dashes,f=mpOliveGreen}{tL,v}
					\fmf{dashes,f=mpOliveGreen}{v,tR}
                    \fmflabel{$q$}{tL}
					\fmflabel{$q'$}{tR}
				\end{fmfgraph*}%
		}}%
		\sim\  \frac{g^2}{v}
	\end{gathered}
$}
\end{equation}

%% file: tikz_figs/ladder_chain_snippet.tex
%
%
%
\makeatletter
\@ifundefined{DefineMPColorFromDvips}{%
  \newcommand{\MPrgb@dvips}{}%
  \newcommand{\DefineMPColorFromDvips}[2]{%
    \convertcolorspec{named}{#2}{rgb}\MPrgb@dvips
    \edef\@temp{\noexpand\fmfcmd{color #1; #1 := (\MPrgb@dvips);}}%
    \@temp
  }%
  \newcommand{\DefineMPColorsFromDvipsList}[1]{%
    \@for\next:=#1\do{%
      \expandafter\DefineMPColorFromDvips@pair\next\@@
    }%
  }%
  \def\DefineMPColorFromDvips@pair#1/#2\@@{%
    \DefineMPColorFromDvips{#1}{#2}%
  }%
}{}%
\makeatother
\DefineMPColorsFromDvipsList{mpRed/BrickRed,mpOliveGreen/OliveGreen}%
\colorlet{BlobOliveFill}{OliveGreen!20!white}%
\colorlet{BlobOliveEdge}{OliveGreen!45!white}%
\DefineMPColorFromDvips{mpBlobFill}{BlobOliveFill}%
\DefineMPColorFromDvips{mpBlobEdge}{BlobOliveEdge}%
\colorlet{colvtx}{OliveGreen!85!black}%

\newcommand{\VtreeBlobLabel}{\mbox{$\bm{V}_{\text{tree}}$}}%
\newcommand{\fmfVTblob}[1]{%
  \fmfv{decor.shape=circle,decor.filled=empty,decor.size=26,%
    f=mpBlobEdge,b=mpBlobFill,%
    l=\color{colvtx}\VtreeBlobLabel,l.a=0,l.d=0}{#1}}%
\begin{equation}
\scalebox{0.92}{$\displaystyle
\begin{aligned}
\lim_{v\rightarrow 0}\qquad
\parbox[c]{26mm}{\centering\resizebox{26mm}{!}{%
  \begin{fmfgraph*}(30,14)
    \fmfstraight
    \fmfleft{i1,i2}\fmfright{o1,o2}
    \fmf{phantom}{i1,v,o1}
    \fmf{phantom}{i2,w,o2}
    \fmffreeze
    \fmf{fermion,f=mpRed}{i1,v,o1}
    \fmf{fermion,f=mpRed}{i2,w,o2}
    \fmf{dashes,f=mpOliveGreen}{v,w}
  \end{fmfgraph*}%
}}
\;+\;
\parbox[c]{32mm}{\centering\resizebox{32mm}{!}{%
  \begin{fmfgraph*}(38,14)
    \fmfstraight
    \fmfleft{i1,i2}\fmfright{o1,o2}
    \fmf{phantom}{i1,v1,v2,o1}
    \fmf{phantom}{i2,w1,w2,o2}
    \fmffreeze
    \fmf{fermion,f=mpRed}{i1,v1,v2,o1}
    \fmf{fermion,f=mpRed}{i2,w1,w2,o2}
    \fmf{dashes,f=mpOliveGreen}{v1,w1}
    \fmf{dashes,f=mpOliveGreen}{v2,w2}
  \end{fmfgraph*}%
}}
\;+\;
\parbox[c]{38mm}{\centering\resizebox{38mm}{!}{%
  \begin{fmfgraph*}(46,14)
    \fmfstraight
    \fmfleft{i1,i2}\fmfright{o1,o2}
    \fmf{phantom}{i1,v1,v2,v3,o1}
    \fmf{phantom}{i2,w1,w2,w3,o2}
    \fmffreeze
    \fmf{fermion,f=mpRed}{i1,v1,v2,v3,o1}
    \fmf{fermion,f=mpRed}{i2,w1,w2,w3,o2}
    \fmf{dashes,f=mpOliveGreen}{v1,w1}
    \fmf{dashes,f=mpOliveGreen}{v2,w2}
    \fmf{dashes,f=mpOliveGreen}{v3,w3}
  \end{fmfgraph*}%
}}
\;+\;\cdots\;
\\[1.5ex]
{}=\;
\parbox[c]{26mm}{\centering\resizebox{26mm}{!}{%
  \begin{fmfgraph*}(28,24)
    \fmfstraight
    \fmfleft{i1,i2}\fmfright{o1,o2}
    \fmf{phantom}{i1,v,o1}
    \fmf{phantom}{i2,v,o2}
    \fmffreeze
    \fmf{fermion,f=mpRed}{i1,v}
    \fmf{fermion,f=mpRed}{i2,v}
    \fmf{fermion,f=mpRed}{v,o1}
    \fmf{fermion,f=mpRed}{v,o2}
    \fmfVTblob{v}
  \end{fmfgraph*}%
}}
\;+\;
\parbox[c]{34mm}{\centering\resizebox{34mm}{!}{%
  \begin{fmfgraph*}(36,24)
    \fmfstraight
    \fmfleft{i1,i2}\fmfright{o1,o2}
    \fmf{fermion,f=mpRed}{i1,v1}
    \fmf{fermion,f=mpRed}{i2,v1}
    \fmf{fermion,f=mpRed,left,tension=0.5}{v1,v2}
    \fmf{fermion,f=mpRed,right,tension=0.5}{v1,v2}
    \fmf{fermion,f=mpRed}{v2,o1}
    \fmf{fermion,f=mpRed}{v2,o2}
    \fmfVTblob{v1}
    \fmfVTblob{v2}
  \end{fmfgraph*}%
}}
\;+\;
\parbox[c]{46mm}{\centering\resizebox{46mm}{!}{%
  \begin{fmfgraph*}(54,30)
    \fmfstraight
    \fmfleft{i1,i2}\fmfright{o1,o2}
    \fmf{fermion,f=mpRed}{i1,v1}
    \fmf{fermion,f=mpRed}{i2,v1}
    \fmf{fermion,f=mpRed,left,tension=0.42}{v1,v2}
    \fmf{fermion,f=mpRed,right,tension=0.42}{v1,v2}
    \fmf{fermion,f=mpRed,left,tension=0.42}{v2,v3}
    \fmf{fermion,f=mpRed,right,tension=0.42}{v2,v3}
    \fmf{fermion,f=mpRed}{v3,o1}
    \fmf{fermion,f=mpRed}{v3,o2}
    \fmfVTblob{v1}
    \fmfVTblob{v2}
    \fmfVTblob{v3}
  \end{fmfgraph*}%
}}
\;+\;\cdots
\end{aligned}$}
\end{equation}

%% file: tikz_figs/crossed_exchange_snippet.tex
%
%
%
\makeatletter\@ifundefined{DefineMPColorFromDvips}{\input{dvips_mp_colors}}{}\makeatother
\DefineMPColorsFromDvipsList{mpRed/BrickRed,mpOliveGreen/OliveGreen}%
\begin{equation}
\lim_{v\rightarrow 0}\qquad \;
\parbox[c]{42mm}{\centering\resizebox{42mm}{!}{%
  \begin{fmfgraph*}(50,24)
    \fmfstraight
    \fmfleft{i1,i2}\fmfright{o1,o2}
    \fmf{phantom}{i1,v1,v2,o1}
    \fmf{phantom}{i2,w1,w2,o2}
    \fmffreeze
    \fmf{fermion,f=mpRed}{i1,v1,v2,o1}
    \fmf{fermion,f=mpRed}{i2,w1,w2,o2}
    \fmf{dashes,f=mpOliveGreen}{v1,w2}
    \fmf{dashes,f=mpOliveGreen}{v2,w1}
  \end{fmfgraph*}%
}}
\end{equation}

%% file: tikz_figs/tree_amp_snippet.tex
%
%
\makeatletter\@ifundefined{DefineMPColorFromDvips}{\input{dvips_mp_colors}}{}\makeatother
\DefineMPColorsFromDvipsList{mpRed/BrickRed,mpPurple/RoyalPurple}%
\providecommand{\NreftQlabel}{\mbox{$l^\mu\!\sim\!(mv^2,mv)$}}
\begin{equation}\label{eq:tree_diagram_ps}
i\mathcal{A}_{\text{tree}} \qquad
= \qquad \lim_{v\rightarrow 0}\qquad
\parbox[c]{35mm}{\centering\resizebox{35mm}{!}{%
  \begin{fmfgraph*}(35,25)
    \fmfstraight
    \fmfleft{i1,i2}\fmfright{o1,o2}
    \fmf{phantom}{i1,v,o1}
    \fmf{phantom}{i2,w,o2}
    \fmffreeze
    \fmf{fermion,f=mpRed}{i1,v,o1}
    \fmf{fermion,f=mpRed}{i2,w,o2}
    \fmf{dashes,f=mpPurple, label=\NreftQlabel}{v,w}
  \end{fmfgraph*}%
}}
\end{equation}

%% file: tikz_figs/ladder_chain_ps.tex
%
%
%
\makeatletter
\@ifundefined{DefineMPColorFromDvips}{%
  \newcommand{\MPrgb@dvips}{}%
  \newcommand{\DefineMPColorFromDvips}[2]{%
    \convertcolorspec{named}{#2}{rgb}\MPrgb@dvips
    \edef\@temp{\noexpand\fmfcmd{color #1; #1 := (\MPrgb@dvips);}}%
    \@temp
  }%
  \newcommand{\DefineMPColorsFromDvipsList}[1]{%
    \@for\next:=#1\do{%
      \expandafter\DefineMPColorFromDvips@pair\next\@@
    }%
  }%
  \def\DefineMPColorFromDvips@pair#1/#2\@@{%
    \DefineMPColorFromDvips{#1}{#2}%
  }%
}{}%
\makeatother
\DefineMPColorsFromDvipsList{mpRed/BrickRed,mpRoyalPurple/RoyalPurple}%
\colorlet{BlobRoyalPurple}{RoyalPurple!20!white}%
\colorlet{BlobRoyalPurpleEdge}{RoyalPurple!45!white}%
\DefineMPColorFromDvips{mpBlobFill}{BlobRoyalPurple}%
\DefineMPColorFromDvips{mpBlobEdge}{BlobRoyalPurpleEdge}%
\colorlet{colvtx}{RoyalPurple!85!black}%

\begin{equation}\label{eq:ladder_ps}
\scalebox{0.92}{$\displaystyle
\begin{aligned}
\lim_{v\rightarrow 0}\qquad
\parbox[c]{26mm}{\centering\resizebox{26mm}{!}{%
  \begin{fmfgraph*}(30,14)
    \fmfstraight
    \fmfleft{i1,i2}\fmfright{o1,o2}
    \fmf{phantom}{i1,v,o1}
    \fmf{phantom}{i2,w,o2}
    \fmffreeze
    \fmf{fermion,f=mpRed}{i1,v,o1}
    \fmf{fermion,f=mpRed}{i2,w,o2}
    \fmf{dashes,f=mpRoyalPurple}{v,w}
  \end{fmfgraph*}%
}}
\;+\;
\parbox[c]{32mm}{\centering\resizebox{32mm}{!}{%
  \begin{fmfgraph*}(38,14)
    \fmfstraight
    \fmfleft{i1,i2}\fmfright{o1,o2}
    \fmf{phantom}{i1,v1,v2,o1}
    \fmf{phantom}{i2,w1,w2,o2}
    \fmffreeze
    \fmf{fermion,f=mpRed}{i1,v1,v2,o1}
    \fmf{fermion,f=mpRed}{i2,w1,w2,o2}
    \fmf{dashes,f=mpRoyalPurple}{v1,w1}
    \fmf{dashes,f=mpRoyalPurple}{v2,w2}
  \end{fmfgraph*}%
}}
\;+\;
\parbox[c]{38mm}{\centering\resizebox{38mm}{!}{%
  \begin{fmfgraph*}(46,14)
    \fmfstraight
    \fmfleft{i1,i2}\fmfright{o1,o2}
    \fmf{phantom}{i1,v1,v2,v3,o1}
    \fmf{phantom}{i2,w1,w2,w3,o2}
    \fmffreeze
    \fmf{fermion,f=mpRed}{i1,v1,v2,v3,o1}
    \fmf{fermion,f=mpRed}{i2,w1,w2,w3,o2}
    \fmf{dashes,f=mpRoyalPurple}{v1,w1}
    \fmf{dashes,f=mpRoyalPurple}{v2,w2}
    \fmf{dashes,f=mpRoyalPurple}{v3,w3}
  \end{fmfgraph*}%
}}
\;+\;\cdots\;
\\[1.5ex]
{}=\;
\parbox[c]{26mm}{\centering\resizebox{26mm}{!}{%
  \begin{fmfgraph*}(28,24)
    \fmfstraight
    \fmfleft{i1,i2}\fmfright{o1,o2}
    \fmf{phantom}{i1,v,o1}
    \fmf{phantom}{i2,v,o2}
    \fmffreeze
    \fmf{fermion,f=mpRed}{i1,v}
    \fmf{fermion,f=mpRed}{i2,v}
    \fmf{fermion,f=mpRed}{v,o1}
    \fmf{fermion,f=mpRed}{v,o2}
    \fmfVTblob{v}
  \end{fmfgraph*}%
}}
\;+\;
\parbox[c]{34mm}{\centering\resizebox{34mm}{!}{%
  \begin{fmfgraph*}(36,24)
    \fmfstraight
    \fmfleft{i1,i2}\fmfright{o1,o2}
    \fmf{fermion,f=mpRed}{i1,v1}
    \fmf{fermion,f=mpRed}{i2,v1}
    \fmf{fermion,f=mpRed,left,tension=0.5}{v1,v2}
    \fmf{fermion,f=mpRed,right,tension=0.5}{v1,v2}
    \fmf{fermion,f=mpRed}{v2,o1}
    \fmf{fermion,f=mpRed}{v2,o2}
    \fmfVTblob{v1}
    \fmfVTblob{v2}
  \end{fmfgraph*}%
}}
\;+\;
\parbox[c]{46mm}{\centering\resizebox{46mm}{!}{%
  \begin{fmfgraph*}(54,30)
    \fmfstraight
    \fmfleft{i1,i2}\fmfright{o1,o2}
    \fmf{fermion,f=mpRed}{i1,v1}
    \fmf{fermion,f=mpRed}{i2,v1}
    \fmf{fermion,f=mpRed,left,tension=0.42}{v1,v2}
    \fmf{fermion,f=mpRed,right,tension=0.42}{v1,v2}
    \fmf{fermion,f=mpRed,left,tension=0.42}{v2,v3}
    \fmf{fermion,f=mpRed,right,tension=0.42}{v2,v3}
    \fmf{fermion,f=mpRed}{v3,o1}
    \fmf{fermion,f=mpRed}{v3,o2}
    \fmfVTblob{v1}
    \fmfVTblob{v2}
    \fmfVTblob{v3}
  \end{fmfgraph*}%
}}
\;+\;\cdots
\end{aligned}$}
\end{equation}

%% file: tikz_figs/soft_2.tex
\makeatletter
\@ifundefined{DefineMPColorFromDvips}{%
  \newcommand{\MPrgb@dvips}{}%
  \newcommand{\DefineMPColorFromDvips}[2]{%
    \convertcolorspec{named}{#2}{rgb}\MPrgb@dvips
    \edef\@temp{\noexpand\fmfcmd{color #1; #1 := (\MPrgb@dvips);}}%
    \@temp
  }%
  \newcommand{\DefineMPColorsFromDvipsList}[1]{%
    \@for\next:=#1\do{%
      \expandafter\DefineMPColorFromDvips@pair\next\@@
    }%
  }%
  \def\DefineMPColorFromDvips@pair#1/#2\@@{%
    \DefineMPColorFromDvips{#1}{#2}%
  }%
}{}%
\makeatother
\DefineMPColorsFromDvipsList{mpRed/BrickRed,mpRoyalPurple/RoyalPurple,mpBlack/Black}%
%
\providecommand{\LabEll}{\mbox{$l^\mu\!\sim\!m(v,v)$}}%
\providecommand{\LabEllp}{\mbox{$l'{}^\mu\!\sim\!m(v,v)$}}%
%
\begin{equation} \label{eq:comp_ps_amp}
i\cu{A}_{\rm s\chi s\chi} =\qquad
\scalebox{0.73}{$\displaystyle
	\begin{gathered}
		\lim_{v\rightarrow 0}\qquad 
		\parbox[c]{42mm}{\centering\resizebox{42mm}{!}{%
				\begin{fmfgraph*}(54,18)
					\fmfstraight
					\fmftop{tL,tR}
					\fmfbottom{bL,bR}
					\fmf{phantom,tension=1}{bL,v1,v2,bR}
					\fmffreeze
					\fmf{phantom,tension=1}{tL,v1}
					\fmf{phantom,tension=1}{v2,tR}
					\fmffreeze
					\fmf{fermion,f=mpRed,tension=1.5}{bL,v1}
					\fmf{fermion,f=mpRed,tension=0.22,label=\LabEll,label.side=bottom}{v1,v2}
					\fmf{fermion,f=mpRed,tension=1.5}{v2,bR}
					\fmf{dashes,f=mpRoyalPurple,tension=1}{v1,tL}
					\fmf{dashes,f=mpRoyalPurple,tension=1}{v2,tR}
					\fmflabel{$q$}{tL}
					\fmflabel{$q'$}{tR}
				\end{fmfgraph*}%
		}}%
		\qquad\;+\;\qquad
		\parbox[c]{42mm}{\centering\resizebox{42mm}{!}{%
				\begin{fmfgraph*}(54,18)
					\fmfstraight
					\fmftop{tL,tR}
					\fmfbottom{bL,bR}
					\fmf{phantom,tension=1}{bL,v1,v2,bR}
					\fmffreeze
					\fmf{phantom,tension=1}{tL,v2}
					\fmf{phantom,tension=1}{v1,tR}
					\fmffreeze
					\fmf{fermion,f=mpRed,tension=1.5}{bL,v1}
					\fmf{fermion,f=mpRed,tension=0.22,label=\LabEllp,label.side=bottom}{v1,v2}
					\fmf{fermion,f=mpRed,tension=1.5}{v2,bR}
					\fmf{dashes,f=mpRoyalPurple,tension=1}{v2,tL}
					\fmf{dashes,f=mpRoyalPurple,tension=1}{v1,tR}
					\fmflabel{$q$}{tL}
					\fmflabel{$q'$}{tR}
				\end{fmfgraph*}%
		}}%
		\,
		\qquad \;=\; \qquad
		\parbox[c]{24mm}{\centering\resizebox{24mm}{!}{%
				\begin{fmfgraph*}(26,22)
					\fmfstraight
					\fmftop{tL,tR}
					\fmfbottom{bL,bR}
					\fmf{phantom}{tL,v,tR}
					\fmf{phantom}{bL,v,bR}
					\fmffreeze
					\fmf{fermion,f=mpRed}{bL,v}
					\fmf{fermion,f=mpRed}{v,bR}
					\fmf{dashes,f=mpRoyalPurple}{tL,v}
					\fmf{dashes,f=mpRoyalPurple}{v,tR}
                    \fmflabel{$q$}{tL}
					\fmflabel{$q'$}{tR}
				\end{fmfgraph*}%
		}}%
	\end{gathered}
$}
\end{equation}

%% file: tikz_figs/double_dash_bubble_snippet.tex
%
%
%
\makeatletter\@ifundefined{DefineMPColorFromDvips}{\input{dvips_mp_colors}}{}\makeatother
\DefineMPColorsFromDvipsList{mpRed/BrickRed,mpRoyalPurple/RoyalPurple}%
\begin{equation}\label{eq:soft_loop}
\parbox[c]{28mm}{\centering\resizebox{28mm}{!}{%
  \begin{fmfgraph*}(32,18)
    \providecommand{\DoubleDashBubbleLsr}{0.58}%
    \edef\DoubleDashBubbleOptsL{dashes,foreground=mpRoyalPurple,left=\DoubleDashBubbleLsr}%
    \edef\DoubleDashBubbleOptsR{dashes,foreground=mpRoyalPurple,right=\DoubleDashBubbleLsr}%
    \fmfstraight
    \fmfleft{i1,i2}\fmfright{o1,o2}
    \fmf{phantom}{i1,v,o1}
    \fmf{phantom}{i2,w,o2}
    \fmffreeze
    \fmf{fermion,foreground=mpRed}{i1,v,o1}
    \fmf{fermion,foreground=mpRed}{i2,w,o2}
    \expandafter\fmf\expandafter{\DoubleDashBubbleOptsL}{v,w}
    \expandafter\fmf\expandafter{\DoubleDashBubbleOptsR}{v,w}
  \end{fmfgraph*}%
}}
\end{equation}

%% file: tikz_figs/tree_amp_pg.tex
%
%
\makeatletter\@ifundefined{DefineMPColorFromDvips}{\input{dvips_mp_colors}}{}\makeatother
\DefineMPColorsFromDvipsList{mpRed/BrickRed,mpPurple/RoyalPurple}%
\providecommand{\NreftQlabel}{\mbox{$l^\mu\!\sim\!(mv^2,mv)$}}
\begin{equation}\label{eq:tree_diagram_pg}
i\mathcal{A}_{\text{tree}} \qquad
= \qquad \lim_{v\rightarrow 0}\qquad
\parbox[c]{35mm}{\centering\resizebox{35mm}{!}{%
  \begin{fmfgraph*}(35,25)
    \fmfstraight
    \fmfleft{i1,i2}\fmfright{o1,o2}
    \fmf{phantom}{i1,v,o1}
    \fmf{phantom}{i2,w,o2}
    \fmffreeze
    \fmf{fermion,f=mpRed}{i1,v,o1}
    \fmf{fermion,f=mpRed}{i2,w,o2}
    \fmf{dashes,f=mpPurple, label=\NreftQlabel}{v,w}
  \end{fmfgraph*}%
}}
\end{equation}